\documentclass[aip,jcp,amsmath,amssymb,reprint,superscriptaddress]{revtex4-2}

\usepackage[dvipdfmx]{graphicx}
\usepackage{color}
\usepackage{epsf}
\usepackage{bm}
\usepackage{lipsum}
\usepackage[normalem]{ulem}
\usepackage{txfonts}

\usepackage{physics}
\usepackage{float}
\usepackage{booktabs}
\usepackage{multirow}
\newcommand{\ow}{$\mathrm{O}^\mathrm{w}$}

\newcommand{\oxy}[1]{$\mathrm{O}^\mathrm{#1}$}

\newcommand{\mr}[1]{\mathrm{#1}}
\newcommand{\hb}{_\mathrm{HB}}
\newcommand{\phb}{P_\mathrm{HB}}
\newcommand{\kww}{_\mathrm{KWW}}
\newcommand{\thb}{\tau_\mathrm{HB}}

\newcommand{\br}{\bm{r}}

\newcommand{\fig}[1]{Fig.~\ref{fig:#1}}

\usepackage{ulem}
\usepackage{xcolor}
\DeclareRobustCommand{\erase}{\bgroup\markoverwith{\textcolor{red}{\rule[.5ex]{2pt}{1pt}}}\ULon}

\graphicspath{{./}{./fig/}{../fig/}}

\begin{document}
\title{Influence of cholesterol on hydrogen-bond dynamics of water molecules in lipid-bilayer systems at varying temperatures}

\author{Kokoro Shikata}
\affiliation{Division of Chemical Engineering, Department of Materials Engineering Science, Graduate School of Engineering Science, Osaka University, Toyonaka, Osaka 560-8531, Japan}

\author{Kento Kasahara}
\email{kasahara@cheng.es.osaka-u.ac.jp}
\affiliation{Division of Chemical Engineering, Department of Materials Engineering Science, Graduate School of Engineering Science, Osaka University, Toyonaka, Osaka 560-8531, Japan}

\author{Nozomi Morishita Watanabe}
\affiliation{Division of Chemical Engineering, Department of Materials Engineering Science, Graduate School of Engineering Science, Osaka University, Toyonaka, Osaka 560-8531, Japan}

\author{Hiroshi Umakoshi}
\affiliation{Division of Chemical Engineering, Department of Materials Engineering Science, Graduate School of Engineering Science, Osaka University, Toyonaka, Osaka 560-8531, Japan}

\author{Kang Kim}
\email{kk@cheng.es.osaka-u.ac.jp}
\affiliation{Division of Chemical Engineering, Department of Materials Engineering Science, Graduate School of Engineering Science, Osaka University, Toyonaka, Osaka 560-8531, Japan}

\author{Nobuyuki Matubayasi}
\email{nobuyuki@cheng.es.osaka-u.ac.jp}
\affiliation{Division of Chemical Engineering, Department of Materials Engineering Science, Graduate School of Engineering Science, Osaka University, Toyonaka, Osaka 560-8531, Japan}

\date{\today}

\begin{abstract}
Cholesterol (Chol) plays a crucial role in shaping the intricate
 physicochemical attributes of biomembranes, exerting considerable
 influence on water molecules proximal to the membrane interface.
In this study, we conducted molecular dynamics simulations on the bilayers of two lipid species,
 dipalmitoyl phosphatidylcholine (DPPC) and palmitoyl sphingomyelin (PSM); 
they are distinct with respect to the structures of the hydrogen-bond (H-bond) acceptors.
Our investigation focuses on the dynamic properties and H-bonds of water
 molecules in the lipid-membrane systems, with particular emphasis on the influence
 of Chol at varying temperatures.
Notably, in the gel phase at 303 K, the presence of Chol
 extends the lifetimes of H-bonds of the  oxygen atoms acting
 as H-bond acceptors within DPPC with
 water molecules by a factor of 1.5 to 2.5. 
In the liquid-crystalline phase at 323 K, on the other hand, H-bonding
 dynamics with lipid membranes
 remain largely unaffected by Chol.
This observed shift in H-bonding states serves as a crucial key
 to unraveling the subtle control mechanisms governing water dynamics in lipid-membrane systems. 
\end{abstract}
\maketitle

\section{Introduction}

Lipid bilayers, serving as the fundamental architectural frameworks of
biological membranes, form stable aggregates
through the
amphiphilic effect inherent to lipid molecules. 
The attributes of lipid bilayers 
vary diversely with
the specific type and composition of lipid molecules, giving rise to
distinctive structures, such as gel and liquid-crystalline
phases.~\cite{nagle2000Structure}
The interactions with water are also a key to
self-organization,\cite{israelachvili2011Intermolecular} and water
properties are affected by the states of lipid molecules in turn.

Cholesterol (Chol), extensively investigated as a constituent
of bio-related membranes,
exhibits significant influence on 
structures of lipid bilayers.~\cite{mouritsen2004What, demeyer2009Effect}
In particular, Chol enhances the packing density and
rigidity of the lipid, thereby modulating membrane fluidity.
Furthermore, 
water molecules play a crucial role in the structure and function of biological
membranes, influencing electrostatic properties, solute
exchange, and protein function.~\cite{marrink1996Membranes,
pratt2002Hydrophobic, higgins2006Structured, raschke2006Water, 
ziegler2008Interface, disalvo2008Structural, cheng2013Hydration, zhou2013Influences,
disalvo2015Membrane, jungwirth2015Biological, laage2017Water, chattopadhyay2021Hydration}
Thus, it is imperative to elucidate the structure and dynamics of water molecules at the
membrane interface, which is expected to differ from those in the bulk 
owing to the interaction with hydrophilic groups on the lipid
head.

Molecular dynamics (MD) simulation is a valuable tool for investigating
lipid bilayers, providing 
molecular-level insights into 
not only lipid properties but also their interactions with other molecules.~
\cite{berkowitz1991Computer, pastor1994Molecular,
marrink1994Simulation, zhou1995Molecular, jakobsson1997Computer, pandit2003Algorithm, 
berkowitz2006Aqueous, matubayasi2008Freeenergy,
marrink2019Computational, karathanou2022Algorithm}
Numerous investigations have also been conducted for water in 
the interface region, 
encompassing  
the distribution of water molecules,
reorientation dynamics, mean square displacement, and hydrogen-bond (H-bond)
dynamics.~\cite{alper1993Limiting, pasenkiewicz-gierula1997Hydrogen,
feller2000Molecular, 
lopez2004Hydrogen, 
bhide2005Structure, volkov2006Hydration, vonhansen2013Anomalous, 
srivastava2018Hydrationa, calero2019Membranes, lee2019Watera,
an2021Interfacial, 
higuchi2021Rotational, malik2021Dehydration, malik2023Relaxation}
Specifically, the slowdown of the H-bond dynamics from the
bulk to the center of the membrane has been demonstrated.~\cite{malik2021Dehydration, malik2023Relaxation}
In addition, 
MD simulations have been used to study the structure and dynamics of
lipid bilayers containing phospholipids and
Chol.~\cite{tu1998ConstantPressure, chiu2002CholesterolInduced,
hofsass2003Molecular, pandit2004Complexation, alwarawrah2010Molecular, 
saito2011Cholesterol, sodt2015Hexagonal, boughter2016Influence, elola2018Influence, pantelopulos2018Regimes,
paslack2019Hydrationmediated, kumari2019Countereffects,
elkins2021Direct, antila2022Rotational}
These simulations have provided insights into the influence of
Chol on a variety of phospholipids differing in the headgroup and
tail.

Despite the numerous studies mentioned above, comprehending the water state proximal to 
lipid membranes, 
as well as understanding its connection with the membrane state in the presence of
Chol, remains a significant problem.
Interestingly, experimental observations have indicated that 
Chol was found to accelerate the water dynamics 
in the dipalmitoyl phosphatidylcholine (DPPC) membrane
interface.~\cite{cheng2014Cholesterol, pyne2022Addition}
Conversely, it has been found that water dynamics decelerate within the
interior region of lipid
bilayers with increasing Chol concentration.~\cite{cheng2014Cholesterol}

MD simulations have elucidated that 
the acceleration of water dynamics at the interface, particularly notable at high Chol
concentrations up to 50\%, arises from
the inhibition of H-bonds between two oxygen atoms of lipid molecules.~\cite{elola2018Influence}
A more recent MD study conducted a detailed analysis of the
H-bond network of water within the DPPC membrane in the
presence of Chol. 
The results unveiled that 
Chol fosters more bulk-like water at the membrane interface,
leading to increased local water density and accelerated water dynamics.~\cite{oh2020Effect}

In this study, we 
conducted MD simulations of two types of lipid bilayers comprising of
DPPC and 
palmitoyl sphingomyelin (PSM) with 
the presence of Chol.
While the Chol concentrations investigated were 0 and 
10\%, the temperature effect was examined at 303 K and 323 K.
The DPPC and PSM membranes are at the gel and liquid-crystalline phases at 303 K and 323 K, respectively.
We investigate the microscopic hydration structure and dynamics 
by considering 
acceptor sites of lipid molecules, which form H-bonds with the hydrogen atoms
of water molecules.
Between DPPC and PSM, the choline and phosphate groups are identical, as shown in \fig{snap}. 
There are differences in the degree of carbon chain saturation and
the functional group acting as the H-bonding site.
Thus, our MD investigations provide insights into H-bonds
influenced by 
Chol, taking into account the molecular structures of the lipids and 
the environmental effects from the membrane composition and temperature.

\section{Simulation details}

\begin{figure}[t]
  \centering
  \includegraphics[width=1.0\linewidth]{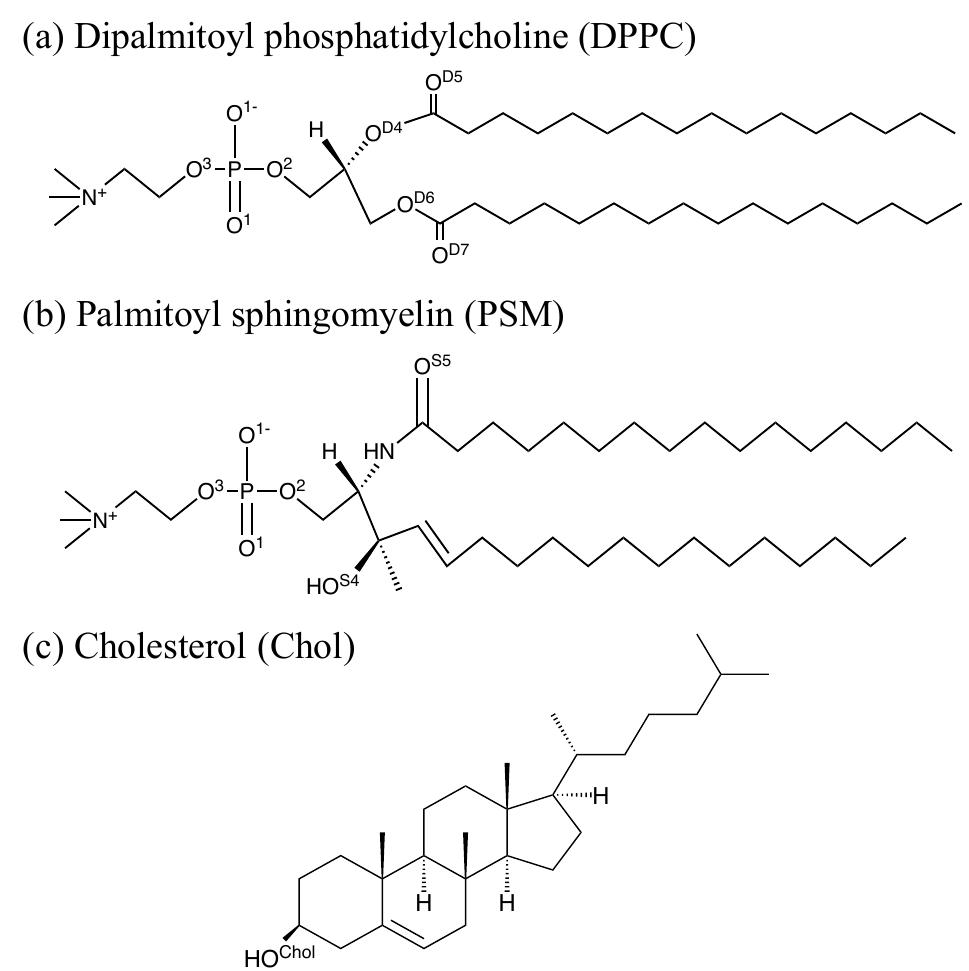}
  \caption{Structures of the lipid and Chol molecules studied in this paper.}
  \label{fig:snap}
\end{figure}

\begin{table}[t]
  \caption{Numbers of lipid (DPPC or PSM), Chol, and water molecules in mixture and pure membrane systems.}
  \label{tab:sys}
  \begin{center}
    \begin{tabular}{ccc} 
      \toprule
      & mixture & pure 
      \\ \midrule
       DPPC / PSM & 200 & 200 \\
       Chol & 22 & -\\
       Water & 22000 & 20000 \\ \bottomrule
    \end{tabular}
  \end{center}
\end{table}

\begin{figure}[t]
    \centering
    \includegraphics[width=1.0\linewidth]{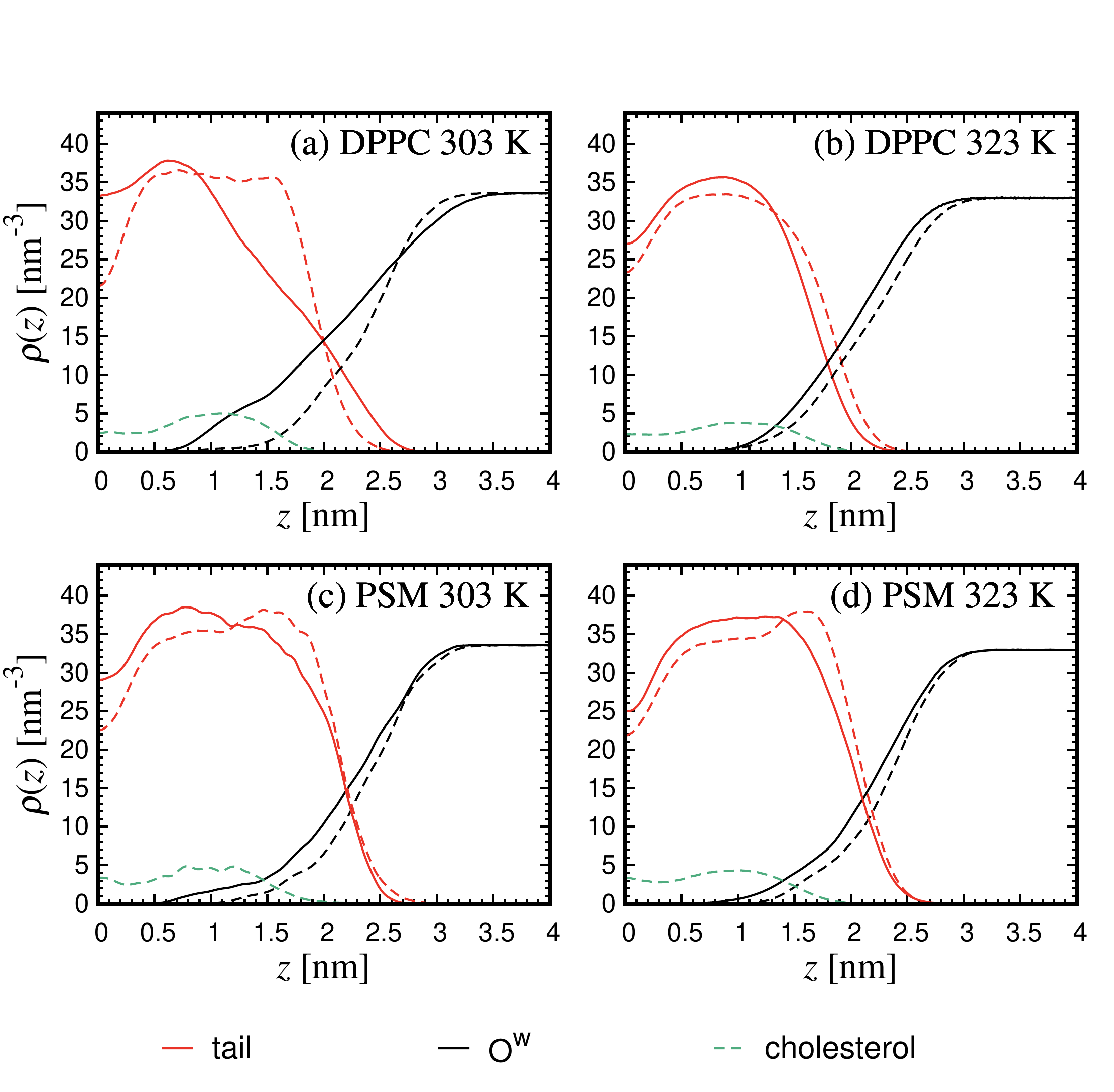}
    \caption{Number density distributions of lipid carbon chain (tail),
 water molecule oxygen (\oxy{w}), and Chol along the
 $z$-direction. Solid lines represent pure membrane systems, while dashed
 lines represent systems containing Chol. (a) and (b) correspond
 to DPPC, and (c) and (d) to PSM.}
     \label{fig:zp}
\end{figure}

The structures of the lipid molecules, DPPC, PSM, and Chol are depicted in \fig{snap}.
The hydrophilic moiety is common between DPPC and PSM,  
while
they are different for the hydrophobic portion and the distributions of oxygen and nitrogen atoms.

The lipid bilayer system was constructed using
CHARMM-GUI,~\cite{jo2008CHARMM, jo2009CHARMMGUI, brooks2009CHARMM,
wu2014CHARMMGUI, lee2016CHARMMGUI}
incorporating
200 lipid molecules with 10\% of Chol if present, as listed in Table~\ref{tab:sys}.
For each lipid and Chol composition, 100 water molecules per
molecule of lipid and Chol were added to create the lipid bilayer
system. 
Three different initial configurations were prepared
for each composition, employing the CHARMM36 force field for DPPC, PSM,
and Chol,~\cite{huang2013CHARMM36} and the CHARMM-compatible TIP3P model for water
molecules.~\cite{jorgensen1983Comparison}
All the MD simulations were performed using
Gromacs 2022.4. ~\cite{abraham2015GROMACS}

\begin{figure*}[t]
    \centering
    \includegraphics[width=1.0\linewidth]{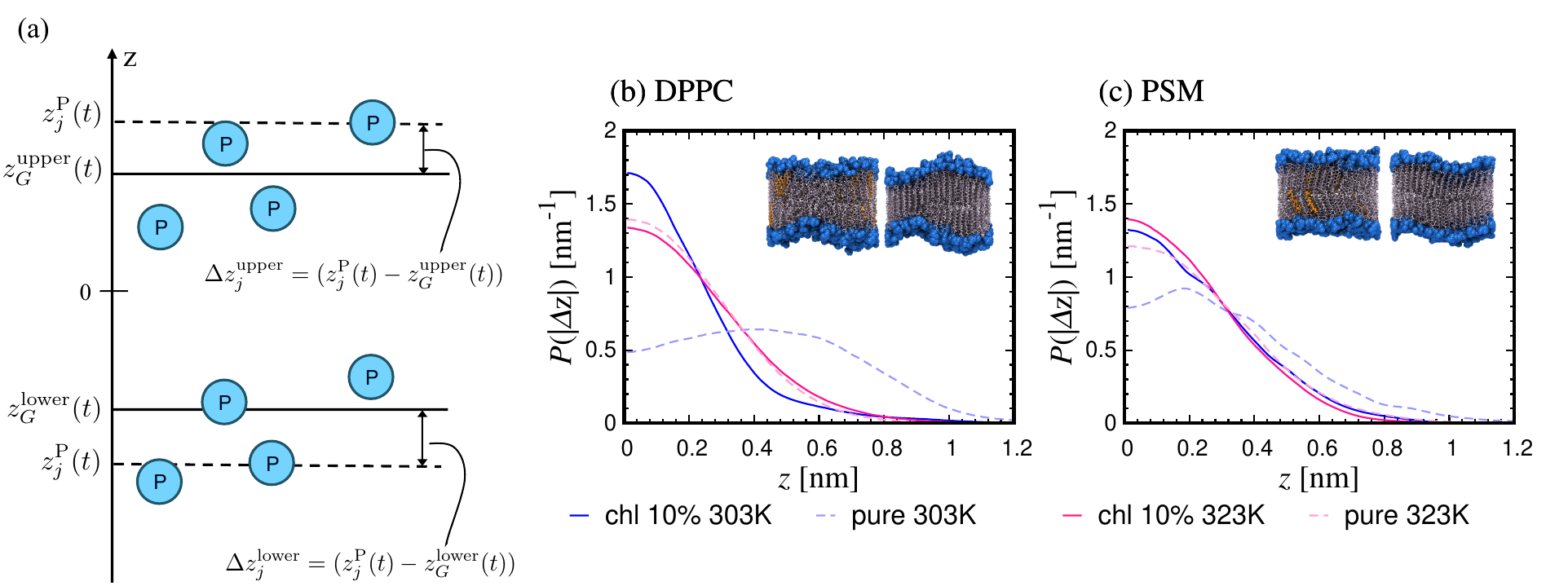}
    \caption{(a) Schematic illustration of lipid phosphorus atom position relative to 
the instantaneous average in the upper and lower leaflet, denoted as
 $z^\mathrm{upper}_G(t)$ and $z^\mathrm{lower}_G(t)$, respectively.
$z^\mathrm{P}_j(t) - z^\mathrm{upper}_G(t)$ and $z^\mathrm{P}_j(t) -
 z^\mathrm{lower}_G(t)$ represent the distances of the P 
 atom of lipid $j$, $z^\mr{P}_j(t)$, from
 $z^\mathrm{upper}_G(t)$ and $z^\mathrm{lower}_G(t)$, respectively,
 along the $z$-direction. 
(b) and (c) depict the distributions of $P(|\Delta z|)$ for DPPC and PSM, respectively. 
Snapshots in each panel are taken at 303~K (Left: with Chol, Right: pure).}
    \label{fig:pz2}
\end{figure*}

The equilibration process is described in Table S1 of the supplementary
material.
In accordance with CHARMM-GUI guidelines,
the process involved gradually relaxing restraints imposed on the phosphorus atom and the chiral
carbon center of the lipid molecule (Nos.~1-6).
The constants of the restraining forces on the $z$-coordinate of the
phosphorus atom and
on dihedral angle concerning the asymmetric center and double bond are denoted as $k_z$ and
$k_\mathrm{dih}$, respectively, in Table S1.
Subsequently, 
further equilibration steps were carried out 
(Nos. 7-11); 
the computational stability was checked in each of nos.~7-11 and the MD
length was gradually increased from no.~7 to 11.
Finally, 
three production runs under $NPT$ conditions for 10 ns each were
performed (No.~12).
To examine how the effect of Chol depends on the phase of the lipid membrane (gel vs liquid-crystalline), 
MD simulations were conducted at 303 K and 323 K for each system.
The coordinate system was set so that the $z$-axis is normal to the
membrane surface, which spans over the $x$- and $y$-directions.

To confirm the adequacy of the equilibration process, we examined the time
evolution of surface area $S$ in
the $x$-$y$ plane.
Figures~S1 and S2 of the supplementary material illustrate these
results during the 3 $\mu$s equilibration at 303 K
and 323 K, respectively.
While noticeable fluctuations are observed around 1.5
$\mu$s in some systems, the area $S$ converges
to a stable value at approximately 3 $\mu$s across all systems. 
Consequently, equilibration for 3 $\mu$s
is considered adequate, and a production run was carried out 
after this equilibration period.

\begin{figure*}[t]
    \centering
    \includegraphics[width=1.0\textwidth]{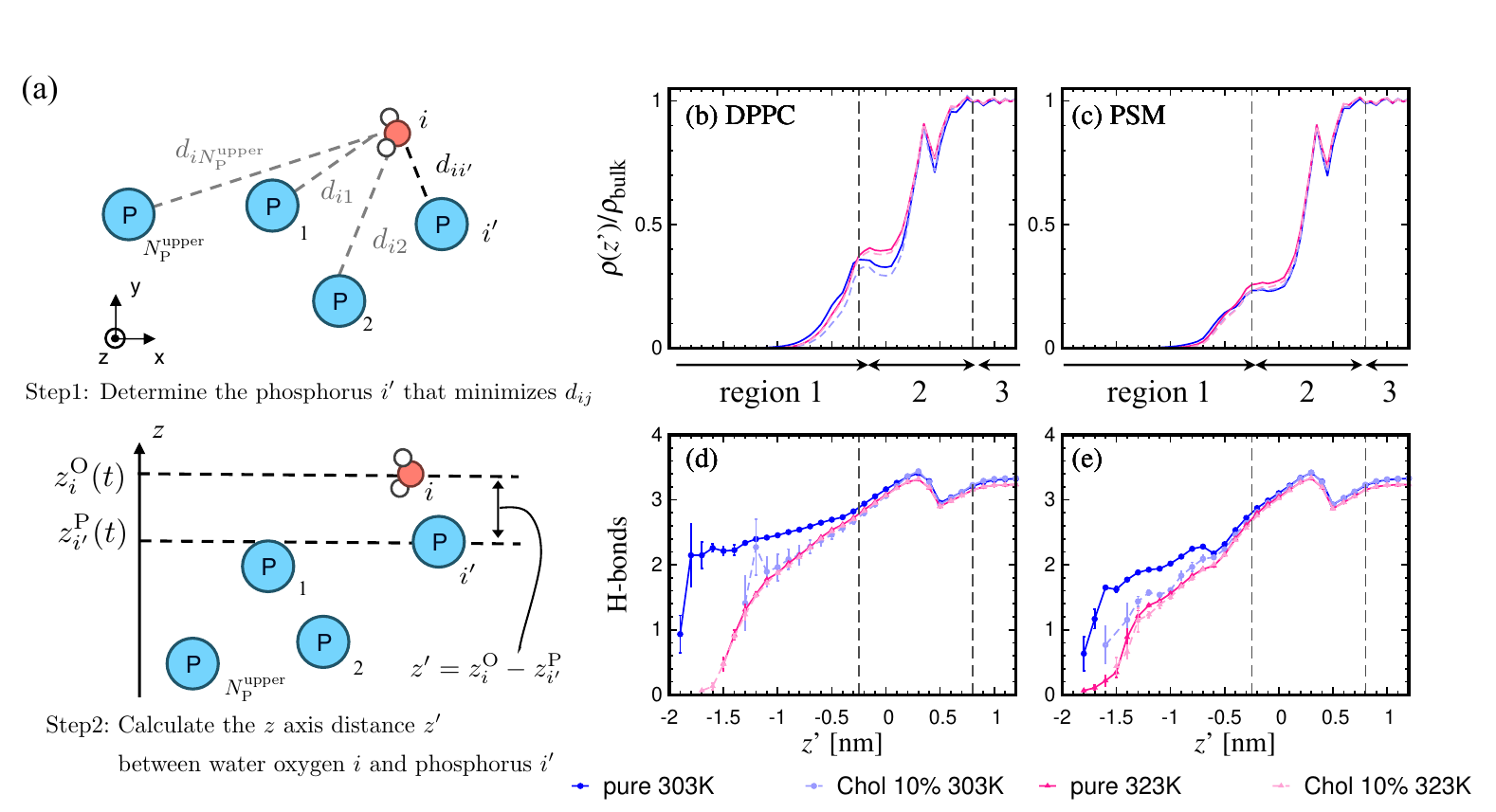}
    \caption{
(a) Schematic illustration of calculations of water molecule
 distribution $\rho(z')$.
(b) and (c) show the ratios of the water molecule distribution $\rho(z')$, to the number density of bulk
 water, $\rho_\mathrm{bulk}$, for DPPC and PSM, respectively.
The values of $\rho_\mathrm{bulk}$, determined from $\rho(z)$ as
 illustrated in Fig.~\ref{fig:zp}, are 33.50 nm$^{-3}$ and 32.95 nm$^{-3}$ for DPPC and PSM, respectively.
(d) and (e) depict the average numbers of H-bonds using the same
 $z'$-axis as (b) and (c), for DPPC and PSM, respectively.
The dashed lines at -0.25 nm and 0.8 nm define the boundaries between
 region 1 and 2 and between region 2 and 3, respectively.}
    \label{fig:zphb}
\end{figure*}

\section{Results and discussion}

\subsection{Density distributions of lipid, water, and Chol}
\label{section:density}

Initially, we analyzed the structure of the lipid bilayer, as well as the
configurations of Chol and water.
Figure~\ref{fig:zp} illustrates 
the number density distributions $\rho(z)$ of the lipid carbon chain
(tail), water molecular oxygen (\oxy{w}), and center of mass of Chol along the $z$-direction. 
Here, 
$z$ denotes the distance from the bilayer center, and the center of the
bilayer is the center of mass of the lipid molecules.
Note that the number density distributions of the upper and lower leaflets were
found to be coincident with each other within the margin of errors (data not shown). 
$\rho(z)$ in Fig.~\ref{fig:zp} represents the average of the profiles of both leaflets.
As depicted in Fig.~\ref{fig:zp}, Chol is predominantly situated near the carbon chain of the
lipid, leading to a broadening of the tail 
distribution along the $z$-direction, particularly evident at
303 K in DPPC. 
To examine this in detail, 
the number density distributions of
\oxy{D7} in DPPC and \oxy{S5} in PSM are plotted in Fig.~S3 of the
supplementary material.
These oxygen atoms are part of the hydrophilic functional groups within each
lipid molecule, which are located in
the innermost part of the lipid membranes (see Fig.~\ref{fig:snap}).
Figure~S3 of the
supplementary material demonstrates that the tail of water density distribution
overlaps with those of \oxy{D7} in DPPC and \oxy{S5} in PSM.
In the presence of Chol, particularly for DPPC at 303 K, the
distribution of \oxy{D7} becomes narrower, 
hindering the penetration of water molecules into the
membrane's inner regions, as observed in
Fig.~\ref{fig:zp}(a).
Additionally, Fig.~S4 of the supplementary material displays 
the number density distributions of nitrogen (N$^+$) in the choline group
and phosphorus (P) atoms along the $z$-direction.
The peak intensities of distributions for N$^+$ and P atoms were enhanced, 
particularly at 303 K in DPPC. 

\subsection{Fluctuation of the membrane interface}
The surfaces of lipid membranes are soft and fluctuate with time.
To elucidate the structure of the membrane interface, our focus was
directed towards the lipid head, where we examined the distribution
of lipid phosphorus atom position relative to the instantaneous interface defined below. 
The fluctuation of the interface between the membrane and water is seen evidently 
by employing the instantaneous interface method.~\cite{willard2010Instantaneous}

We assign each lipid to either the
upper or lower leaflet at each time.
Here, $N^\mathrm{upper}_\mathrm{p}$ denotes the number of lipid
molecules in the upper leaflet, $z_{j}^{\mathrm{P}}\left(t\right)$ is the $z$-coordinate 
at time $t$ of the $j$th lipid molecule, and
$z^\mathrm{upper}_G(t) =(1/{N^\mathrm{upper}_\mathrm{p}}) \sum_{j \in{\mathrm{upper}}}
z^\mathrm{P}_j(t)$ represents
the average of the $z$-coordinates of phosphorus atoms in the upper leaflet
of the lipid bilayer at time $t$.
Similarly, $N^\mathrm{lower}_\mathrm{p}$ and $z^\mathrm{lower}_G(t)$ can be computed for the lower leaflet.

As shown in Fig.~\ref{fig:pz2}(a), the deviation of the $z$-coordinate of the phosphorus atom of
lipid $j$ from $z^\mathrm{upper}_G(t)$ is expressed as 
$\Delta z^\mathrm{upper}_j(t) = z^\mathrm{P}_j(t) -
z^\mathrm{upper}_G(t)$ in the upper leaflet.
Similarly, for lipids in the lower leaflet, $\Delta
z^\mathrm{lower}_j(t)$ is defined.
Then, the time-averaged distribution function of the absolute values $\left|\Delta z\right|$ of 
$\Delta z^\mathrm{upper}_j(t)$ and $\Delta z^\mathrm{lower}_j(t)$ can be
assessed and is denoted as $P(|\Delta z|)$.

Figures~\ref{fig:pz2}(b) and \ref{fig:pz2}(c) illustrate the results of $P(|\Delta
z|)$ for DPPC and PSM, respectively.
In both DPPC and PSM, 
Chol does not exert discernible effects on $P(|\Delta z|)$ at 323 K.
Snapshots captured at 323 K are depicted in Fig.~S5 of the
supplementary material, revealing 
a disordered orientation of carbon chains within lipids.
This observation signifies a high degree of membrane fluidity, 
with weak influence of Chol.
In contrast, at 303 K, the distribution of $P(|\Delta z|)$ 
is broader in the pure lipid membrane systems, 
suggesting that Chol enhances membrane
stability and maintains the interface position.
The effect of Chol to suppress the interface fluctuations is particularly evident
in DPPC, as illustrated in Fig.~\ref{fig:pz2}(b) (see also snapshots 
captured at 303 K 
in the insets of Figs.~\ref{fig:pz2}(b) and \ref{fig:pz2}(c)).

\subsection{Classification of water molecules}

The analysis of water molecule distribution 
near the rugged membrane interface was conducted.
A precise description of the local water distribution relative to the
lipid interface was proposed~\cite{pandit2003Algorithm, berkowitz2006Aqueous}
The location of a water molecule is provided by $z'$, which is defined
as the distance from the interface
using Voronoi
tessellation.
Unlike $z$, $z'$ takes into account the effects of the fluctuation of
the lipid/water interfaces, and 
the distribution $\rho(z')$ 
characterizes the layered structures of
water molecules.

We propose a simpler method closely resembling Voronoi
tessellation.
The schematic illustration of the method is described in Fig.~\ref{fig:zphb}(a) and
the detail is provided as follows:
Step 1: Project the oxygen atoms of water molecules and the lipid phosphorus atoms onto the $x$-$y$ plane.
Calculate the distance $d_{ij}=|\br^\mathrm{O}_i(t)-\br^\mathrm{P}_j(t)|$ 
between the oxygen atom of water molecule $i$ and the phosphorus
atom $j$ in the $x$-$y$ plane.
Identify the lipid phosphorus atom $i'$ that gives the smallest
$d_{ij}$ for water molecule $i$.
Step 2: 
Determine $z^\mathrm{O}_i(t)$ as the $z$-coordinate of oxygen atom
of water molecule $i$
and $z^\mathrm{P}_{i'}(t)$ as the $z$-coordinate of the
phosphorus atom $i'$ nearest to $i$ in the $x$-$y$ plane.
Define the water molecule
density of $z'=z^\mathrm{O}_i - z^\mathrm{P}_{i'}$ as $\rho(z')$.
Note that the negative $z'$ indicates that 
the water molecule is located in more inner positions of the membrane than the phosphorus atom. 

\begin{figure*}[t]
  \centering
  \includegraphics[width=1.0\textwidth]{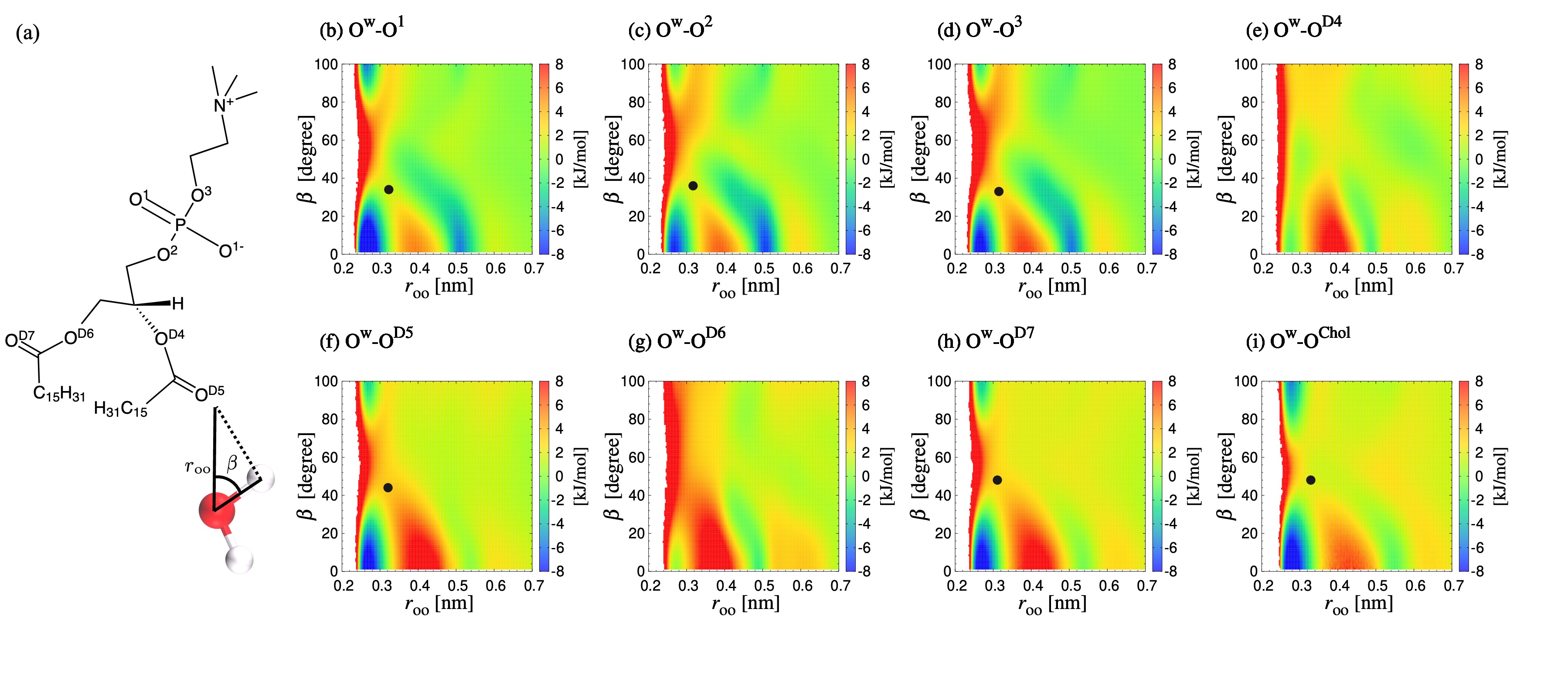}
  \caption{
    (a) Schematic illustration of $r_\mr{oo}$ and $\beta$ 
    for H-bond formed between the oxygen atom (\oxy{D5}) of a functional group within DPPC and
    a water molecule.
    (b)-(i) 2D PMF, $W(r_\mr{oo},\beta)$,
    between water oxygen (\ow) and each acceptor oxygen in DPPC
    with Chol at 303 K.
    Black points represent saddle points.}
  \label{fig:pmfdppc_303chl}
\end{figure*}

Figures~\ref{fig:zphb}(b) and \ref{fig:zphb}(c)
illustrates the ratio of $\rho(z')$
to the number density of bulk water, $\rho_\mathrm{bulk}$,
for DPPC and PSM, respectively.
In contrast to the water molecule density profile $\rho(z)$ in
Fig.~\ref{fig:zp}, which is influenced by the instantaneous
fluctuations of lipid membranes, 
$\rho(z')$ can be a more faithful representation of the water
distribution 
near the rugged surface.

From the profile of $\rho(z')$, the water molecules can be categorized
into three regions (regions 1-3), as depicted in Figs.~\ref{fig:zphb}(b) and \ref{fig:zphb}(c).
Specifically, 
region 1 represents the region inside the membrane at $z'<$ -0.25 nm, region 2 denotes the
interface region at -0.25 nm $<z'<$ 0.8 nm, and region 3 encompasses the bulk region at
$z'>$ 0.8 nm.
These classifications align with previous studies.~\cite{pandit2003Algorithm, berkowitz2006Aqueous}
Note that the number of DPPC molecules was 128 in
Ref.~\onlinecite{pandit2003Algorithm}, which is slightly smaller than
that of our system.
Furthermore, similar results of $\rho(z')$ were reported by Elola
\textit{et al},
where 968 DPPC molecules with the united-atom model were simulated.~\cite{elola2018Influence}

The water content in region 1 decreases progressively towards the center of the membrane.
In the interface region (region 2), a minimum was observed near $z'=0$ nm,
with a peak occurring around $z'=0.4$ nm, for both DPPC and PSM. 
This peak stems from the tendency for H-bond formation around phosphate groups.
A further elucidation on the H-bond rearrangement will be provided in subsequent Sec.~\ref{section:H-bond}.
Remarkably, as shown in Fig.~\ref{fig:zphb}(b), at 303 K in DPPC, the water content in the
interface region 
(region 2) is larger in the pure lipid membrane system than in the presence of Chol. 
This observation aligns with the variation in $\rho(z)$ of
water molecules due to Chol, as shown in Fig.~\ref{fig:zp}(a).
The pronounced stabilization of the DPPC membrane by 
Chol at 303 K highlights the significant impact on the hydration structure
near the interface.
On the contrary, 
at 323 K in DPPC, but the effect of Chol is less significant. 
Moreover, for PSM, the impacts of both temperature variations and Chol 
on $\rho(z')$ are not appreciable, as demonstrated in Fig.~\ref{fig:zphb}(c).

\begin{figure*}[t]
    \centering
    \includegraphics[width=1.0\textwidth]{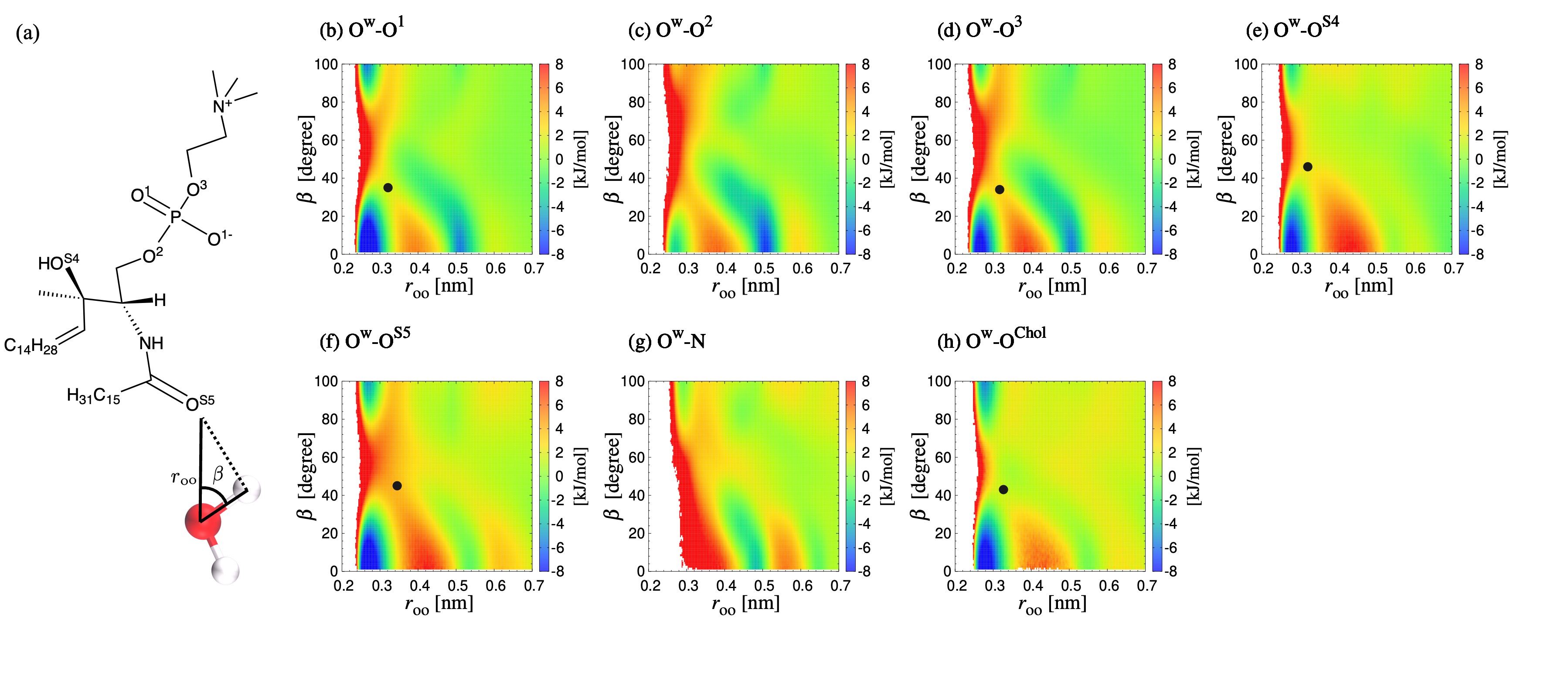}
    \caption{
      (a) Schematic illustration of $r_\mr{oo}$ and $\beta$ 
      for H-bond formed between the oxygen atom (\oxy{S5}) of a functional group within PSM and
      a water molecule.
      (b)-(i) 2D PMF, $W(r_\mr{oo},\beta)$,
      between water oxygen (\ow) and each acceptor oxygen in PSM
      with Chol at 303 K.
      Black points represent saddle points.}
    \label{fig:pmfpsm_303chl}
  \end{figure*}

\subsection{H-bond arrangement}
\label{section:H-bond}

The H-bonding states at each of the donor and acceptor sites were analyzed.
When investigating H-bond state in MD
simulations, a commonly employed approach involves applying a geometric
criterion to identify an H-bond between two water molecules. 
The
predominant definition often adopts the distance between oxygen
atoms (referred to as $r_\mathrm{oo}$) and the angle formed by the oxygen atom and the
oxygen-hydrogen bond (referred to as $\beta$) within a water
dimer.~\cite{luzar1996Effect, luzar1996Hydrogenbond, laage2006Molecular}

A more comprehensive understanding of the H-bond state can be obtained
by analyzing 
the distribution function of $r_\mathrm{oo}$ and $\beta$, 
denoted as $g(r_\mr{oo},\beta)$.~\cite{kumar2007Hydrogen,
kikutsuji2018How, kikutsuji2019Consistency, kikutsuji2021Transition}
In this context, 
$2\pi\rho r_\mathrm{oo}^2 \sin\beta g(r_\mathrm{oo},
\beta)\dd{r_\mathrm{oo}}\dd{\beta}$ 
represents the average number of oxygen atoms acting as H-bond acceptors within the
partial spherical shell volume characterized by $\mathrm{d}r_{\mathrm{oo}}$ and $\mathrm{d}\beta$ at the
position $(r_{\mathrm{oo}},\beta)$, with the average number density of
water molecules, $\rho$.
The logarithm form $W(r_\mathrm{oo},\beta) =
-k_\mathrm{B}T\ln{g(r_\mathrm{oo},\beta)}$ can be interpreted as the
two-dimensional potential of mean force (2D PMF).
For reference, the 2D PMF $W(r_\mathrm{oo},\beta)$ of bulk water at 303 K and 323 K
with a density of 1 g/cm$^3$ is depicted in Fig.~S6 of the supplementary material.
The temperature-independent energetically stable state is characterized by $r_\mathrm{oo} < 0.35$
nm and $0^{\circ}< \beta < 30^{\circ}$, which can be considered indicative of
H-bond state.

Figures~\ref{fig:pmfdppc_303chl}(a) and \ref{fig:pmfpsm_303chl}(a) provide
schematic illustrations of $r_{\mathrm{oo}}$ and $\beta$ for
H-bond formed between 
a water molecule and 
the oxygen atom of a functional group within DPPC
and PSM, respectively.
Given the presence of H-bonds between water molecules and those with
acceptors within lipid molecules, the analysis of 2D PMF was conducted
for water molecules and potential acceptors, including the oxygen atoms in
DPPC and the oxygen and nitrogen atoms in PSM. 
Additionally, in the presence of Chol, the analysis was also performed for  
H-bonds between water molecule and oxygen atom of hydroxy group, denoted
as \oxy{Chol}.
A similar 2D PMF analysis was previously done 
in polymer-water mixtures.~\cite{shikata2023Revealing}

Figure~\ref{fig:pmfdppc_303chl} illustrates the 2D PMF,
$W(r_\mathrm{oo},\beta)$, representing the interaction between water molecule as 
donors and DPPC oxygen atoms as acceptors at 303 K
in the presence of Chol.
Other results at 303 K in the absence of Chol and at 323 K both in the
presence and absence of Chol are displayed in Figs.~S7-S9 of the
supplementary material.
Similarly, in Fig.~\ref{fig:pmfpsm_303chl} and Figs.~S10-S12 of the supplementary material provide the 2D PMF,
$W(r_\mathrm{oo},\beta)$, for PSM systems.
Note that the 2D PMF between water molecules is omitted since
the overall profile remains unchanged for
both DPPC and PSM, when compared to that of bulk water (see 
Fig.~S6 of the supplementary material).
Based on the 2D PMF analysis, we identify potential H-bond acceptors as \oxy{1}, \oxy{2},
\oxy{3}, \oxy{D5}, \oxy{D7} and \oxy{Chol} for DPPC, and
\oxy{1}, \oxy{3}, \oxy{S4}, \oxy{S5}, and \oxy{Chol} for PSM,
respectively. 
See \fig{snap} for the notations of the acceptor sites in the lipids.
As illustrated in Figs.~\ref{fig:pmfdppc_303chl} and
\ref{fig:pmfpsm_303chl} and Figs.~S7-S12 of the supplementary material,
the H-bond region is characterized by 
$r_\mathrm{oo} < 3.5$
nm and $0^{\circ}< \beta < 30^{\circ}$ in the 2D PMF, irrespective of
the acceptor.
Furthermore, 
the H-bond regions remain unchanged regardless of the presence of
Chol or variations in temperature for both DPPC and PSM.
Nevertheless, 
specific oxygen atoms such as \oxy{1}, \oxy{D5}, and \oxy{D7} in DPPC, and \oxy{1} and \oxy{S5} in
PSM, which possess higher negative charges, form 
more energetically stable states compared to other acceptors.
In contrast, 
oxygen atoms \oxy{D4} and \oxy{D6} in DPPC, and oxygen atom \oxy{2} and
nitrogen (N) in PSM, form 
a second coordination region outside the defined H-bond
region.
Consequently, these are excluded from further
H-bond analysis due to their indeterminate bond characteristics.

Figures~\ref{fig:zphb}(d) and \ref{fig:zphb}(e) illustrate the distributions of the
average number of H-bonds formed by water molecules at each position,
using the $z'$-axis corresponding to $\rho(z')$, for DPPC and PSM, respectively.
The average number of H-bonds in region 3 converged to 
3.33 at 303 K and 3.23 at 323 K, respectively, corresponding to those
observed in bulk
water at each temperature.
In region 2, the average number of H-bonds reaches a maximum value higher than the
average observed in the bulk, gradually decreasing 
towards the interior of the bilayer.
This peak position corresponds to that in $\rho(z')$, as observed in
Figs.~\ref{fig:zphb}(b) and \ref{fig:zphb}(c).
These observations suggest that 
around phosphate groups, the oxygen atoms within the
lipid head group, such as \oxy{1}, \oxy{2}, and \oxy{3} in
DPPC, and \oxy{1} and \oxy{3} in PSM, act as acceptors, promoting the formation of
H-bonds with water molecules.
Remarkably, at 303 K, 
the average number of H-bonds in region 1 
increases with removal of Chol 
for both DPPC and PSM, with this trend being particularly notable in the
DPPC system.
However, at 323 K, the average number of H-bonds remains unchanged regardless
of the presence or absence of Chol, for both DPPC and PSM.
These findings indicate
variations in membrane structure induced by 
Chol impact the propensity for
H-bond formation within the membrane.

\begin{figure}[t]
    \centering
    \includegraphics[width=1.0\linewidth]{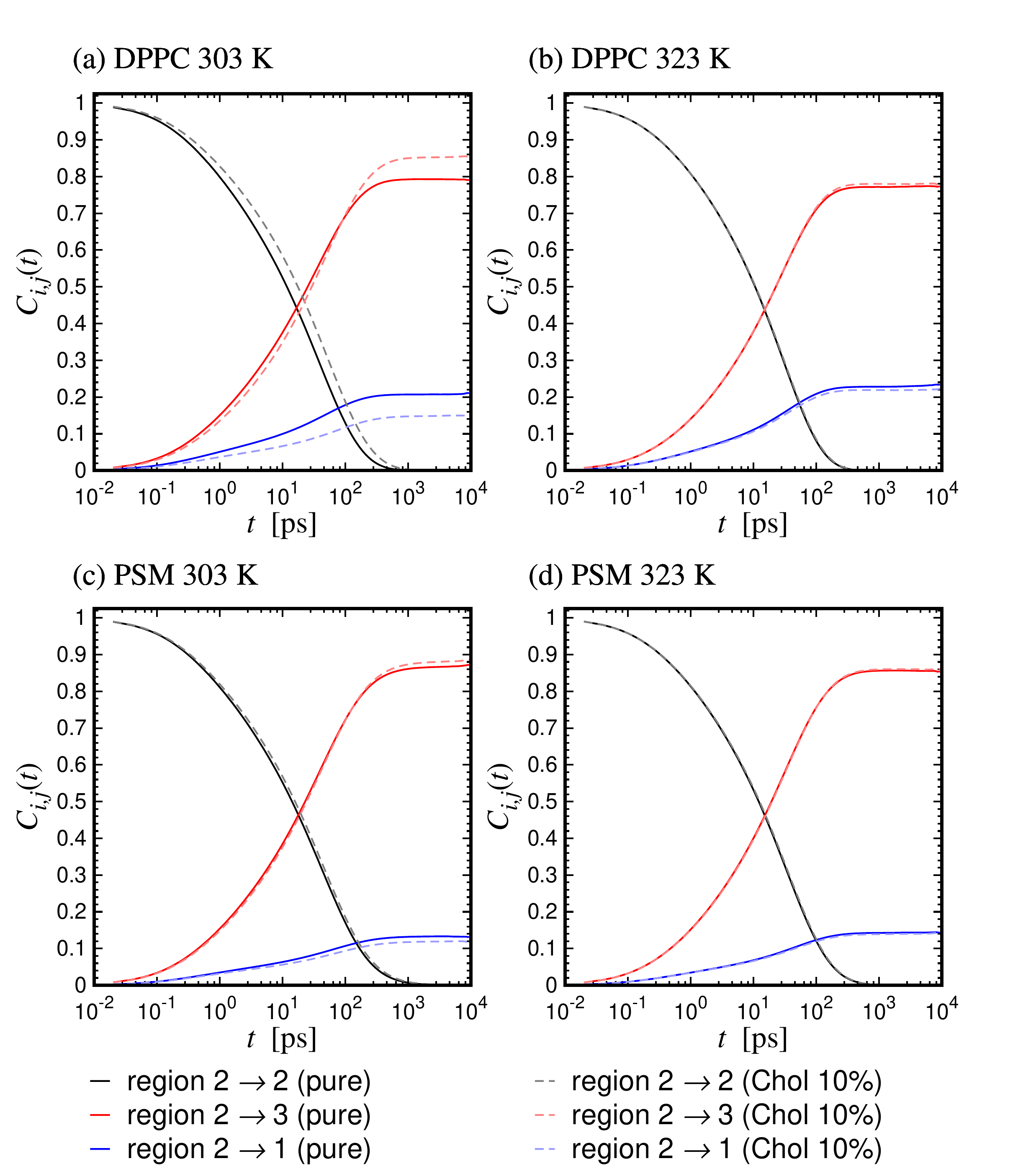}
    \caption{Conditional probability $C_{2,j}(t)$, representing 
transition dynamics from region $2$ at the initial time $t=0$ to either
 regions 1 or 
3 at time $t$ or remaining within the
same region 2 in the time interval $t$ [(a) DPPC at 303 K, (b) DPPC at 323 K, (c)
 PSM at 303 K, and (d) PSM at 323 K].
}
    \label{fig:tcf}
\end{figure}

\subsection{Water molecule rearrangement dynamics}
\label{section:water_transition}

We explore the transition dynamics of water molecules
among the three regions.
We address the dynamics of transition by defining $C_{i,j}(t)$ ($i \ne
j$) as the conditional probability that when a water molecule is in
region $i$ at time 0, it visits region $j$ with the number of passing the
boundaries of the regions being unity by time $t$.
Further, $C_{i,i}(t)$ is the probability that the water molecule stays in
region $i$ without visiting the other regions during the time interval
between 0 and $t$.
Note that the summation over all possible $j$ states ensures $\sum_{j}
C_{i,j}(t) = 1$, conserving the number of water molecules.
In practice, the trajectories of water molecules are continuously
monitored from $t=0$, tracking the subsequent transitions from
region $i$ to region $j$ at each time $t$.

Figure~\ref{fig:tcf} presents the results of $C_{2,j}(t)$, illustrating the
transition dynamics of water molecules originating from region 2 at $t=0$.
Note that the sum $C_{2,1}(t)+C_{2,2}(t)+C_{2,3}(t)=1$ holds at all
times $t$, as explained in the definition of $C_{i,j}(t)$.
Except for DPPC at 303 K,
$C_{2,j}(t)$ is not affected by the presence or absence of Chol and the transition rates 
from region 2 to regions 1 and 3 are common between the systems with and without Chol. 
In contrast, for DPPC at 303 K, the decay of $C_{2,2}(t)$ exhibits a
slower rate 
in the presence of Chol.
Furthermore, Chol alters the fraction of water molecules
transitioning to their respective destination.
Specifically, the population of water molecules transitioning from
region 2 to region 1 decreases by approximately 5\% in the presence of
Chol, while the transition to region 3 increases by a similar proportion.
At 323 K, the saturated values of $C_{2,1}(t)$ and
$C_{2,3}(t)$ resemble those of DPPC 
without Chol at 303 K.
These observations indicate the significant impact of 
membrane structure variations on water molecule dynamics.
For DPPC at 303 K, in particular, Chol enhance the tendency of keeping water
molecules in the interface region
(region 2) and relocating them 
towards the bulk region (region 3).
This suggests the reduction of 
water molecule exchanges between the
interface region (region 2) and the
inner side of the membrane (region 1).

Figure~S13 of the supplementary material illustrates the results of
$C_{1,j}(t)$, which represent
the transition dynamics from region $1$ at the initial time $t=0$ to 
 regions 2 or remaining within the
same region $1$ at subsequent time $t$.
Here, $C_{1,3}(t)$ is excluded given that the transition from
region 1 to region 3 inevitably passes through
region 2, ensuring the relationship, $C_{1,1}(t)+C_{1,2}(t)=1$.
Interestingly,  
the transition from region 1 to region 2 exhibits slower dynamics in
PSM compared to in DPPC at 303 K and 323 K.
This observation may be linked to the shorter H-bond lifetime $\thb$
of water molecules in the PSM system than with DPPC, as elucidated in
the subsequent Sec.~\ref{section:tau_HB}.
In addition, Chol further retards these dynamics, 
particularly evident at 303 K for both DPPC and PSM.
This observation suggests that 
Chol, situated within the membrane interior, 
exerts 
a notable influence on the dynamics of water molecules within region 1.

\begin{figure}[t]
    \centering
    \includegraphics[width=1.0\linewidth]{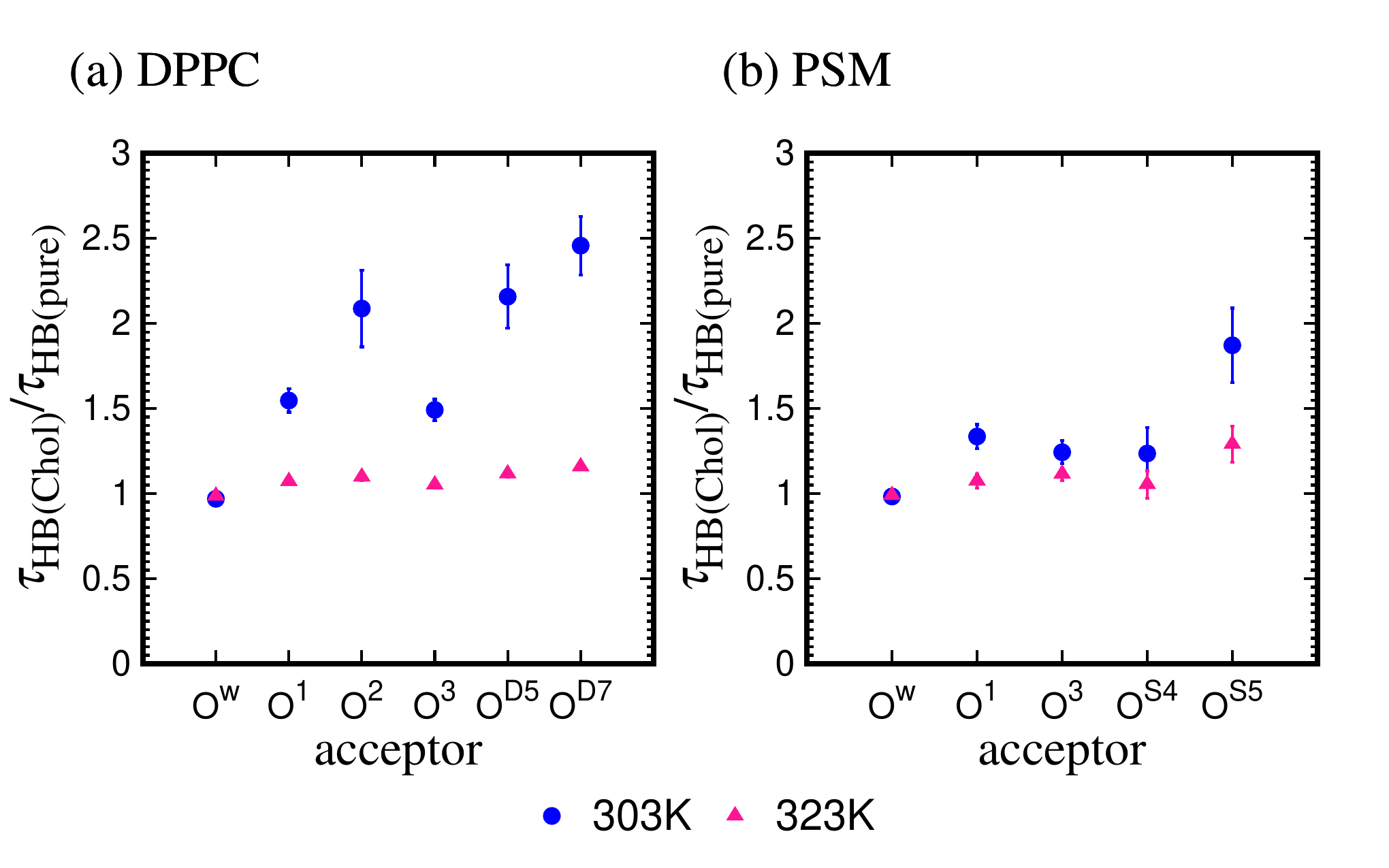}
    \caption{
Dependency of the ratio of H-bond lifetimes, $\thb$, 
with and without Chol, denoted as 
$\tau_\mr{HB(Chol)}/\tau_\mr{HB(pure)}$ for (a) DPPC and (b) PSM.}
    \label{fig:ratio}
\end{figure}

\subsection{Chol influence on H-bond lifetime}
\label{section:tau_HB}

Finally, we conducted an analysis of the H-bonding dynamics involving lipid
molecules, Chol, and water molecules to elucidate the timescale
of H-bond lifetime, by focusing on the acceptor oxygen atoms, such as
\oxy{1}, \oxy{2}, \oxy{3}, \oxy{D5}, and \oxy{D7} in DPPC, and \oxy{1}, \oxy{3},
\oxy{S4}, and \oxy{S5} in PSM, and \oxy{Chol} within Chol.
The H-bond time correlation function $\phb(t)$ is defined as
\begin{align}
    \phb(t) = \dfrac{\left\langle h_{i,j}(t)h_{i,j}(0)\right\rangle}{\left\langle h_{i,j}(0)\right\rangle},
    \label{eq:phb}
\end{align}
where
$h_{i,j}(t)$ equals 1 if water molecule $i$ is H-bonded with acceptor
oxygen $j$ at time $t$, otherwise 0.~\cite{rapaport1983Hydrogen, luzar1996Hydrogenbond, luzar1996Effect}
We computed $\phb(t)$ using the Monte-Carlo bootstrap method, which employs
a non-parametric approach to statistical
inference.~\cite{efron1992Bootstrap}

Figures~S14 and S15  of the supplementary material show the $\phb(t)$ results for acceptor oxygen
atoms in the DPPC and PSM systems, respectively.
Notably, as the temperature decreases, 
the decay of $\phb(t)$ slows down for
each acceptor oxygen atom, with a significant effect
observed for \oxy{Chol} in Chol.
However, the impact of temperature variation on H-bond breakages between
water molecules is negligible for both DPPC and PSM, owing to the
abundance 
of H-bonding partners of \oxy{w} in the bulk.
Furthermore, the influence of Chol is more pronounced in DPPC
compared to PSM, particularly at 303 K.
As illustrated in Figs.~S15 and S16, the
H-bond correlation between \oxy{w} and \oxy{Chol} exhibits a slower
dynamics in PSM
compared to DPPC.

The H-bond time correlation function,
$\phb(t)$, is approximated using the Kohlrausch–Williams–Watts (KWW)
function, $\phb(t) \simeq
\exp[-(t/\tau_{\mr{KWW}})^{\beta_{\mr{KWW}}}]$.
The fitting results are depicted in Figs.~S14 and S15 of the
supplementary material for
DPPC and PSM, respectively.
The H-bond lifetime $\tau_\mathrm{HB}$ is evaluated by integrating $\phb(t)$, yielding
\begin{align}
    \tau\hb = \int_0^\infty  \phb(t) \dd{t} =
 \frac{\tau_\mathrm{KWW}}{\beta_\mathrm{KWW}} \Gamma\left(\frac{1}{\beta_\mathrm{KWW}}\right),
    \label{eq:intkww}
\end{align}
where $\Gamma(\cdots)$ denotes the Gamma function.
The raw data of $\tau_\mathrm{KWW}$, $\beta_\mathrm{KWW}$, and $\thb$ are
summarized in Tables~S2-S7 of the supplementary material for both DPPC
and PSM.
To highlight the influence of Chol on H-bond lifetime $\thb$, the
dependency of 
the ratio between $\tau_{\mr{HB}}$ 
with and without Chol, denoted as 
$\tau_\mr{HB(Chol)}/\tau_\mr{HB(pure)}$, on acceptor oxygen at 303 K and 323 K is shown in Fig.~\ref{fig:ratio}.

Figure~\ref{fig:ratio}(a) illustrates the ratio
$\tau_\mr{HB(Chol)}/\tau_\mr{HB(pure)}$, ranging from 1.5 to 2.5, excluding \oxy{w}, at
303 K for DPPC.
Moreover, $\thb$ becomes large with the internal
oxygen atoms within the membrane, such as \oxy{D5} and \oxy{D7}.
In contrast, \oxy{1} and \oxy{3} exhibits
a relatively faster H-bond lifetime, 
showing that the H-bond dynamics is less susceptible to the presence or absence of Chol near the aqueous region.
However, in the case of PSM, 
the influence of cholesterol on H-bond lifetime
is limited, 
except for \oxy{S5}, at both 303 K and 323 K, as observed in
Fig.~\ref{fig:ratio}(b).
The slower dynamics observed for \oxy{S5} in PSM can be attributed to
its position as the innermost oxygen atom within the membrane, rendering
it more susceptible to Chol than other oxygen atoms.

Finally, we examine the energetic aspect of the H-bond lifetime
$\tau_\mathrm{HB}$.
Specifically, the activation energy $\Delta G$ of the H-bond is
estimated by the free energy difference between the most stable and
saddle points on the 2D PMF, $W(r_\mathrm{OO}, \beta)$ (see
Figs.~\ref{fig:pmfdppc_303chl} and ~\ref{fig:pmfpsm_303chl}, and
Figs.~S7-S12 of the supplementary material).
We plotted the relationship between
$\tau_\mathrm{HB}$ and $\Delta G/k_\mathrm{B}T$ in Fig.~S16 of the
supplementary material. 
Assuming the Arrhenius equation, 
$\tau_\text{HB} = A \exp({\Delta G}/{k_\text{B}T})$, 
we determined the constant $A$ by fitting.
The straight lines with $A=0.183$ ps and $A=0.385$ ps are
shown for DPPC and PSM systems, respectively, in Figs.~S16(a) and
S16(b) of the supplementary material.
The positive correlation between $\tau_\text{HB}$ and $\Delta
G/k_\mathrm{B}T$ is demonstrated for both DPPC and PSM systems, although
the deviation from the Arrhenius
equation is noticeable.
This result suggests that a more accurate energetic description of
H-bond breakage involving a large number of molecules necessitates
additional detailed variables beyond the two variables of distance
$r_\mathrm{OO}$ and angle $\beta$.

\section{Conclusions}

In this study, we employed MD simulations to investigate the influence
of Chol on water molecule behavior within lipid
membranes, with a specific focus on systems comprising DPPC and PSM.
While lipid membrane structures are not susceptible to the presence of Chol at 323 K, 
Chol at 303 K serves to stabilize
carbon chains, thereby reducing structural fluctuations at the membrane
interface, particularly for DPPC.

The spatial distribution of water molecules surrounding the membrane can
be classified into three distinct regions: the membrane interior, the
interface, and the bulk. 
The transition of water from the interface to the bulk is facilitated by Chol for the DPPC system at 303 K.
In contrast, the presence of Chol induces entrapment of water 
molecules within the membrane, leading to reduced rates of transition to 
the interface region from the interior region.

Our exploration into the dynamic attributes of water molecules 
in lipid-membrane systems,
considering the influence of Chol and
temperature variations
has yielded insights into the
intricate interplay at the membrane interface. 
DPPC is more susceptible to modifications at
303 K, significantly influencing H-bond dynamics within the
membrane. Specifically, at 303 K in DPPC, Chol was found to
markedly increase the H-bonding lifetime,
particularly impacting
internal oxygen atoms.

It is important to note the discrepancy in Chol density between
our study, which utilized up to 10\%, and 50\% employed by Elola
\textit{et al}.~\cite{elola2018Influence} 
This emphasizes the need to further investigate H-bonding dynamics
in lipid-membrane systems under conditions compatible to real Chol
contents in future research.
While Elola \textit{et al}. reported accelerated water dynamics
near the interface region, our findings have not directly corroborated
these observations.
However, we observed a notable migration tendency of water molecules
from region 2 (interface region) to region 3 (bulk region) in 
DPPC in the presence of Chol, as illustrated in Fig.~\ref{fig:tcf}(a).
Therefore, future studies should focus on systematically varying both
temperature and Chol content to provide a comprehensive
understanding of H-bonding dynamics.

\section*{Supplementary material}

The supplementary material include
equilibration scheme of MD simulations (Table~S1), time evolution of
surface area $S$ in the $x$-$y$ plane during the equilibration (Fig.~S1
and S2), 
number density distribution of \oxy{D7} in DPPC and \oxy{S5} in PSM
along the $z$-direction (Fig.~S3), 
number density distribution of nitrogen and phosphorus atoms
along the $z$-direction (Fig.~S4), MD snapshots taken at 323 K (Fig.~S5), 
2D PMF between water molecules in bulk water (Fig.~S6), 
2D PMF between water oxygen and acceptor oxygen atoms in lipid molecules
(Fig.~S7-S12), 
conditional probability $C_{1,j}(t)$ (Fig.~S13), 
H-bond time correlation function $\phb(t)$ (Figs.~S14 and S15), raw data
of $\tau_\mathrm{KWW}$, $\beta_\mathrm{KWW}$, and $\thb$ (Table~S2-S7),
and the relationship between the H-bond lifetime $\tau_\mathrm{HB}$ and
activation energy normalized by thermal energy, $\Delta G/k_\mathrm{B}T$
(Fig.~S16).

\begin{acknowledgments}
This work was supported by 
JSPS KAKENHI Grant-in-Aid 
Grant Nos.~\mbox{JP21H04628}, \mbox{JP21H05249}, \mbox{JP22H04542},
 \mbox{JP22K03550}, \mbox{JP23H01924}, \mbox{JP23K26617}, 
 \mbox{JP23H02622}, \mbox{JP23K27313}, and \mbox{JP24H01719}.
We are grateful to 
the Fugaku Supercomputing Project (Nos.~JPMXP1020230325 and JPMXP1020230327) and 
the Data-Driven Material Research Project (No.~\mbox{JPMXP1122714694})
from the
Ministry of Education, Culture, Sports, Science, and Technology and to Maruho Collaborative Project for Theoretical Pharmaceutics.
The numerical calculations were performed at Research Center for
Computational Science, Okazaki Research Facilities, National Institutes
of Natural Sciences (Project: \mbox{24-IMS-C051}) and at the Cybermedia Center, Osaka University.
\end{acknowledgments}

\section*{AUTHOR DECLARATIONS}

\section*{Conflict of Interest}
The authors have no conflicts to disclose.

\section*{Data availability statement}
The program codes and data that support the findings of this study
are openly available in Zenodo at https://doi.org/10.5281/zenodo.11273701. 
Further data are available from the corresponding author upon reasonable
request.


\begin{thebibliography}{76}%
\makeatletter
\providecommand \@ifxundefined [1]{%
 \@ifx{#1\undefined}
}%
\providecommand \@ifnum [1]{%
 \ifnum #1\expandafter \@firstoftwo
 \else \expandafter \@secondoftwo
 \fi
}%
\providecommand \@ifx [1]{%
 \ifx #1\expandafter \@firstoftwo
 \else \expandafter \@secondoftwo
 \fi
}%
\providecommand \natexlab [1]{#1}%
\providecommand \enquote  [1]{``#1''}%
\providecommand \bibnamefont  [1]{#1}%
\providecommand \bibfnamefont [1]{#1}%
\providecommand \citenamefont [1]{#1}%
\providecommand \href@noop [0]{\@secondoftwo}%
\providecommand \href [0]{\begingroup \@sanitize@url \@href}%
\providecommand \@href[1]{\@@startlink{#1}\@@href}%
\providecommand \@@href[1]{\endgroup#1\@@endlink}%
\providecommand \@sanitize@url [0]{\catcode `\\12\catcode `\$12\catcode
  `\&12\catcode `\#12\catcode `\^12\catcode `\_12\catcode `\%12\relax}%
\providecommand \@@startlink[1]{}%
\providecommand \@@endlink[0]{}%
\providecommand \url  [0]{\begingroup\@sanitize@url \@url }%
\providecommand \@url [1]{\endgroup\@href {#1}{\urlprefix }}%
\providecommand \urlprefix  [0]{URL }%
\providecommand \Eprint [0]{\href }%
\providecommand \doibase [0]{https://doi.org/}%
\providecommand \selectlanguage [0]{\@gobble}%
\providecommand \bibinfo  [0]{\@secondoftwo}%
\providecommand \bibfield  [0]{\@secondoftwo}%
\providecommand \translation [1]{[#1]}%
\providecommand \BibitemOpen [0]{}%
\providecommand \bibitemStop [0]{}%
\providecommand \bibitemNoStop [0]{.\EOS\space}%
\providecommand \EOS [0]{\spacefactor3000\relax}%
\providecommand \BibitemShut  [1]{\csname bibitem#1\endcsname}%
\let\auto@bib@innerbib\@empty
\bibitem [{\citenamefont {Nagle}\ and\ \citenamefont
  {{Tristram-Nagle}}(2000)}]{nagle2000Structure}%
  \BibitemOpen
  \bibfield  {author} {\bibinfo {author} {\bibfnamefont {J.~F.}\ \bibnamefont
  {Nagle}}\ and\ \bibinfo {author} {\bibfnamefont {S.}~\bibnamefont
  {{Tristram-Nagle}}},\ }\bibfield  {title} {\enquote {\bibinfo {title}
  {Structure of lipid bilayers},}\ }\href
  {https://doi.org/10.1016/S0304-4157(00)00016-2} {\bibfield  {journal}
  {\bibinfo  {journal} {Biochim. Biophys. Acta Biomembr.}\ }\textbf {\bibinfo
  {volume} {1469}},\ \bibinfo {pages} {159--195} (\bibinfo {year}
  {2000})}\BibitemShut {NoStop}%
\bibitem [{\citenamefont
  {Israelachvili}(2011)}]{israelachvili2011Intermolecular}%
  \BibitemOpen
  \bibfield  {author} {\bibinfo {author} {\bibfnamefont {J.~N.}\ \bibnamefont
  {Israelachvili}},\ }\href@noop {} {\emph {\bibinfo {title} {Intermolecular
  and Surface Forces}}},\ \bibinfo {edition} {3rd}\ ed.\ (\bibinfo  {publisher}
  {Academic Press},\ \bibinfo {address} {{Burlington, MA}},\ \bibinfo {year}
  {2011})\BibitemShut {NoStop}%
\bibitem [{\citenamefont {Mouritsen}\ and\ \citenamefont
  {Zuckermann}(2004)}]{mouritsen2004What}%
  \BibitemOpen
  \bibfield  {author} {\bibinfo {author} {\bibfnamefont {O.~G.}\ \bibnamefont
  {Mouritsen}}\ and\ \bibinfo {author} {\bibfnamefont {M.~J.}\ \bibnamefont
  {Zuckermann}},\ }\bibfield  {title} {\enquote {\bibinfo {title} {What's so
  special about cholesterol?}}\ }\href
  {https://doi.org/10.1007/s11745-004-1336-x} {\bibfield  {journal} {\bibinfo
  {journal} {Lipids}\ }\textbf {\bibinfo {volume} {39}},\ \bibinfo {pages}
  {1101--1113} (\bibinfo {year} {2004})}\BibitemShut {NoStop}%
\bibitem [{\citenamefont {{de Meyer}}\ and\ \citenamefont
  {Smit}(2009)}]{demeyer2009Effect}%
  \BibitemOpen
  \bibfield  {author} {\bibinfo {author} {\bibfnamefont {F.}~\bibnamefont {{de
  Meyer}}}\ and\ \bibinfo {author} {\bibfnamefont {B.}~\bibnamefont {Smit}},\
  }\bibfield  {title} {\enquote {\bibinfo {title} {Effect of cholesterol on the
  structure of a phospholipid bilayer},}\ }\href
  {https://doi.org/10.1073/pnas.0809959106} {\bibfield  {journal} {\bibinfo
  {journal} {Proc. Natl. Acad. Sci. U.S.A.}\ }\textbf {\bibinfo {volume}
  {106}},\ \bibinfo {pages} {3654--3658} (\bibinfo {year} {2009})}\BibitemShut
  {NoStop}%
\bibitem [{\citenamefont {Marrink}\ \emph {et~al.}(1996)\citenamefont
  {Marrink}, \citenamefont {Tieleman}, \citenamefont {van Buuren},\ and\
  \citenamefont {Berendsen}}]{marrink1996Membranes}%
  \BibitemOpen
  \bibfield  {author} {\bibinfo {author} {\bibfnamefont {S.-J.}\ \bibnamefont
  {Marrink}}, \bibinfo {author} {\bibfnamefont {D.~P.}\ \bibnamefont
  {Tieleman}}, \bibinfo {author} {\bibfnamefont {A.~R.}\ \bibnamefont {van
  Buuren}},\ and\ \bibinfo {author} {\bibfnamefont {H.~J.~C.}\ \bibnamefont
  {Berendsen}},\ }\bibfield  {title} {\enquote {\bibinfo {title} {Membranes and
  water: An interesting relationship},}\ }\href
  {https://doi.org/10.1039/FD9960300191} {\bibfield  {journal} {\bibinfo
  {journal} {Faraday Discuss.}\ }\textbf {\bibinfo {volume} {103}},\ \bibinfo
  {pages} {191--201} (\bibinfo {year} {1996})}\BibitemShut {NoStop}%
\bibitem [{\citenamefont {Pratt}\ and\ \citenamefont
  {Pohorille}(2002)}]{pratt2002Hydrophobic}%
  \BibitemOpen
  \bibfield  {author} {\bibinfo {author} {\bibfnamefont {L.~R.}\ \bibnamefont
  {Pratt}}\ and\ \bibinfo {author} {\bibfnamefont {A.}~\bibnamefont
  {Pohorille}},\ }\bibfield  {title} {\enquote {\bibinfo {title} {Hydrophobic
  {{Effects}} and {{Modeling}} of {{Biophysical Aqueous Solution
  Interfaces}}},}\ }\href {https://doi.org/10.1021/cr000692+} {\bibfield
  {journal} {\bibinfo  {journal} {Chem. Rev.}\ }\textbf {\bibinfo {volume}
  {102}},\ \bibinfo {pages} {2671--2692} (\bibinfo {year} {2002})}\BibitemShut
  {NoStop}%
\bibitem [{\citenamefont {Higgins}\ \emph {et~al.}(2006)\citenamefont
  {Higgins}, \citenamefont {Polcik}, \citenamefont {Fukuma}, \citenamefont
  {Sader}, \citenamefont {Nakayama},\ and\ \citenamefont
  {Jarvis}}]{higgins2006Structured}%
  \BibitemOpen
  \bibfield  {author} {\bibinfo {author} {\bibfnamefont {M.~J.}\ \bibnamefont
  {Higgins}}, \bibinfo {author} {\bibfnamefont {M.}~\bibnamefont {Polcik}},
  \bibinfo {author} {\bibfnamefont {T.}~\bibnamefont {Fukuma}}, \bibinfo
  {author} {\bibfnamefont {J.~E.}\ \bibnamefont {Sader}}, \bibinfo {author}
  {\bibfnamefont {Y.}~\bibnamefont {Nakayama}},\ and\ \bibinfo {author}
  {\bibfnamefont {S.~P.}\ \bibnamefont {Jarvis}},\ }\bibfield  {title}
  {\enquote {\bibinfo {title} {Structured {{Water Layers Adjacent}} to
  {{Biological Membranes}}},}\ }\href
  {https://doi.org/10.1529/biophysj.106.085688} {\bibfield  {journal} {\bibinfo
   {journal} {Biophys. J.}\ }\textbf {\bibinfo {volume} {91}},\ \bibinfo
  {pages} {2532--2542} (\bibinfo {year} {2006})}\BibitemShut {NoStop}%
\bibitem [{\citenamefont {Raschke}(2006)}]{raschke2006Water}%
  \BibitemOpen
  \bibfield  {author} {\bibinfo {author} {\bibfnamefont {T.~M.}\ \bibnamefont
  {Raschke}},\ }\bibfield  {title} {\enquote {\bibinfo {title} {Water structure
  and interactions with protein surfaces},}\ }\href
  {https://doi.org/10.1016/j.sbi.2006.03.002} {\bibfield  {journal} {\bibinfo
  {journal} {Curr. Opin. Struct. Biol.}\ }\bibinfo {series} {Theory and
  Simulation/{{Macromolecular}} Assemblages},\ \textbf {\bibinfo {volume}
  {16}},\ \bibinfo {pages} {152--159} (\bibinfo {year} {2006})}\BibitemShut
  {NoStop}%
\bibitem [{\citenamefont {Ziegler}\ and\ \citenamefont
  {Vernier}(2008)}]{ziegler2008Interface}%
  \BibitemOpen
  \bibfield  {author} {\bibinfo {author} {\bibfnamefont {M.~J.}\ \bibnamefont
  {Ziegler}}\ and\ \bibinfo {author} {\bibfnamefont {P.~T.}\ \bibnamefont
  {Vernier}},\ }\bibfield  {title} {\enquote {\bibinfo {title} {Interface
  {{Water Dynamics}} and {{Porating Electric Fields}} for {{Phospholipid
  Bilayers}}},}\ }\href {https://doi.org/10.1021/jp8027726} {\bibfield
  {journal} {\bibinfo  {journal} {J. Phys. Chem. B}\ }\textbf {\bibinfo
  {volume} {112}},\ \bibinfo {pages} {13588--13596} (\bibinfo {year}
  {2008})}\BibitemShut {NoStop}%
\bibitem [{\citenamefont {Disalvo}\ \emph {et~al.}(2008)\citenamefont
  {Disalvo}, \citenamefont {Lairion}, \citenamefont {Martini}, \citenamefont
  {Tymczyszyn}, \citenamefont {Fr{\'i}as}, \citenamefont {Almaleck},\ and\
  \citenamefont {Gordillo}}]{disalvo2008Structural}%
  \BibitemOpen
  \bibfield  {author} {\bibinfo {author} {\bibfnamefont {E.~A.}\ \bibnamefont
  {Disalvo}}, \bibinfo {author} {\bibfnamefont {F.}~\bibnamefont {Lairion}},
  \bibinfo {author} {\bibfnamefont {F.}~\bibnamefont {Martini}}, \bibinfo
  {author} {\bibfnamefont {E.}~\bibnamefont {Tymczyszyn}}, \bibinfo {author}
  {\bibfnamefont {M.}~\bibnamefont {Fr{\'i}as}}, \bibinfo {author}
  {\bibfnamefont {H.}~\bibnamefont {Almaleck}},\ and\ \bibinfo {author}
  {\bibfnamefont {G.~J.}\ \bibnamefont {Gordillo}},\ }\bibfield  {title}
  {\enquote {\bibinfo {title} {Structural and functional properties of
  hydration and confined water in membrane interfaces},}\ }\href
  {https://doi.org/10.1016/j.bbamem.2008.08.025} {\bibfield  {journal}
  {\bibinfo  {journal} {Biochim. Biophys. Acta Biomembr.}\ }\textbf {\bibinfo
  {volume} {1778}},\ \bibinfo {pages} {2655--2670} (\bibinfo {year}
  {2008})}\BibitemShut {NoStop}%
\bibitem [{\citenamefont {Cheng}\ \emph {et~al.}(2013)\citenamefont {Cheng},
  \citenamefont {Varkey}, \citenamefont {Ambroso}, \citenamefont {Langen},\
  and\ \citenamefont {Han}}]{cheng2013Hydration}%
  \BibitemOpen
  \bibfield  {author} {\bibinfo {author} {\bibfnamefont {C.-Y.}\ \bibnamefont
  {Cheng}}, \bibinfo {author} {\bibfnamefont {J.}~\bibnamefont {Varkey}},
  \bibinfo {author} {\bibfnamefont {M.~R.}\ \bibnamefont {Ambroso}}, \bibinfo
  {author} {\bibfnamefont {R.}~\bibnamefont {Langen}},\ and\ \bibinfo {author}
  {\bibfnamefont {S.}~\bibnamefont {Han}},\ }\bibfield  {title} {\enquote
  {\bibinfo {title} {Hydration dynamics as an intrinsic ruler for refining
  protein structure at lipid membrane interfaces},}\ }\href
  {https://doi.org/10.1073/pnas.1307678110} {\bibfield  {journal} {\bibinfo
  {journal} {Proc. Natl. Acad. Sci. U.S.A.}\ }\textbf {\bibinfo {volume}
  {110}},\ \bibinfo {pages} {16838--16843} (\bibinfo {year}
  {2013})}\BibitemShut {NoStop}%
\bibitem [{\citenamefont {Zhou}\ and\ \citenamefont
  {Cross}(2013)}]{zhou2013Influences}%
  \BibitemOpen
  \bibfield  {author} {\bibinfo {author} {\bibfnamefont {H.-X.}\ \bibnamefont
  {Zhou}}\ and\ \bibinfo {author} {\bibfnamefont {T.~A.}\ \bibnamefont
  {Cross}},\ }\bibfield  {title} {\enquote {\bibinfo {title} {Influences of
  {{Membrane Mimetic Environments}} on {{Membrane Protein Structures}}},}\
  }\href {https://doi.org/10.1146/annurev-biophys-083012-130326} {\bibfield
  {journal} {\bibinfo  {journal} {Annu. Rev. Biophys.}\ }\textbf {\bibinfo
  {volume} {42}},\ \bibinfo {pages} {361--392} (\bibinfo {year}
  {2013})}\BibitemShut {NoStop}%
\bibitem [{\citenamefont {Disalvo}(2015)}]{disalvo2015Membrane}%
  \BibitemOpen
  \bibinfo {editor} {\bibfnamefont {E.~A.}\ \bibnamefont {Disalvo}},\ ed.,\
  \href {https://doi.org/10.1007/978-3-319-19060-0} {\emph {\bibinfo {title}
  {Membrane {{Hydration}}: {{The Role}} of {{Water}} in the {{Structure}} and
  {{Function}} of {{Biological Membranes}}}}},\ \bibinfo {series} {Subcellular
  {{Biochemistry}}}, Vol.~\bibinfo {volume} {71}\ (\bibinfo  {publisher}
  {Springer, Cham},\ \bibinfo {address} {{Cham}},\ \bibinfo {year}
  {2015})\BibitemShut {NoStop}%
\bibitem [{\citenamefont {Jungwirth}(2015)}]{jungwirth2015Biological}%
  \BibitemOpen
  \bibfield  {author} {\bibinfo {author} {\bibfnamefont {P.}~\bibnamefont
  {Jungwirth}},\ }\bibfield  {title} {\enquote {\bibinfo {title} {Biological
  {{Water}} or {{Rather Water}} in {{Biology}}?}}\ }\href
  {https://doi.org/10.1021/acs.jpclett.5b01143} {\bibfield  {journal} {\bibinfo
   {journal} {J. Phys. Chem. Lett.}\ }\textbf {\bibinfo {volume} {6}},\
  \bibinfo {pages} {2449--2451} (\bibinfo {year} {2015})}\BibitemShut {NoStop}%
\bibitem [{\citenamefont {Laage}, \citenamefont {Elsaesser},\ and\
  \citenamefont {Hynes}(2017)}]{laage2017Water}%
  \BibitemOpen
  \bibfield  {author} {\bibinfo {author} {\bibfnamefont {D.}~\bibnamefont
  {Laage}}, \bibinfo {author} {\bibfnamefont {T.}~\bibnamefont {Elsaesser}},\
  and\ \bibinfo {author} {\bibfnamefont {J.~T.}\ \bibnamefont {Hynes}},\
  }\bibfield  {title} {\enquote {\bibinfo {title} {Water {{Dynamics}} in the
  {{Hydration Shells}} of {{Biomolecules}}},}\ }\href
  {https://doi.org/10.1021/acs.chemrev.6b00765} {\bibfield  {journal} {\bibinfo
   {journal} {Chem. Rev.}\ }\textbf {\bibinfo {volume} {117}},\ \bibinfo
  {pages} {10694--10725} (\bibinfo {year} {2017})}\BibitemShut {NoStop}%
\bibitem [{\citenamefont {Chattopadhyay}\ \emph {et~al.}(2021)\citenamefont
  {Chattopadhyay}, \citenamefont {Krok}, \citenamefont {Orlikowska},
  \citenamefont {Schwille}, \citenamefont {Franquelim},\ and\ \citenamefont
  {Piatkowski}}]{chattopadhyay2021Hydration}%
  \BibitemOpen
  \bibfield  {author} {\bibinfo {author} {\bibfnamefont {M.}~\bibnamefont
  {Chattopadhyay}}, \bibinfo {author} {\bibfnamefont {E.}~\bibnamefont {Krok}},
  \bibinfo {author} {\bibfnamefont {H.}~\bibnamefont {Orlikowska}}, \bibinfo
  {author} {\bibfnamefont {P.}~\bibnamefont {Schwille}}, \bibinfo {author}
  {\bibfnamefont {H.~G.}\ \bibnamefont {Franquelim}},\ and\ \bibinfo {author}
  {\bibfnamefont {L.}~\bibnamefont {Piatkowski}},\ }\bibfield  {title}
  {\enquote {\bibinfo {title} {Hydration {{Layer}} of {{Only}} a {{Few
  Molecules Controls Lipid Mobility}} in {{Biomimetic Membranes}}},}\ }\href
  {https://doi.org/10.1021/jacs.1c04314} {\bibfield  {journal} {\bibinfo
  {journal} {J. Am. Chem. Soc.}\ }\textbf {\bibinfo {volume} {143}},\ \bibinfo
  {pages} {14551--14562} (\bibinfo {year} {2021})}\BibitemShut {NoStop}%
\bibitem [{\citenamefont {Berkowitz}\ and\ \citenamefont
  {Raghavan}(1991)}]{berkowitz1991Computer}%
  \BibitemOpen
  \bibfield  {author} {\bibinfo {author} {\bibfnamefont {M.~L.}\ \bibnamefont
  {Berkowitz}}\ and\ \bibinfo {author} {\bibfnamefont {K.}~\bibnamefont
  {Raghavan}},\ }\bibfield  {title} {\enquote {\bibinfo {title} {Computer
  simulation of a water/membrane interface},}\ }\href
  {https://doi.org/10.1021/la00054a002} {\bibfield  {journal} {\bibinfo
  {journal} {Langmuir}\ }\textbf {\bibinfo {volume} {7}},\ \bibinfo {pages}
  {1042--1044} (\bibinfo {year} {1991})}\BibitemShut {NoStop}%
\bibitem [{\citenamefont {Pastor}(1994)}]{pastor1994Molecular}%
  \BibitemOpen
  \bibfield  {author} {\bibinfo {author} {\bibfnamefont {R.~W.}\ \bibnamefont
  {Pastor}},\ }\bibfield  {title} {\enquote {\bibinfo {title} {Molecular
  dynamics and {{Monte Carlo}} simulations of lipid bilayers},}\ }\href
  {https://doi.org/10.1016/S0959-440X(94)90209-7} {\bibfield  {journal}
  {\bibinfo  {journal} {Curr. Opin. Struct. Biol.}\ }\textbf {\bibinfo {volume}
  {4}},\ \bibinfo {pages} {486--492} (\bibinfo {year} {1994})}\BibitemShut
  {NoStop}%
\bibitem [{\citenamefont {Marrink}\ and\ \citenamefont
  {Berendsen}(1994)}]{marrink1994Simulation}%
  \BibitemOpen
  \bibfield  {author} {\bibinfo {author} {\bibfnamefont {S.-J.}\ \bibnamefont
  {Marrink}}\ and\ \bibinfo {author} {\bibfnamefont {H.~J.~C.}\ \bibnamefont
  {Berendsen}},\ }\bibfield  {title} {\enquote {\bibinfo {title} {Simulation of
  water transport through a lipid membrane},}\ }\href
  {https://doi.org/10.1021/j100066a040} {\bibfield  {journal} {\bibinfo
  {journal} {J. Phys. Chem.}\ }\textbf {\bibinfo {volume} {98}},\ \bibinfo
  {pages} {4155--4168} (\bibinfo {year} {1994})}\BibitemShut {NoStop}%
\bibitem [{\citenamefont {Zhou}\ and\ \citenamefont
  {Schulten}(1995)}]{zhou1995Molecular}%
  \BibitemOpen
  \bibfield  {author} {\bibinfo {author} {\bibfnamefont {F.}~\bibnamefont
  {Zhou}}\ and\ \bibinfo {author} {\bibfnamefont {K.}~\bibnamefont
  {Schulten}},\ }\bibfield  {title} {\enquote {\bibinfo {title} {Molecular
  {{Dynamics Study}} of a {{Membrane-Water Interface}}},}\ }\href
  {https://doi.org/10.1021/j100007a059} {\bibfield  {journal} {\bibinfo
  {journal} {J. Phys. Chem.}\ }\textbf {\bibinfo {volume} {99}},\ \bibinfo
  {pages} {2194--2207} (\bibinfo {year} {1995})}\BibitemShut {NoStop}%
\bibitem [{\citenamefont {Jakobsson}(1997)}]{jakobsson1997Computer}%
  \BibitemOpen
  \bibfield  {author} {\bibinfo {author} {\bibfnamefont {E.}~\bibnamefont
  {Jakobsson}},\ }\bibfield  {title} {\enquote {\bibinfo {title} {Computer
  simulation studies of biological membranes: Progress, promise and
  pitfalls},}\ }\href {https://doi.org/10.1016/S0968-0004(97)01096-7}
  {\bibfield  {journal} {\bibinfo  {journal} {Trends Biochem. Sci.}\ }\textbf
  {\bibinfo {volume} {22}},\ \bibinfo {pages} {339--344} (\bibinfo {year}
  {1997})}\BibitemShut {NoStop}%
\bibitem [{\citenamefont {Pandit}, \citenamefont {Bostick},\ and\ \citenamefont
  {Berkowitz}(2003)}]{pandit2003Algorithm}%
  \BibitemOpen
  \bibfield  {author} {\bibinfo {author} {\bibfnamefont {S.~A.}\ \bibnamefont
  {Pandit}}, \bibinfo {author} {\bibfnamefont {D.}~\bibnamefont {Bostick}},\
  and\ \bibinfo {author} {\bibfnamefont {M.~L.}\ \bibnamefont {Berkowitz}},\
  }\bibfield  {title} {\enquote {\bibinfo {title} {An algorithm to describe
  molecular scale rugged surfaces and its application to the study of a
  water/lipid bilayer interface},}\ }\href {https://doi.org/10.1063/1.1582833}
  {\bibfield  {journal} {\bibinfo  {journal} {J. Chem. Phys.}\ }\textbf
  {\bibinfo {volume} {119}},\ \bibinfo {pages} {2199--2205} (\bibinfo {year}
  {2003})}\BibitemShut {NoStop}%
\bibitem [{\citenamefont {Berkowitz}, \citenamefont {Bostick},\ and\
  \citenamefont {Pandit}(2006)}]{berkowitz2006Aqueous}%
  \BibitemOpen
  \bibfield  {author} {\bibinfo {author} {\bibfnamefont {M.~L.}\ \bibnamefont
  {Berkowitz}}, \bibinfo {author} {\bibfnamefont {D.~L.}\ \bibnamefont
  {Bostick}},\ and\ \bibinfo {author} {\bibfnamefont {S.}~\bibnamefont
  {Pandit}},\ }\bibfield  {title} {\enquote {\bibinfo {title} {Aqueous
  {{Solutions}} next to {{Phospholipid Membrane Surfaces}}: {{Insights}} from
  {{Simulations}}},}\ }\href {https://doi.org/10.1021/cr0403638} {\bibfield
  {journal} {\bibinfo  {journal} {Chem. Rev.}\ }\textbf {\bibinfo {volume}
  {106}},\ \bibinfo {pages} {1527--1539} (\bibinfo {year} {2006})}\BibitemShut
  {NoStop}%
\bibitem [{\citenamefont {Matubayasi}, \citenamefont {Shinoda},\ and\
  \citenamefont {Nakahara}(2008)}]{matubayasi2008Freeenergy}%
  \BibitemOpen
  \bibfield  {author} {\bibinfo {author} {\bibfnamefont {N.}~\bibnamefont
  {Matubayasi}}, \bibinfo {author} {\bibfnamefont {W.}~\bibnamefont
  {Shinoda}},\ and\ \bibinfo {author} {\bibfnamefont {M.}~\bibnamefont
  {Nakahara}},\ }\bibfield  {title} {\enquote {\bibinfo {title} {Free-energy
  analysis of the molecular binding into lipid membrane with the method of
  energy representation},}\ }\href {https://doi.org/10.1063/1.2919117}
  {\bibfield  {journal} {\bibinfo  {journal} {J. Chem. Phys.}\ }\textbf
  {\bibinfo {volume} {128}},\ \bibinfo {pages} {195107} (\bibinfo {year}
  {2008})}\BibitemShut {NoStop}%
\bibitem [{\citenamefont {Marrink}\ \emph {et~al.}(2019)\citenamefont
  {Marrink}, \citenamefont {Corradi}, \citenamefont {Souza}, \citenamefont
  {Ing{\'o}lfsson}, \citenamefont {Tieleman},\ and\ \citenamefont
  {Sansom}}]{marrink2019Computational}%
  \BibitemOpen
  \bibfield  {author} {\bibinfo {author} {\bibfnamefont {S.~J.}\ \bibnamefont
  {Marrink}}, \bibinfo {author} {\bibfnamefont {V.}~\bibnamefont {Corradi}},
  \bibinfo {author} {\bibfnamefont {P.~C.}\ \bibnamefont {Souza}}, \bibinfo
  {author} {\bibfnamefont {H.~I.}\ \bibnamefont {Ing{\'o}lfsson}}, \bibinfo
  {author} {\bibfnamefont {D.~P.}\ \bibnamefont {Tieleman}},\ and\ \bibinfo
  {author} {\bibfnamefont {M.~S.}\ \bibnamefont {Sansom}},\ }\bibfield  {title}
  {\enquote {\bibinfo {title} {Computational {{Modeling}} of {{Realistic Cell
  Membranes}}},}\ }\href {https://doi.org/10.1021/acs.chemrev.8b00460}
  {\bibfield  {journal} {\bibinfo  {journal} {Chem. Rev.}\ }\textbf {\bibinfo
  {volume} {119}},\ \bibinfo {pages} {6184--6226} (\bibinfo {year}
  {2019})}\BibitemShut {NoStop}%
\bibitem [{\citenamefont {Karathanou}\ and\ \citenamefont
  {Bondar}(2022)}]{karathanou2022Algorithm}%
  \BibitemOpen
  \bibfield  {author} {\bibinfo {author} {\bibfnamefont {K.}~\bibnamefont
  {Karathanou}}\ and\ \bibinfo {author} {\bibfnamefont {A.-N.}\ \bibnamefont
  {Bondar}},\ }\bibfield  {title} {\enquote {\bibinfo {title} {Algorithm to
  catalogue topologies of dynamic lipid hydrogen-bond networks},}\ }\href
  {https://doi.org/10.1016/j.bbamem.2022.183859} {\bibfield  {journal}
  {\bibinfo  {journal} {Biochim. Biophys. Acta}\ }\textbf {\bibinfo {volume}
  {1864}},\ \bibinfo {pages} {183859} (\bibinfo {year} {2022})}\BibitemShut
  {NoStop}%
\bibitem [{\citenamefont {Alper}, \citenamefont {Bassolino-Klimas},\ and\
  \citenamefont {Stouch}(1993)}]{alper1993Limiting}%
  \BibitemOpen
  \bibfield  {author} {\bibinfo {author} {\bibfnamefont {H.~E.}\ \bibnamefont
  {Alper}}, \bibinfo {author} {\bibfnamefont {D.}~\bibnamefont
  {Bassolino-Klimas}},\ and\ \bibinfo {author} {\bibfnamefont {T.~R.}\
  \bibnamefont {Stouch}},\ }\bibfield  {title} {\enquote {\bibinfo {title} {The
  limiting behavior of water hydrating a phospholipid monolayer: {{A}} computer
  simulation study},}\ }\href {https://doi.org/10.1063/1.465947} {\bibfield
  {journal} {\bibinfo  {journal} {J. Chem. Phys.}\ }\textbf {\bibinfo {volume}
  {99}},\ \bibinfo {pages} {5547--5559} (\bibinfo {year} {1993})}\BibitemShut
  {NoStop}%
\bibitem [{\citenamefont {{Pasenkiewicz-Gierula}}\ \emph
  {et~al.}(1997)\citenamefont {{Pasenkiewicz-Gierula}}, \citenamefont
  {Takaoka}, \citenamefont {Miyagawa}, \citenamefont {Kitamura},\ and\
  \citenamefont {Kusumi}}]{pasenkiewicz-gierula1997Hydrogen}%
  \BibitemOpen
  \bibfield  {author} {\bibinfo {author} {\bibfnamefont {M.}~\bibnamefont
  {{Pasenkiewicz-Gierula}}}, \bibinfo {author} {\bibfnamefont {Y.}~\bibnamefont
  {Takaoka}}, \bibinfo {author} {\bibfnamefont {H.}~\bibnamefont {Miyagawa}},
  \bibinfo {author} {\bibfnamefont {K.}~\bibnamefont {Kitamura}},\ and\
  \bibinfo {author} {\bibfnamefont {A.}~\bibnamefont {Kusumi}},\ }\bibfield
  {title} {\enquote {\bibinfo {title} {Hydrogen {{Bonding}} of {{Water}} to
  {{Phosphatidylcholine}} in the {{Membrane As Studied}} by a {{Molecular
  Dynamics Simulation}}:\, {{Location}}, {{Geometry}}, and {{Lipid}}-{{Lipid
  Bridging}} via {{Hydrogen-Bonded Water}}},}\ }\href
  {https://doi.org/10.1021/jp962099v} {\bibfield  {journal} {\bibinfo
  {journal} {J. Phys. Chem. A}\ }\textbf {\bibinfo {volume} {101}},\ \bibinfo
  {pages} {3677--3691} (\bibinfo {year} {1997})}\BibitemShut {NoStop}%
\bibitem [{\citenamefont {Feller}(2000)}]{feller2000Molecular}%
  \BibitemOpen
  \bibfield  {author} {\bibinfo {author} {\bibfnamefont {S.~E.}\ \bibnamefont
  {Feller}},\ }\bibfield  {title} {\enquote {\bibinfo {title} {Molecular
  dynamics simulations of lipid bilayers},}\ }\href
  {https://doi.org/10.1016/S1359-0294(00)00058-3} {\bibfield  {journal}
  {\bibinfo  {journal} {Curr. Opin. Colloid Interface Sci.}\ }\textbf {\bibinfo
  {volume} {5}},\ \bibinfo {pages} {217--223} (\bibinfo {year}
  {2000})}\BibitemShut {NoStop}%
\bibitem [{\citenamefont {Lopez}\ \emph {et~al.}(2004)\citenamefont {Lopez},
  \citenamefont {Nielsen}, \citenamefont {Klein},\ and\ \citenamefont
  {Moore}}]{lopez2004Hydrogen}%
  \BibitemOpen
  \bibfield  {author} {\bibinfo {author} {\bibfnamefont {C.~F.}\ \bibnamefont
  {Lopez}}, \bibinfo {author} {\bibfnamefont {S.~O.}\ \bibnamefont {Nielsen}},
  \bibinfo {author} {\bibfnamefont {M.~L.}\ \bibnamefont {Klein}},\ and\
  \bibinfo {author} {\bibfnamefont {P.~B.}\ \bibnamefont {Moore}},\ }\bibfield
  {title} {\enquote {\bibinfo {title} {Hydrogen {{Bonding Structure}} and
  {{Dynamics}} of {{Water}} at the {{Dimyristoylphosphatidylcholine Lipid
  Bilayer Surface}} from a {{Molecular Dynamics Simulation}}},}\ }\href
  {https://doi.org/10.1021/jp037618q} {\bibfield  {journal} {\bibinfo
  {journal} {J. Phys. Chem. B}\ }\textbf {\bibinfo {volume} {108}},\ \bibinfo
  {pages} {6603--6610} (\bibinfo {year} {2004})}\BibitemShut {NoStop}%
\bibitem [{\citenamefont {Bhide}\ and\ \citenamefont
  {Berkowitz}(2005)}]{bhide2005Structure}%
  \BibitemOpen
  \bibfield  {author} {\bibinfo {author} {\bibfnamefont {S.~Y.}\ \bibnamefont
  {Bhide}}\ and\ \bibinfo {author} {\bibfnamefont {M.~L.}\ \bibnamefont
  {Berkowitz}},\ }\bibfield  {title} {\enquote {\bibinfo {title} {Structure and
  dynamics of water at the interface with phospholipid bilayers},}\ }\href
  {https://doi.org/10.1063/1.2132277} {\bibfield  {journal} {\bibinfo
  {journal} {J. Chem. Phys.}\ }\textbf {\bibinfo {volume} {123}},\ \bibinfo
  {pages} {224702} (\bibinfo {year} {2005})}\BibitemShut {NoStop}%
\bibitem [{\citenamefont {Volkov}\ \emph {et~al.}(2006)\citenamefont {Volkov},
  \citenamefont {Nuti}, \citenamefont {Takaoka}, \citenamefont {Chelli},
  \citenamefont {Papini},\ and\ \citenamefont {Righini}}]{volkov2006Hydration}%
  \BibitemOpen
  \bibfield  {author} {\bibinfo {author} {\bibfnamefont {V.~V.}\ \bibnamefont
  {Volkov}}, \bibinfo {author} {\bibfnamefont {F.}~\bibnamefont {Nuti}},
  \bibinfo {author} {\bibfnamefont {Y.}~\bibnamefont {Takaoka}}, \bibinfo
  {author} {\bibfnamefont {R.}~\bibnamefont {Chelli}}, \bibinfo {author}
  {\bibfnamefont {A.~M.}\ \bibnamefont {Papini}},\ and\ \bibinfo {author}
  {\bibfnamefont {R.}~\bibnamefont {Righini}},\ }\bibfield  {title} {\enquote
  {\bibinfo {title} {Hydration and {{Hydrogen Bonding}} of {{Carbonyls}} in
  {{Dimyristoyl-Phosphatidylcholine Bilayer}}},}\ }\href
  {https://doi.org/10.1021/ja0614621} {\bibfield  {journal} {\bibinfo
  {journal} {J. Am. Chem. Soc.}\ }\textbf {\bibinfo {volume} {128}},\ \bibinfo
  {pages} {9466--9471} (\bibinfo {year} {2006})}\BibitemShut {NoStop}%
\bibitem [{\citenamefont {{von Hansen}}, \citenamefont {Gekle},\ and\
  \citenamefont {Netz}(2013)}]{vonhansen2013Anomalous}%
  \BibitemOpen
  \bibfield  {author} {\bibinfo {author} {\bibfnamefont {Y.}~\bibnamefont {{von
  Hansen}}}, \bibinfo {author} {\bibfnamefont {S.}~\bibnamefont {Gekle}},\ and\
  \bibinfo {author} {\bibfnamefont {R.~R.}\ \bibnamefont {Netz}},\ }\bibfield
  {title} {\enquote {\bibinfo {title} {Anomalous {{Anisotropic Diffusion
  Dynamics}} of {{Hydration Water}} at {{Lipid Membranes}}},}\ }\href
  {https://doi.org/10.1103/PhysRevLett.111.118103} {\bibfield  {journal}
  {\bibinfo  {journal} {Phys. Rev. Lett.}\ }\textbf {\bibinfo {volume} {111}},\
  \bibinfo {pages} {118103} (\bibinfo {year} {2013})}\BibitemShut {NoStop}%
\bibitem [{\citenamefont {Srivastava}\ and\ \citenamefont
  {Debnath}(2018)}]{srivastava2018Hydrationa}%
  \BibitemOpen
  \bibfield  {author} {\bibinfo {author} {\bibfnamefont {A.}~\bibnamefont
  {Srivastava}}\ and\ \bibinfo {author} {\bibfnamefont {A.}~\bibnamefont
  {Debnath}},\ }\bibfield  {title} {\enquote {\bibinfo {title} {Hydration
  dynamics of a lipid membrane: {{Hydrogen}} bond networks and lipid-lipid
  associations},}\ }\href {https://doi.org/10.1063/1.5011803} {\bibfield
  {journal} {\bibinfo  {journal} {J. Chem. Phys.}\ }\textbf {\bibinfo {volume}
  {148}},\ \bibinfo {pages} {094901} (\bibinfo {year} {2018})}\BibitemShut
  {NoStop}%
\bibitem [{\citenamefont {Calero}\ and\ \citenamefont
  {Franzese}(2019)}]{calero2019Membranes}%
  \BibitemOpen
  \bibfield  {author} {\bibinfo {author} {\bibfnamefont {C.}~\bibnamefont
  {Calero}}\ and\ \bibinfo {author} {\bibfnamefont {G.}~\bibnamefont
  {Franzese}},\ }\bibfield  {title} {\enquote {\bibinfo {title} {Membranes with
  different hydration levels: {{The}} interface between bound and unbound
  hydration water},}\ }\href {https://doi.org/10.1016/j.molliq.2018.10.074}
  {\bibfield  {journal} {\bibinfo  {journal} {J. Mol. Liq.}\ }\textbf {\bibinfo
  {volume} {273}},\ \bibinfo {pages} {488--496} (\bibinfo {year}
  {2019})}\BibitemShut {NoStop}%
\bibitem [{\citenamefont {Lee}\ \emph {et~al.}(2019)\citenamefont {Lee},
  \citenamefont {Kundu}, \citenamefont {Jeon},\ and\ \citenamefont
  {Cho}}]{lee2019Watera}%
  \BibitemOpen
  \bibfield  {author} {\bibinfo {author} {\bibfnamefont {E.}~\bibnamefont
  {Lee}}, \bibinfo {author} {\bibfnamefont {A.}~\bibnamefont {Kundu}}, \bibinfo
  {author} {\bibfnamefont {J.}~\bibnamefont {Jeon}},\ and\ \bibinfo {author}
  {\bibfnamefont {M.}~\bibnamefont {Cho}},\ }\bibfield  {title} {\enquote
  {\bibinfo {title} {Water hydrogen-bonding structure and dynamics near lipid
  multibilayer surface: {{Molecular}} dynamics simulation study with direct
  experimental comparison},}\ }\href {https://doi.org/10.1063/1.5120456}
  {\bibfield  {journal} {\bibinfo  {journal} {J. Chem. Phys.}\ }\textbf
  {\bibinfo {volume} {151}},\ \bibinfo {pages} {114705} (\bibinfo {year}
  {2019})}\BibitemShut {NoStop}%
\bibitem [{\citenamefont {An}\ \emph {et~al.}(2021)\citenamefont {An},
  \citenamefont {Majumder}, \citenamefont {McNeely}, \citenamefont {Yang},
  \citenamefont {Puri}, \citenamefont {He}, \citenamefont {Liang},
  \citenamefont {Snyder}, \citenamefont {Straub},\ and\ \citenamefont
  {Reinhard}}]{an2021Interfacial}%
  \BibitemOpen
  \bibfield  {author} {\bibinfo {author} {\bibfnamefont {X.}~\bibnamefont
  {An}}, \bibinfo {author} {\bibfnamefont {A.}~\bibnamefont {Majumder}},
  \bibinfo {author} {\bibfnamefont {J.}~\bibnamefont {McNeely}}, \bibinfo
  {author} {\bibfnamefont {J.}~\bibnamefont {Yang}}, \bibinfo {author}
  {\bibfnamefont {T.}~\bibnamefont {Puri}}, \bibinfo {author} {\bibfnamefont
  {Z.}~\bibnamefont {He}}, \bibinfo {author} {\bibfnamefont {T.}~\bibnamefont
  {Liang}}, \bibinfo {author} {\bibfnamefont {J.~K.}\ \bibnamefont {Snyder}},
  \bibinfo {author} {\bibfnamefont {J.~E.}\ \bibnamefont {Straub}},\ and\
  \bibinfo {author} {\bibfnamefont {B.~M.}\ \bibnamefont {Reinhard}},\
  }\bibfield  {title} {\enquote {\bibinfo {title} {Interfacial hydration
  determines orientational and functional dimorphism of sterol-derived
  {{Raman}} tags in lipid-coated nanoparticles},}\ }\href
  {https://doi.org/10.1073/pnas.2105913118} {\bibfield  {journal} {\bibinfo
  {journal} {Proc. Natl. Acad. Sci. U.S.A.}\ }\textbf {\bibinfo {volume}
  {118}},\ \bibinfo {pages} {e2105913118} (\bibinfo {year} {2021})}\BibitemShut
  {NoStop}%
\bibitem [{\citenamefont {Higuchi}\ \emph {et~al.}(2021)\citenamefont
  {Higuchi}, \citenamefont {Asano}, \citenamefont {Kuwahara},\ and\
  \citenamefont {Hishida}}]{higuchi2021Rotational}%
  \BibitemOpen
  \bibfield  {author} {\bibinfo {author} {\bibfnamefont {Y.}~\bibnamefont
  {Higuchi}}, \bibinfo {author} {\bibfnamefont {Y.}~\bibnamefont {Asano}},
  \bibinfo {author} {\bibfnamefont {T.}~\bibnamefont {Kuwahara}},\ and\
  \bibinfo {author} {\bibfnamefont {M.}~\bibnamefont {Hishida}},\ }\bibfield
  {title} {\enquote {\bibinfo {title} {Rotational {{Dynamics}} of {{Water}} at
  the {{Phospholipid Bilayer Depending}} on the {{Head Groups Studied}} by
  {{Molecular Dynamics Simulations}}},}\ }\href
  {https://doi.org/10.1021/acs.langmuir.1c00417} {\bibfield  {journal}
  {\bibinfo  {journal} {Langmuir}\ }\textbf {\bibinfo {volume} {37}},\ \bibinfo
  {pages} {5329--5338} (\bibinfo {year} {2021})}\BibitemShut {NoStop}%
\bibitem [{\citenamefont {Malik}\ and\ \citenamefont
  {Debnath}(2021)}]{malik2021Dehydration}%
  \BibitemOpen
  \bibfield  {author} {\bibinfo {author} {\bibfnamefont {S.}~\bibnamefont
  {Malik}}\ and\ \bibinfo {author} {\bibfnamefont {A.}~\bibnamefont
  {Debnath}},\ }\bibfield  {title} {\enquote {\bibinfo {title} {Dehydration
  induced dynamical heterogeneity and ordering mechanism of lipid bilayers},}\
  }\href {https://doi.org/10.1063/5.0044614} {\bibfield  {journal} {\bibinfo
  {journal} {J. Chem. Phys.}\ }\textbf {\bibinfo {volume} {154}},\ \bibinfo
  {pages} {174904} (\bibinfo {year} {2021})}\BibitemShut {NoStop}%
\bibitem [{\citenamefont {Malik}, \citenamefont {Karmakar},\ and\ \citenamefont
  {Debnath}(2023)}]{malik2023Relaxation}%
  \BibitemOpen
  \bibfield  {author} {\bibinfo {author} {\bibfnamefont {S.}~\bibnamefont
  {Malik}}, \bibinfo {author} {\bibfnamefont {S.}~\bibnamefont {Karmakar}},\
  and\ \bibinfo {author} {\bibfnamefont {A.}~\bibnamefont {Debnath}},\
  }\bibfield  {title} {\enquote {\bibinfo {title} {Relaxation time scales of
  interfacial water upon fluid to ripple to gel phase transitions of
  bilayers},}\ }\href {https://doi.org/10.1063/5.0138681} {\bibfield  {journal}
  {\bibinfo  {journal} {J. Chem. Phys.}\ }\textbf {\bibinfo {volume} {158}},\
  \bibinfo {pages} {114503} (\bibinfo {year} {2023})}\BibitemShut {NoStop}%
\bibitem [{\citenamefont {Tu}, \citenamefont {Klein},\ and\ \citenamefont
  {Tobias}(1998)}]{tu1998ConstantPressure}%
  \BibitemOpen
  \bibfield  {author} {\bibinfo {author} {\bibfnamefont {K.}~\bibnamefont
  {Tu}}, \bibinfo {author} {\bibfnamefont {M.~L.}\ \bibnamefont {Klein}},\ and\
  \bibinfo {author} {\bibfnamefont {D.~J.}\ \bibnamefont {Tobias}},\ }\bibfield
   {title} {\enquote {\bibinfo {title} {Constant-{{Pressure Molecular Dynamics
  Investigation}} of {{Cholesterol Effects}} in a
  {{Dipalmitoylphosphatidylcholine Bilayer}}},}\ }\href
  {https://doi.org/10.1016/S0006-3495(98)77657-X} {\bibfield  {journal}
  {\bibinfo  {journal} {Biophys. J.}\ }\textbf {\bibinfo {volume} {75}},\
  \bibinfo {pages} {2147--2156} (\bibinfo {year} {1998})}\BibitemShut {NoStop}%
\bibitem [{\citenamefont {Chiu}\ \emph {et~al.}(2002)\citenamefont {Chiu},
  \citenamefont {Jakobsson}, \citenamefont {Mashl},\ and\ \citenamefont
  {Scott}}]{chiu2002CholesterolInduced}%
  \BibitemOpen
  \bibfield  {author} {\bibinfo {author} {\bibfnamefont {S.~W.}\ \bibnamefont
  {Chiu}}, \bibinfo {author} {\bibfnamefont {E.}~\bibnamefont {Jakobsson}},
  \bibinfo {author} {\bibfnamefont {R.~J.}\ \bibnamefont {Mashl}},\ and\
  \bibinfo {author} {\bibfnamefont {H.~L.}\ \bibnamefont {Scott}},\ }\bibfield
  {title} {\enquote {\bibinfo {title} {Cholesterol-{{Induced Modifications}} in
  {{Lipid Bilayers}}: {{A Simulation Study}}},}\ }\href
  {https://doi.org/10.1016/S0006-3495(02)73949-0} {\bibfield  {journal}
  {\bibinfo  {journal} {Biophys. J.}\ }\textbf {\bibinfo {volume} {83}},\
  \bibinfo {pages} {1842--1853} (\bibinfo {year} {2002})}\BibitemShut {NoStop}%
\bibitem [{\citenamefont {Hofs{\"a}{\ss}}, \citenamefont {Lindahl},\ and\
  \citenamefont {Edholm}(2003)}]{hofsass2003Molecular}%
  \BibitemOpen
  \bibfield  {author} {\bibinfo {author} {\bibfnamefont {C.}~\bibnamefont
  {Hofs{\"a}{\ss}}}, \bibinfo {author} {\bibfnamefont {E.}~\bibnamefont
  {Lindahl}},\ and\ \bibinfo {author} {\bibfnamefont {O.}~\bibnamefont
  {Edholm}},\ }\bibfield  {title} {\enquote {\bibinfo {title} {Molecular
  {{Dynamics Simulations}} of {{Phospholipid Bilayers}} with
  {{Cholesterol}}},}\ }\href {https://doi.org/10.1016/S0006-3495(03)75025-5}
  {\bibfield  {journal} {\bibinfo  {journal} {Biophys. J.}\ }\textbf {\bibinfo
  {volume} {84}},\ \bibinfo {pages} {2192--2206} (\bibinfo {year}
  {2003})}\BibitemShut {NoStop}%
\bibitem [{\citenamefont {Pandit}, \citenamefont {Bostick},\ and\ \citenamefont
  {Berkowitz}(2004)}]{pandit2004Complexation}%
  \BibitemOpen
  \bibfield  {author} {\bibinfo {author} {\bibfnamefont {S.~A.}\ \bibnamefont
  {Pandit}}, \bibinfo {author} {\bibfnamefont {D.}~\bibnamefont {Bostick}},\
  and\ \bibinfo {author} {\bibfnamefont {M.~L.}\ \bibnamefont {Berkowitz}},\
  }\bibfield  {title} {\enquote {\bibinfo {title} {Complexation of
  {{Phosphatidylcholine Lipids}} with {{Cholesterol}}},}\ }\href
  {https://doi.org/10.1016/S0006-3495(04)74206-X} {\bibfield  {journal}
  {\bibinfo  {journal} {Biophys. J.}\ }\textbf {\bibinfo {volume} {86}},\
  \bibinfo {pages} {1345--1356} (\bibinfo {year} {2004})}\BibitemShut {NoStop}%
\bibitem [{\citenamefont {Alwarawrah}, \citenamefont {Dai},\ and\ \citenamefont
  {Huang}(2010)}]{alwarawrah2010Molecular}%
  \BibitemOpen
  \bibfield  {author} {\bibinfo {author} {\bibfnamefont {M.}~\bibnamefont
  {Alwarawrah}}, \bibinfo {author} {\bibfnamefont {J.}~\bibnamefont {Dai}},\
  and\ \bibinfo {author} {\bibfnamefont {J.}~\bibnamefont {Huang}},\ }\bibfield
   {title} {\enquote {\bibinfo {title} {A {{Molecular View}} of the
  {{Cholesterol Condensing Effect}} in {{DOPC Lipid Bilayers}}},}\ }\href
  {https://doi.org/10.1021/jp101415g} {\bibfield  {journal} {\bibinfo
  {journal} {J. Phys. Chem. B}\ }\textbf {\bibinfo {volume} {114}},\ \bibinfo
  {pages} {7516--7523} (\bibinfo {year} {2010})}\BibitemShut {NoStop}%
\bibitem [{\citenamefont {Saito}\ and\ \citenamefont
  {Shinoda}(2011)}]{saito2011Cholesterol}%
  \BibitemOpen
  \bibfield  {author} {\bibinfo {author} {\bibfnamefont {H.}~\bibnamefont
  {Saito}}\ and\ \bibinfo {author} {\bibfnamefont {W.}~\bibnamefont
  {Shinoda}},\ }\bibfield  {title} {\enquote {\bibinfo {title} {Cholesterol
  {{Effect}} on {{Water Permeability}} through {{DPPC}} and {{PSM Lipid
  Bilayers}}: {{A Molecular Dynamics Study}}},}\ }\href
  {https://doi.org/10.1021/jp201611p} {\bibfield  {journal} {\bibinfo
  {journal} {J. Phys. Chem. B}\ }\textbf {\bibinfo {volume} {115}},\ \bibinfo
  {pages} {15241--15250} (\bibinfo {year} {2011})}\BibitemShut {NoStop}%
\bibitem [{\citenamefont {Sodt}, \citenamefont {Pastor},\ and\ \citenamefont
  {Lyman}(2015)}]{sodt2015Hexagonal}%
  \BibitemOpen
  \bibfield  {author} {\bibinfo {author} {\bibfnamefont {A.~J.}\ \bibnamefont
  {Sodt}}, \bibinfo {author} {\bibfnamefont {R.~W.}\ \bibnamefont {Pastor}},\
  and\ \bibinfo {author} {\bibfnamefont {E.}~\bibnamefont {Lyman}},\ }\bibfield
   {title} {\enquote {\bibinfo {title} {Hexagonal {{Substructure}} and
  {{Hydrogen Bonding}} in {{Liquid-Ordered Phases Containing Palmitoyl
  Sphingomyelin}}},}\ }\href {https://doi.org/10.1016/j.bpj.2015.07.036}
  {\bibfield  {journal} {\bibinfo  {journal} {Biophys. J.}\ }\textbf {\bibinfo
  {volume} {109}},\ \bibinfo {pages} {948--955} (\bibinfo {year}
  {2015})}\BibitemShut {NoStop}%
\bibitem [{\citenamefont {Boughter}\ \emph {et~al.}(2016)\citenamefont
  {Boughter}, \citenamefont {{Monje-Galvan}}, \citenamefont {Im},\ and\
  \citenamefont {Klauda}}]{boughter2016Influence}%
  \BibitemOpen
  \bibfield  {author} {\bibinfo {author} {\bibfnamefont {C.~T.}\ \bibnamefont
  {Boughter}}, \bibinfo {author} {\bibfnamefont {V.}~\bibnamefont
  {{Monje-Galvan}}}, \bibinfo {author} {\bibfnamefont {W.}~\bibnamefont {Im}},\
  and\ \bibinfo {author} {\bibfnamefont {J.~B.}\ \bibnamefont {Klauda}},\
  }\bibfield  {title} {\enquote {\bibinfo {title} {Influence of {{Cholesterol}}
  on {{Phospholipid Bilayer Structure}} and {{Dynamics}}},}\ }\href
  {https://doi.org/10.1021/acs.jpcb.6b08574} {\bibfield  {journal} {\bibinfo
  {journal} {J. Phys. Chem. B}\ }\textbf {\bibinfo {volume} {120}},\ \bibinfo
  {pages} {11761--11772} (\bibinfo {year} {2016})}\BibitemShut {NoStop}%
\bibitem [{\citenamefont {Elola}\ and\ \citenamefont
  {Rodriguez}(2018)}]{elola2018Influence}%
  \BibitemOpen
  \bibfield  {author} {\bibinfo {author} {\bibfnamefont {M.~D.}\ \bibnamefont
  {Elola}}\ and\ \bibinfo {author} {\bibfnamefont {J.}~\bibnamefont
  {Rodriguez}},\ }\bibfield  {title} {\enquote {\bibinfo {title} {Influence of
  {{Cholesterol}} on the {{Dynamics}} of {{Hydration}} in {{Phospholipid
  Bilayers}}},}\ }\href {https://doi.org/10.1021/acs.jpcb.8b00360} {\bibfield
  {journal} {\bibinfo  {journal} {J. Phys. Chem. B}\ }\textbf {\bibinfo
  {volume} {122}},\ \bibinfo {pages} {5897--5907} (\bibinfo {year}
  {2018})}\BibitemShut {NoStop}%
\bibitem [{\citenamefont {Pantelopulos}\ and\ \citenamefont
  {Straub}(2018)}]{pantelopulos2018Regimes}%
  \BibitemOpen
  \bibfield  {author} {\bibinfo {author} {\bibfnamefont {G.~A.}\ \bibnamefont
  {Pantelopulos}}\ and\ \bibinfo {author} {\bibfnamefont {J.~E.}\ \bibnamefont
  {Straub}},\ }\bibfield  {title} {\enquote {\bibinfo {title} {Regimes of
  {{Complex Lipid Bilayer Phases Induced}} by {{Cholesterol Concentration}} in
  {{MD Simulation}}},}\ }\href {https://doi.org/10.1016/j.bpj.2018.10.011}
  {\bibfield  {journal} {\bibinfo  {journal} {Biophys. J.}\ }\textbf {\bibinfo
  {volume} {115}},\ \bibinfo {pages} {2167--2178} (\bibinfo {year}
  {2018})}\BibitemShut {NoStop}%
\bibitem [{\citenamefont {P{\"a}slack}\ \emph {et~al.}(2019)\citenamefont
  {P{\"a}slack}, \citenamefont {Smith}, \citenamefont {Heyden},\ and\
  \citenamefont {Sch{\"a}fer}}]{paslack2019Hydrationmediated}%
  \BibitemOpen
  \bibfield  {author} {\bibinfo {author} {\bibfnamefont {C.}~\bibnamefont
  {P{\"a}slack}}, \bibinfo {author} {\bibfnamefont {J.~C.}\ \bibnamefont
  {Smith}}, \bibinfo {author} {\bibfnamefont {M.}~\bibnamefont {Heyden}},\ and\
  \bibinfo {author} {\bibfnamefont {L.~V.}\ \bibnamefont {Sch{\"a}fer}},\
  }\bibfield  {title} {\enquote {\bibinfo {title} {Hydration-mediated
  stiffening of collective membrane dynamics by cholesterol},}\ }\href
  {https://doi.org/10.1039/C9CP01431D} {\bibfield  {journal} {\bibinfo
  {journal} {Phys. Chem. Chem. Phys.}\ }\textbf {\bibinfo {volume} {21}},\
  \bibinfo {pages} {10370--10376} (\bibinfo {year} {2019})}\BibitemShut
  {NoStop}%
\bibitem [{\citenamefont {Kumari}, \citenamefont {Kumari},\ and\ \citenamefont
  {Kashyap}(2019)}]{kumari2019Countereffects}%
  \BibitemOpen
  \bibfield  {author} {\bibinfo {author} {\bibfnamefont {P.}~\bibnamefont
  {Kumari}}, \bibinfo {author} {\bibfnamefont {M.}~\bibnamefont {Kumari}},\
  and\ \bibinfo {author} {\bibfnamefont {H.~K.}\ \bibnamefont {Kashyap}},\
  }\bibfield  {title} {\enquote {\bibinfo {title} {Counter-effects of
  {{Ethanol}} and {{Cholesterol}} on the {{Heterogeneous
  PSM}}{\textendash}{{POPC Lipid Membrane}}: {{A Molecular Dynamics Simulation
  Study}}},}\ }\href {https://doi.org/10.1021/acs.jpcb.9b07107} {\bibfield
  {journal} {\bibinfo  {journal} {J. Phys. Chem. B}\ }\textbf {\bibinfo
  {volume} {123}},\ \bibinfo {pages} {9616--9628} (\bibinfo {year}
  {2019})}\BibitemShut {NoStop}%
\bibitem [{\citenamefont {Elkins}\ \emph {et~al.}(2021)\citenamefont {Elkins},
  \citenamefont {Bandara}, \citenamefont {Pantelopulos}, \citenamefont
  {Straub},\ and\ \citenamefont {Hong}}]{elkins2021Direct}%
  \BibitemOpen
  \bibfield  {author} {\bibinfo {author} {\bibfnamefont {M.~R.}\ \bibnamefont
  {Elkins}}, \bibinfo {author} {\bibfnamefont {A.}~\bibnamefont {Bandara}},
  \bibinfo {author} {\bibfnamefont {G.~A.}\ \bibnamefont {Pantelopulos}},
  \bibinfo {author} {\bibfnamefont {J.~E.}\ \bibnamefont {Straub}},\ and\
  \bibinfo {author} {\bibfnamefont {M.}~\bibnamefont {Hong}},\ }\bibfield
  {title} {\enquote {\bibinfo {title} {Direct {{Observation}} of {{Cholesterol
  Dimers}} and {{Tetramers}} in {{Lipid Bilayers}}},}\ }\href
  {https://doi.org/10.1021/acs.jpcb.0c10631} {\bibfield  {journal} {\bibinfo
  {journal} {J. Phys. Chem. B}\ }\textbf {\bibinfo {volume} {125}},\ \bibinfo
  {pages} {1825--1837} (\bibinfo {year} {2021})}\BibitemShut {NoStop}%
\bibitem [{\citenamefont {Antila}\ \emph {et~al.}(2022)\citenamefont {Antila},
  \citenamefont {Wurl}, \citenamefont {Ollila}, \citenamefont {Miettinen},\
  and\ \citenamefont {Ferreira}}]{antila2022Rotational}%
  \BibitemOpen
  \bibfield  {author} {\bibinfo {author} {\bibfnamefont {H.~S.}\ \bibnamefont
  {Antila}}, \bibinfo {author} {\bibfnamefont {A.}~\bibnamefont {Wurl}},
  \bibinfo {author} {\bibfnamefont {O.~S.}\ \bibnamefont {Ollila}}, \bibinfo
  {author} {\bibfnamefont {M.~S.}\ \bibnamefont {Miettinen}},\ and\ \bibinfo
  {author} {\bibfnamefont {T.~M.}\ \bibnamefont {Ferreira}},\ }\bibfield
  {title} {\enquote {\bibinfo {title} {Rotational decoupling between the
  hydrophilic and hydrophobic regions in lipid membranes},}\ }\href
  {https://doi.org/10.1016/j.bpj.2021.12.003} {\bibfield  {journal} {\bibinfo
  {journal} {Biophys. J.}\ }\textbf {\bibinfo {volume} {121}},\ \bibinfo
  {pages} {68--78} (\bibinfo {year} {2022})}\BibitemShut {NoStop}%
\bibitem [{\citenamefont {Cheng}\ \emph {et~al.}(2014)\citenamefont {Cheng},
  \citenamefont {Olijve}, \citenamefont {Kausik},\ and\ \citenamefont
  {Han}}]{cheng2014Cholesterol}%
  \BibitemOpen
  \bibfield  {author} {\bibinfo {author} {\bibfnamefont {C.-Y.}\ \bibnamefont
  {Cheng}}, \bibinfo {author} {\bibfnamefont {L.~L.~C.}\ \bibnamefont
  {Olijve}}, \bibinfo {author} {\bibfnamefont {R.}~\bibnamefont {Kausik}},\
  and\ \bibinfo {author} {\bibfnamefont {S.}~\bibnamefont {Han}},\ }\bibfield
  {title} {\enquote {\bibinfo {title} {Cholesterol enhances surface water
  diffusion of phospholipid bilayers},}\ }\href
  {https://doi.org/10.1063/1.4897539} {\bibfield  {journal} {\bibinfo
  {journal} {J. Chem. Phys.}\ }\textbf {\bibinfo {volume} {141}},\ \bibinfo
  {pages} {22D513} (\bibinfo {year} {2014})}\BibitemShut {NoStop}%
\bibitem [{\citenamefont {Pyne}, \citenamefont {Pyne},\ and\ \citenamefont
  {Mitra}(2022)}]{pyne2022Addition}%
  \BibitemOpen
  \bibfield  {author} {\bibinfo {author} {\bibfnamefont {S.}~\bibnamefont
  {Pyne}}, \bibinfo {author} {\bibfnamefont {P.}~\bibnamefont {Pyne}},\ and\
  \bibinfo {author} {\bibfnamefont {R.~K.}\ \bibnamefont {Mitra}},\ }\bibfield
  {title} {\enquote {\bibinfo {title} {Addition of cholesterol alters the
  hydration at the surface of model lipids: A spectroscopic investigation},}\
  }\href {https://doi.org/10.1039/D2CP01905A} {\bibfield  {journal} {\bibinfo
  {journal} {Phys. Chem. Chem. Phys.}\ }\textbf {\bibinfo {volume} {24}},\
  \bibinfo {pages} {20381--20389} (\bibinfo {year} {2022})}\BibitemShut
  {NoStop}%
\bibitem [{\citenamefont {Oh}, \citenamefont {Oh},\ and\ \citenamefont
  {Weaver}(2020)}]{oh2020Effect}%
  \BibitemOpen
  \bibfield  {author} {\bibinfo {author} {\bibfnamefont {M.~I.}\ \bibnamefont
  {Oh}}, \bibinfo {author} {\bibfnamefont {C.~I.}\ \bibnamefont {Oh}},\ and\
  \bibinfo {author} {\bibfnamefont {D.~F.}\ \bibnamefont {Weaver}},\ }\bibfield
   {title} {\enquote {\bibinfo {title} {Effect of {{Cholesterol}} on the
  {{Structure}} of {{Networked Water}} at the {{Surface}} of a {{Model Lipid
  Membrane}}},}\ }\href {https://doi.org/10.1021/acs.jpcb.0c01889} {\bibfield
  {journal} {\bibinfo  {journal} {J. Phys. Chem. B}\ }\textbf {\bibinfo
  {volume} {124}},\ \bibinfo {pages} {3686--3694} (\bibinfo {year}
  {2020})}\BibitemShut {NoStop}%
\bibitem [{\citenamefont {Jo}\ \emph {et~al.}(2008)\citenamefont {Jo},
  \citenamefont {Kim}, \citenamefont {Iyer},\ and\ \citenamefont
  {Im}}]{jo2008CHARMM}%
  \BibitemOpen
  \bibfield  {author} {\bibinfo {author} {\bibfnamefont {S.}~\bibnamefont
  {Jo}}, \bibinfo {author} {\bibfnamefont {T.}~\bibnamefont {Kim}}, \bibinfo
  {author} {\bibfnamefont {V.~G.}\ \bibnamefont {Iyer}},\ and\ \bibinfo
  {author} {\bibfnamefont {W.}~\bibnamefont {Im}},\ }\bibfield  {title}
  {\enquote {\bibinfo {title} {{{CHARMM}}-{{GUI}}: {{A}} web-based graphical
  user interface for {{CHARMM}}},}\ }\href {https://doi.org/10.1002/jcc.20945}
  {\bibfield  {journal} {\bibinfo  {journal} {J. Comput. Chem.}\ }\textbf
  {\bibinfo {volume} {29}},\ \bibinfo {pages} {1859--1865} (\bibinfo {year}
  {2008})}\BibitemShut {NoStop}%
\bibitem [{\citenamefont {Jo}\ \emph {et~al.}(2009)\citenamefont {Jo},
  \citenamefont {Lim}, \citenamefont {Klauda},\ and\ \citenamefont
  {Im}}]{jo2009CHARMMGUI}%
  \BibitemOpen
  \bibfield  {author} {\bibinfo {author} {\bibfnamefont {S.}~\bibnamefont
  {Jo}}, \bibinfo {author} {\bibfnamefont {J.~B.}\ \bibnamefont {Lim}},
  \bibinfo {author} {\bibfnamefont {J.~B.}\ \bibnamefont {Klauda}},\ and\
  \bibinfo {author} {\bibfnamefont {W.}~\bibnamefont {Im}},\ }\bibfield
  {title} {\enquote {\bibinfo {title} {{{CHARMM-GUI Membrane Builder}} for
  {{Mixed Bilayers}} and {{Its Application}} to {{Yeast Membranes}}},}\ }\href
  {https://doi.org/10.1016/j.bpj.2009.04.013} {\bibfield  {journal} {\bibinfo
  {journal} {Biophys. J.}\ }\textbf {\bibinfo {volume} {97}},\ \bibinfo {pages}
  {50--58} (\bibinfo {year} {2009})}\BibitemShut {NoStop}%
\bibitem [{\citenamefont {Brooks}\ \emph {et~al.}(2009)\citenamefont {Brooks},
  \citenamefont {Brooks}, \citenamefont {Mackerell}, \citenamefont {Nilsson},
  \citenamefont {Petrella}, \citenamefont {Roux}, \citenamefont {Won},
  \citenamefont {Archontis}, \citenamefont {Bartels}, \citenamefont {Boresch},
  \citenamefont {Caflisch}, \citenamefont {Caves}, \citenamefont {Cui},
  \citenamefont {Dinner}, \citenamefont {Feig}, \citenamefont {Fischer},
  \citenamefont {Gao}, \citenamefont {Hodoscek}, \citenamefont {Im},
  \citenamefont {Kuczera}, \citenamefont {Lazaridis}, \citenamefont {Ma},
  \citenamefont {Ovchinnikov}, \citenamefont {Paci}, \citenamefont {Pastor},
  \citenamefont {Post}, \citenamefont {Pu}, \citenamefont {Schaefer},
  \citenamefont {Tidor}, \citenamefont {Venable}, \citenamefont {Woodcock},
  \citenamefont {Wu}, \citenamefont {Yang}, \citenamefont {York},\ and\
  \citenamefont {Karplus}}]{brooks2009CHARMM}%
  \BibitemOpen
  \bibfield  {author} {\bibinfo {author} {\bibfnamefont {B.~R.}\ \bibnamefont
  {Brooks}}, \bibinfo {author} {\bibfnamefont {C.~L.}\ \bibnamefont {Brooks}},
  \bibinfo {author} {\bibfnamefont {A.~D.}\ \bibnamefont {Mackerell}}, \bibinfo
  {author} {\bibfnamefont {L.}~\bibnamefont {Nilsson}}, \bibinfo {author}
  {\bibfnamefont {R.~J.}\ \bibnamefont {Petrella}}, \bibinfo {author}
  {\bibfnamefont {B.}~\bibnamefont {Roux}}, \bibinfo {author} {\bibfnamefont
  {Y.}~\bibnamefont {Won}}, \bibinfo {author} {\bibfnamefont {G.}~\bibnamefont
  {Archontis}}, \bibinfo {author} {\bibfnamefont {C.}~\bibnamefont {Bartels}},
  \bibinfo {author} {\bibfnamefont {S.}~\bibnamefont {Boresch}}, \bibinfo
  {author} {\bibfnamefont {A.}~\bibnamefont {Caflisch}}, \bibinfo {author}
  {\bibfnamefont {L.}~\bibnamefont {Caves}}, \bibinfo {author} {\bibfnamefont
  {Q.}~\bibnamefont {Cui}}, \bibinfo {author} {\bibfnamefont {A.~R.}\
  \bibnamefont {Dinner}}, \bibinfo {author} {\bibfnamefont {M.}~\bibnamefont
  {Feig}}, \bibinfo {author} {\bibfnamefont {S.}~\bibnamefont {Fischer}},
  \bibinfo {author} {\bibfnamefont {J.}~\bibnamefont {Gao}}, \bibinfo {author}
  {\bibfnamefont {M.}~\bibnamefont {Hodoscek}}, \bibinfo {author}
  {\bibfnamefont {W.}~\bibnamefont {Im}}, \bibinfo {author} {\bibfnamefont
  {K.}~\bibnamefont {Kuczera}}, \bibinfo {author} {\bibfnamefont
  {T.}~\bibnamefont {Lazaridis}}, \bibinfo {author} {\bibfnamefont
  {J.}~\bibnamefont {Ma}}, \bibinfo {author} {\bibfnamefont {V.}~\bibnamefont
  {Ovchinnikov}}, \bibinfo {author} {\bibfnamefont {E.}~\bibnamefont {Paci}},
  \bibinfo {author} {\bibfnamefont {R.~W.}\ \bibnamefont {Pastor}}, \bibinfo
  {author} {\bibfnamefont {C.~B.}\ \bibnamefont {Post}}, \bibinfo {author}
  {\bibfnamefont {J.~Z.}\ \bibnamefont {Pu}}, \bibinfo {author} {\bibfnamefont
  {M.}~\bibnamefont {Schaefer}}, \bibinfo {author} {\bibfnamefont
  {B.}~\bibnamefont {Tidor}}, \bibinfo {author} {\bibfnamefont {R.~M.}\
  \bibnamefont {Venable}}, \bibinfo {author} {\bibfnamefont {H.~L.}\
  \bibnamefont {Woodcock}}, \bibinfo {author} {\bibfnamefont {X.}~\bibnamefont
  {Wu}}, \bibinfo {author} {\bibfnamefont {W.}~\bibnamefont {Yang}}, \bibinfo
  {author} {\bibfnamefont {D.~M.}\ \bibnamefont {York}},\ and\ \bibinfo
  {author} {\bibfnamefont {M.}~\bibnamefont {Karplus}},\ }\bibfield  {title}
  {\enquote {\bibinfo {title} {{{CHARMM}}: {{The}} biomolecular simulation
  program},}\ }\href {https://doi.org/10.1002/jcc.21287} {\bibfield  {journal}
  {\bibinfo  {journal} {J. Comput. Chem.}\ }\textbf {\bibinfo {volume} {30}},\
  \bibinfo {pages} {1545--1614} (\bibinfo {year} {2009})}\BibitemShut {NoStop}%
\bibitem [{\citenamefont {Wu}\ \emph {et~al.}(2014)\citenamefont {Wu},
  \citenamefont {Cheng}, \citenamefont {Jo}, \citenamefont {Rui}, \citenamefont
  {Song}, \citenamefont {{D{\'a}vila-Contreras}}, \citenamefont {Qi},
  \citenamefont {Lee}, \citenamefont {{Monje-Galvan}}, \citenamefont {Venable},
  \citenamefont {Klauda},\ and\ \citenamefont {Im}}]{wu2014CHARMMGUI}%
  \BibitemOpen
  \bibfield  {author} {\bibinfo {author} {\bibfnamefont {E.~L.}\ \bibnamefont
  {Wu}}, \bibinfo {author} {\bibfnamefont {X.}~\bibnamefont {Cheng}}, \bibinfo
  {author} {\bibfnamefont {S.}~\bibnamefont {Jo}}, \bibinfo {author}
  {\bibfnamefont {H.}~\bibnamefont {Rui}}, \bibinfo {author} {\bibfnamefont
  {K.~C.}\ \bibnamefont {Song}}, \bibinfo {author} {\bibfnamefont {E.~M.}\
  \bibnamefont {{D{\'a}vila-Contreras}}}, \bibinfo {author} {\bibfnamefont
  {Y.}~\bibnamefont {Qi}}, \bibinfo {author} {\bibfnamefont {J.}~\bibnamefont
  {Lee}}, \bibinfo {author} {\bibfnamefont {V.}~\bibnamefont {{Monje-Galvan}}},
  \bibinfo {author} {\bibfnamefont {R.~M.}\ \bibnamefont {Venable}}, \bibinfo
  {author} {\bibfnamefont {J.~B.}\ \bibnamefont {Klauda}},\ and\ \bibinfo
  {author} {\bibfnamefont {W.}~\bibnamefont {Im}},\ }\bibfield  {title}
  {\enquote {\bibinfo {title} {{{CHARMM-GUI}} {{{\emph{Membrane Builder}}}}
  toward realistic biological membrane simulations},}\ }\href
  {https://doi.org/10.1002/jcc.23702} {\bibfield  {journal} {\bibinfo
  {journal} {J. Comput. Chem.}\ }\textbf {\bibinfo {volume} {35}},\ \bibinfo
  {pages} {1997--2004} (\bibinfo {year} {2014})}\BibitemShut {NoStop}%
\bibitem [{\citenamefont {Lee}\ \emph {et~al.}(2016)\citenamefont {Lee},
  \citenamefont {Cheng}, \citenamefont {Swails}, \citenamefont {Yeom},
  \citenamefont {Eastman}, \citenamefont {Lemkul}, \citenamefont {Wei},
  \citenamefont {Buckner}, \citenamefont {Jeong}, \citenamefont {Qi},
  \citenamefont {Jo}, \citenamefont {Pande}, \citenamefont {Case},
  \citenamefont {Brooks}, \citenamefont {MacKerell}, \citenamefont {Klauda},\
  and\ \citenamefont {Im}}]{lee2016CHARMMGUI}%
  \BibitemOpen
  \bibfield  {author} {\bibinfo {author} {\bibfnamefont {J.}~\bibnamefont
  {Lee}}, \bibinfo {author} {\bibfnamefont {X.}~\bibnamefont {Cheng}}, \bibinfo
  {author} {\bibfnamefont {J.~M.}\ \bibnamefont {Swails}}, \bibinfo {author}
  {\bibfnamefont {M.~S.}\ \bibnamefont {Yeom}}, \bibinfo {author}
  {\bibfnamefont {P.~K.}\ \bibnamefont {Eastman}}, \bibinfo {author}
  {\bibfnamefont {J.~A.}\ \bibnamefont {Lemkul}}, \bibinfo {author}
  {\bibfnamefont {S.}~\bibnamefont {Wei}}, \bibinfo {author} {\bibfnamefont
  {J.}~\bibnamefont {Buckner}}, \bibinfo {author} {\bibfnamefont {J.~C.}\
  \bibnamefont {Jeong}}, \bibinfo {author} {\bibfnamefont {Y.}~\bibnamefont
  {Qi}}, \bibinfo {author} {\bibfnamefont {S.}~\bibnamefont {Jo}}, \bibinfo
  {author} {\bibfnamefont {V.~S.}\ \bibnamefont {Pande}}, \bibinfo {author}
  {\bibfnamefont {D.~A.}\ \bibnamefont {Case}}, \bibinfo {author}
  {\bibfnamefont {C.~L.}\ \bibnamefont {Brooks}}, \bibinfo {author}
  {\bibfnamefont {A.~D.}\ \bibnamefont {MacKerell}}, \bibinfo {author}
  {\bibfnamefont {J.~B.}\ \bibnamefont {Klauda}},\ and\ \bibinfo {author}
  {\bibfnamefont {W.}~\bibnamefont {Im}},\ }\bibfield  {title} {\enquote
  {\bibinfo {title} {{{CHARMM-GUI Input Generator}} for {{NAMD}}, {{GROMACS}},
  {{AMBER}}, {{OpenMM}}, and {{CHARMM}}/{{OpenMM Simulations Using}} the
  {{CHARMM36 Additive Force Field}}},}\ }\href
  {https://doi.org/10.1021/acs.jctc.5b00935} {\bibfield  {journal} {\bibinfo
  {journal} {J. Chem. Theory Comput.}\ }\textbf {\bibinfo {volume} {12}},\
  \bibinfo {pages} {405--413} (\bibinfo {year} {2016})}\BibitemShut {NoStop}%
\bibitem [{\citenamefont {Huang}\ and\ \citenamefont
  {MacKerell~Jr}(2013)}]{huang2013CHARMM36}%
  \BibitemOpen
  \bibfield  {author} {\bibinfo {author} {\bibfnamefont {J.}~\bibnamefont
  {Huang}}\ and\ \bibinfo {author} {\bibfnamefont {A.~D.}\ \bibnamefont
  {MacKerell~Jr}},\ }\bibfield  {title} {\enquote {\bibinfo {title}
  {{{CHARMM36}} all-atom additive protein force field: {{Validation}} based on
  comparison to {{NMR}} data},}\ }\href {https://doi.org/10.1002/jcc.23354}
  {\bibfield  {journal} {\bibinfo  {journal} {J. Comput. Chem.}\ }\textbf
  {\bibinfo {volume} {34}},\ \bibinfo {pages} {2135--2145} (\bibinfo {year}
  {2013})}\BibitemShut {NoStop}%
\bibitem [{\citenamefont {Jorgensen}\ \emph {et~al.}(1983)\citenamefont
  {Jorgensen}, \citenamefont {Chandrasekhar}, \citenamefont {Madura},
  \citenamefont {Impey},\ and\ \citenamefont
  {Klein}}]{jorgensen1983Comparison}%
  \BibitemOpen
  \bibfield  {author} {\bibinfo {author} {\bibfnamefont {W.~L.}\ \bibnamefont
  {Jorgensen}}, \bibinfo {author} {\bibfnamefont {J.}~\bibnamefont
  {Chandrasekhar}}, \bibinfo {author} {\bibfnamefont {J.~D.}\ \bibnamefont
  {Madura}}, \bibinfo {author} {\bibfnamefont {R.~W.}\ \bibnamefont {Impey}},\
  and\ \bibinfo {author} {\bibfnamefont {M.~L.}\ \bibnamefont {Klein}},\
  }\bibfield  {title} {\enquote {\bibinfo {title} {Comparison of simple
  potential functions for simulating liquid water},}\ }\href
  {https://doi.org/10.1063/1.445869} {\bibfield  {journal} {\bibinfo  {journal}
  {J. Chem. Phys.}\ }\textbf {\bibinfo {volume} {79}},\ \bibinfo {pages}
  {926--935} (\bibinfo {year} {1983})}\BibitemShut {NoStop}%
\bibitem [{\citenamefont {Abraham}\ \emph {et~al.}(2015)\citenamefont
  {Abraham}, \citenamefont {Murtola}, \citenamefont {Schulz}, \citenamefont
  {P{\'a}ll}, \citenamefont {Smith}, \citenamefont {Hess},\ and\ \citenamefont
  {Lindahl}}]{abraham2015GROMACS}%
  \BibitemOpen
  \bibfield  {author} {\bibinfo {author} {\bibfnamefont {M.~J.}\ \bibnamefont
  {Abraham}}, \bibinfo {author} {\bibfnamefont {T.}~\bibnamefont {Murtola}},
  \bibinfo {author} {\bibfnamefont {R.}~\bibnamefont {Schulz}}, \bibinfo
  {author} {\bibfnamefont {S.}~\bibnamefont {P{\'a}ll}}, \bibinfo {author}
  {\bibfnamefont {J.~C.}\ \bibnamefont {Smith}}, \bibinfo {author}
  {\bibfnamefont {B.}~\bibnamefont {Hess}},\ and\ \bibinfo {author}
  {\bibfnamefont {E.}~\bibnamefont {Lindahl}},\ }\bibfield  {title} {\enquote
  {\bibinfo {title} {{{GROMACS}}: {{High}} performance molecular simulations
  through multi-level parallelism from laptops to supercomputers},}\ }\href
  {https://doi.org/10.1016/j.softx.2015.06.001} {\bibfield  {journal} {\bibinfo
   {journal} {SoftwareX}\ }\textbf {\bibinfo {volume} {1--2}},\ \bibinfo
  {pages} {19--25} (\bibinfo {year} {2015})}\BibitemShut {NoStop}%
\bibitem [{\citenamefont {Willard}\ and\ \citenamefont
  {Chandler}(2010)}]{willard2010Instantaneous}%
  \BibitemOpen
  \bibfield  {author} {\bibinfo {author} {\bibfnamefont {A.~P.}\ \bibnamefont
  {Willard}}\ and\ \bibinfo {author} {\bibfnamefont {D.}~\bibnamefont
  {Chandler}},\ }\bibfield  {title} {\enquote {\bibinfo {title} {Instantaneous
  {{Liquid Interfaces}}},}\ }\href {https://doi.org/10.1021/jp909219k}
  {\bibfield  {journal} {\bibinfo  {journal} {J. Phys. Chem. B}\ }\textbf
  {\bibinfo {volume} {114}},\ \bibinfo {pages} {1954--1958} (\bibinfo {year}
  {2010})}\BibitemShut {NoStop}%
\bibitem [{\citenamefont {Luzar}\ and\ \citenamefont
  {Chandler}(1996{\natexlab{a}})}]{luzar1996Effect}%
  \BibitemOpen
  \bibfield  {author} {\bibinfo {author} {\bibfnamefont {A.}~\bibnamefont
  {Luzar}}\ and\ \bibinfo {author} {\bibfnamefont {D.}~\bibnamefont
  {Chandler}},\ }\bibfield  {title} {\enquote {\bibinfo {title} {Effect of
  {{Environment}} on {{Hydrogen Bond Dynamics}} in {{Liquid Water}}},}\ }\href
  {https://doi.org/10.1103/PhysRevLett.76.928} {\bibfield  {journal} {\bibinfo
  {journal} {Phys. Rev. Lett.}\ }\textbf {\bibinfo {volume} {76}},\ \bibinfo
  {pages} {928--931} (\bibinfo {year} {1996}{\natexlab{a}})}\BibitemShut
  {NoStop}%
\bibitem [{\citenamefont {Luzar}\ and\ \citenamefont
  {Chandler}(1996{\natexlab{b}})}]{luzar1996Hydrogenbond}%
  \BibitemOpen
  \bibfield  {author} {\bibinfo {author} {\bibfnamefont {A.}~\bibnamefont
  {Luzar}}\ and\ \bibinfo {author} {\bibfnamefont {D.}~\bibnamefont
  {Chandler}},\ }\bibfield  {title} {\enquote {\bibinfo {title} {Hydrogen-bond
  kinetics in liquid water},}\ }\href {https://doi.org/10.1038/379055a0}
  {\bibfield  {journal} {\bibinfo  {journal} {Nature}\ }\textbf {\bibinfo
  {volume} {379}},\ \bibinfo {pages} {55--57} (\bibinfo {year}
  {1996}{\natexlab{b}})}\BibitemShut {NoStop}%
\bibitem [{\citenamefont {Laage}\ and\ \citenamefont
  {Hynes}(2006)}]{laage2006Molecular}%
  \BibitemOpen
  \bibfield  {author} {\bibinfo {author} {\bibfnamefont {D.}~\bibnamefont
  {Laage}}\ and\ \bibinfo {author} {\bibfnamefont {J.~T.}\ \bibnamefont
  {Hynes}},\ }\bibfield  {title} {\enquote {\bibinfo {title} {A {{Molecular
  Jump Mechanism}} of {{Water Reorientation}}},}\ }\href
  {https://doi.org/10.1126/science.1122154} {\bibfield  {journal} {\bibinfo
  {journal} {Science}\ }\textbf {\bibinfo {volume} {311}},\ \bibinfo {pages}
  {832--835} (\bibinfo {year} {2006})}\BibitemShut {NoStop}%
\bibitem [{\citenamefont {Kumar}, \citenamefont {Schmidt},\ and\ \citenamefont
  {Skinner}(2007)}]{kumar2007Hydrogen}%
  \BibitemOpen
  \bibfield  {author} {\bibinfo {author} {\bibfnamefont {R.}~\bibnamefont
  {Kumar}}, \bibinfo {author} {\bibfnamefont {J.~R.}\ \bibnamefont {Schmidt}},\
  and\ \bibinfo {author} {\bibfnamefont {J.~L.}\ \bibnamefont {Skinner}},\
  }\bibfield  {title} {\enquote {\bibinfo {title} {Hydrogen bonding definitions
  and dynamics in liquid water},}\ }\href {https://doi.org/10.1063/1.2742385}
  {\bibfield  {journal} {\bibinfo  {journal} {J. Chem. Phys.}\ }\textbf
  {\bibinfo {volume} {126}},\ \bibinfo {pages} {204107} (\bibinfo {year}
  {2007})}\BibitemShut {NoStop}%
\bibitem [{\citenamefont {Kikutsuji}, \citenamefont {Kim},\ and\ \citenamefont
  {Matubayasi}(2018)}]{kikutsuji2018How}%
  \BibitemOpen
  \bibfield  {author} {\bibinfo {author} {\bibfnamefont {T.}~\bibnamefont
  {Kikutsuji}}, \bibinfo {author} {\bibfnamefont {K.}~\bibnamefont {Kim}},\
  and\ \bibinfo {author} {\bibfnamefont {N.}~\bibnamefont {Matubayasi}},\
  }\bibfield  {title} {\enquote {\bibinfo {title} {How do hydrogen bonds break
  in supercooled water?: {{Detecting}} pathways not going through saddle point
  of two-dimensional potential of mean force},}\ }\href
  {https://doi.org/10.1063/1.5033419} {\bibfield  {journal} {\bibinfo
  {journal} {J. Chem. Phys.}\ }\textbf {\bibinfo {volume} {148}},\ \bibinfo
  {pages} {244501} (\bibinfo {year} {2018})}\BibitemShut {NoStop}%
\bibitem [{\citenamefont {Kikutsuji}, \citenamefont {Kim},\ and\ \citenamefont
  {Matubayasi}(2019)}]{kikutsuji2019Consistency}%
  \BibitemOpen
  \bibfield  {author} {\bibinfo {author} {\bibfnamefont {T.}~\bibnamefont
  {Kikutsuji}}, \bibinfo {author} {\bibfnamefont {K.}~\bibnamefont {Kim}},\
  and\ \bibinfo {author} {\bibfnamefont {N.}~\bibnamefont {Matubayasi}},\
  }\bibfield  {title} {\enquote {\bibinfo {title} {Consistency of geometrical
  definitions of hydrogen bonds based on the two-dimensional potential of mean
  force with respect to the time correlation in liquid water over a wide range
  of temperatures},}\ }\href {https://doi.org/10.1016/j.molliq.2019.111603}
  {\bibfield  {journal} {\bibinfo  {journal} {J. Mol. Liq.}\ }\textbf {\bibinfo
  {volume} {294}},\ \bibinfo {pages} {111603} (\bibinfo {year}
  {2019})}\BibitemShut {NoStop}%
\bibitem [{\citenamefont {Kikutsuji}, \citenamefont {Kim},\ and\ \citenamefont
  {Matubayasi}(2021)}]{kikutsuji2021Transition}%
  \BibitemOpen
  \bibfield  {author} {\bibinfo {author} {\bibfnamefont {T.}~\bibnamefont
  {Kikutsuji}}, \bibinfo {author} {\bibfnamefont {K.}~\bibnamefont {Kim}},\
  and\ \bibinfo {author} {\bibfnamefont {N.}~\bibnamefont {Matubayasi}},\
  }\bibfield  {title} {\enquote {\bibinfo {title} {Transition pathway of
  hydrogen bond switching in supercooled water analyzed by the {{Markov}} state
  model},}\ }\href {https://doi.org/10.1063/5.0055531} {\bibfield  {journal}
  {\bibinfo  {journal} {J. Chem. Phys.}\ }\textbf {\bibinfo {volume} {154}},\
  \bibinfo {pages} {234501} (\bibinfo {year} {2021})}\BibitemShut {NoStop}%
\bibitem [{\citenamefont {Shikata}\ \emph {et~al.}(2023)\citenamefont
  {Shikata}, \citenamefont {Kikutsuji}, \citenamefont {Yasoshima},
  \citenamefont {Kim},\ and\ \citenamefont
  {Matubayasi}}]{shikata2023Revealing}%
  \BibitemOpen
  \bibfield  {author} {\bibinfo {author} {\bibfnamefont {K.}~\bibnamefont
  {Shikata}}, \bibinfo {author} {\bibfnamefont {T.}~\bibnamefont {Kikutsuji}},
  \bibinfo {author} {\bibfnamefont {N.}~\bibnamefont {Yasoshima}}, \bibinfo
  {author} {\bibfnamefont {K.}~\bibnamefont {Kim}},\ and\ \bibinfo {author}
  {\bibfnamefont {N.}~\bibnamefont {Matubayasi}},\ }\bibfield  {title}
  {\enquote {\bibinfo {title} {Revealing the hidden dynamics of confined water
  in acrylate polymers: {{Insights}} from hydrogen-bond lifetime analysis},}\
  }\href {https://doi.org/10.1063/5.0148753} {\bibfield  {journal} {\bibinfo
  {journal} {J. Chem. Phys.}\ }\textbf {\bibinfo {volume} {158}},\ \bibinfo
  {pages} {174901} (\bibinfo {year} {2023})}\BibitemShut {NoStop}%
\bibitem [{\citenamefont {Rapaport}(1983)}]{rapaport1983Hydrogen}%
  \BibitemOpen
  \bibfield  {author} {\bibinfo {author} {\bibfnamefont {D.}~\bibnamefont
  {Rapaport}},\ }\bibfield  {title} {\enquote {\bibinfo {title} {Hydrogen bonds
  in water: {{Network}} organization and lifetimes},}\ }\href
  {https://doi.org/10.1080/00268978300102931} {\bibfield  {journal} {\bibinfo
  {journal} {Mol. Phys.}\ }\textbf {\bibinfo {volume} {50}},\ \bibinfo {pages}
  {1151--1162} (\bibinfo {year} {1983})}\BibitemShut {NoStop}%
\bibitem [{\citenamefont {Efron}(1992)}]{efron1992Bootstrap}%
  \BibitemOpen
  \bibfield  {author} {\bibinfo {author} {\bibfnamefont {B.}~\bibnamefont
  {Efron}},\ }\bibfield  {title} {\enquote {\bibinfo {title} {Bootstrap
  {{Methods}}: {{Another Look}} at the {{Jackknife}}},}\ }in\ \href
  {https://doi.org/10.1007/978-1-4612-4380-9_41} {\emph {\bibinfo {booktitle}
  {Breakthroughs in {{Statistics}}}}},\ \bibinfo {editor} {edited by\ \bibinfo
  {editor} {\bibfnamefont {S.}~\bibnamefont {Kotz}}\ and\ \bibinfo {editor}
  {\bibfnamefont {N.~L.}\ \bibnamefont {Johnson}}}\ (\bibinfo  {publisher}
  {Springer},\ \bibinfo {address} {{New York, NY}},\ \bibinfo {year} {1992})\
  pp.\ \bibinfo {pages} {569--593}\BibitemShut {NoStop}%
\end{thebibliography}
%

\clearpage
\widetext

\setcounter{equation}{0}
\setcounter{figure}{0}
\setcounter{table}{0}
\setcounter{page}{1}

\renewcommand{\theequation}{S.\arabic{equation}}
\renewcommand{\thefigure}{S\arabic{figure}}
\renewcommand{\thetable}{S\arabic{table}}
\renewcommand{\bibnumfmt}[1]{[S#1]}
\renewcommand{\citenumfont}[1]{S#1}

\noindent{\bf\Large Supplementary Material}
\vspace{5mm}
\begin{center}
\textbf{\large  Influence of cholesterol on hydrogen-bond dynamics of water molecules in lipid-bilayer systems at varying temperatures}
\\

\vspace{5mm}
 
{Kokoro Shikata, Kento Kasahara, Nozomi Morishita Watanabe, Hiroshi Umakoshi, Kang Kim,
and Nobuyuki Matubayasi}
\\

\vspace{5mm}

\noindent
\textit{Division of Chemical Engineering, Department of Materials Engineering Science, Graduate School of Engineering Science, Osaka University, Toyonaka, Osaka 560-8531, Japan}

\end{center}

\begin{table}[H]
  \caption{Equilibration scheme.}
  \label{tab:equil}
  \begin{center}
    \begin{tabular}{ccccccc} 
      \toprule
      No. & process & $dt$~[fs] &time~[ns] & integrater & $k_z~[\mr{kJ/mol\,nm^2}]$ & $k_{\mr{dih}} ~\mr{[kJ/mol\,rad^2]}$
      \\ \midrule
       1& Energy Minimization & - &5000~(steps)& steepest descent &1000 & 1000 \\
       2& $NVT$ & 1 & 0.125 &Leap-Flog &1000 & 1000 \\
       3& $NVT$ & 1 & 0.125 &Leap-Flog &400  & 400 \\
       4& $NPT$ & 1 & 0.125 &Leap-Flog &400  & 200 \\
       5& $NPT$ & 2 & 0.5   &Leap-Flog & 200 & 200 \\
       6& $NPT$ & 2 & 0.5   &Leap-Flog & 40  & 100 \\
       7& $NPT$ & 2 & 0.5   &Leap-Flog & 0   & 0 \\
       8& $NPT$ & 2 & 10   &Leap-Flog & 0   & 0 \\
       9& $NPT$ & 2 & 100   &Leap-Flog & 0   & 0 \\
       10& $NPT$ & 2 & 500   &Leap-Flog & 0   & 0 \\
       11& $NPT$ & 2 & 3000   &Leap-Flog & 0   & 0 \\
       12& Production ($NPT$) & 2 & 10   &Leap-Flog & 0   & 0 \\
       \bottomrule
    \end{tabular}
  \end{center}
\end{table}

\newpage

\begin{figure}[H]
  \centering
  \includegraphics[width=0.75\textwidth]{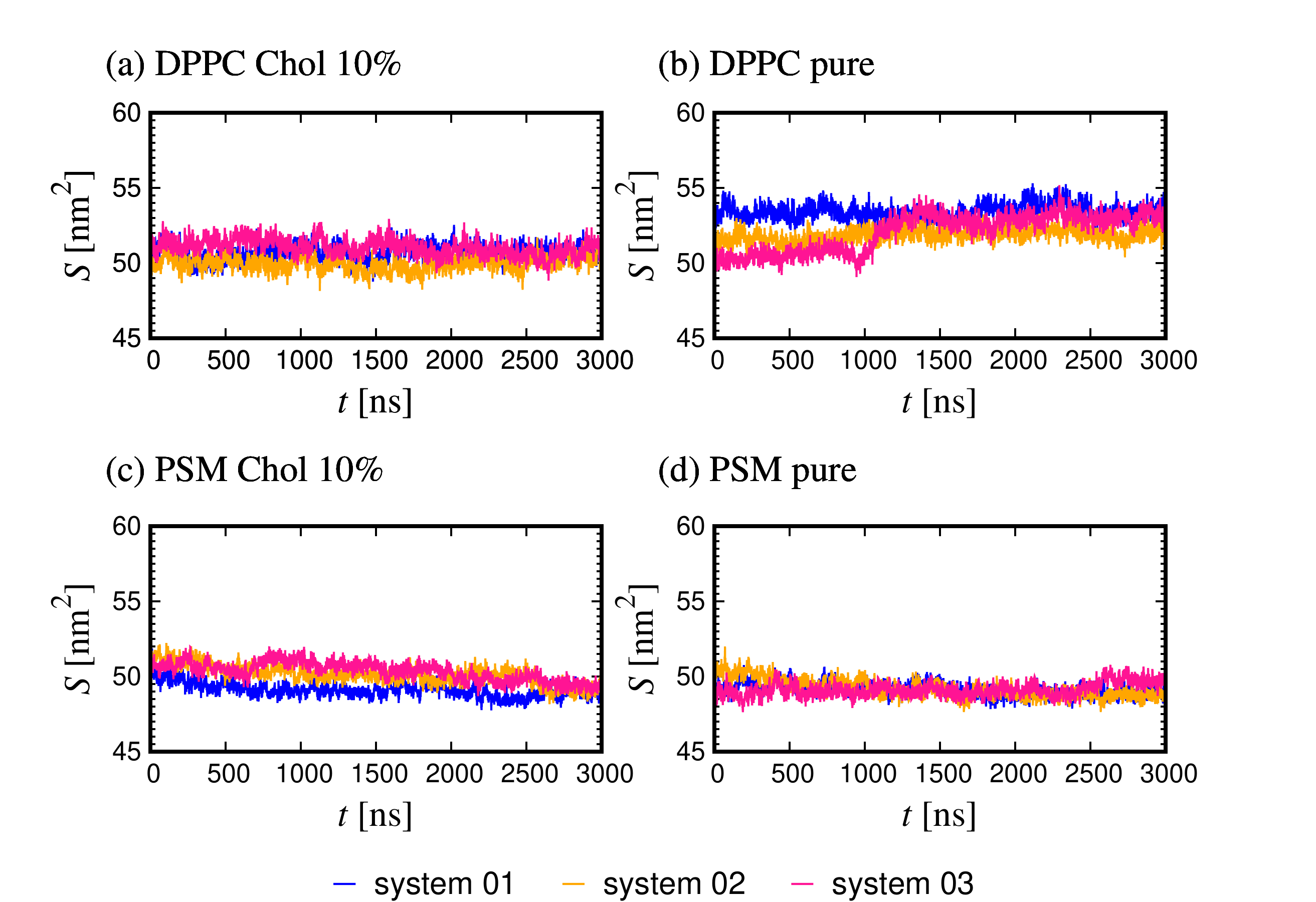}
  \caption{Time evolution of surface area $S$ in the $x$-$y$ plane during
 the equilibration at 303K. (a) and
 (b): DPPC; (c) and (d): PSM.
(a) and (b) refer to the systems with Chol, while (c) and (d) 
 are for those without Chol.
Systems 01, 02, and 03 correspond to three distinct runs with different initial configurations.
}
  \label{fig:xy303}
\end{figure}

\begin{figure}[H]
  \centering
  \includegraphics[width=0.75\textwidth]{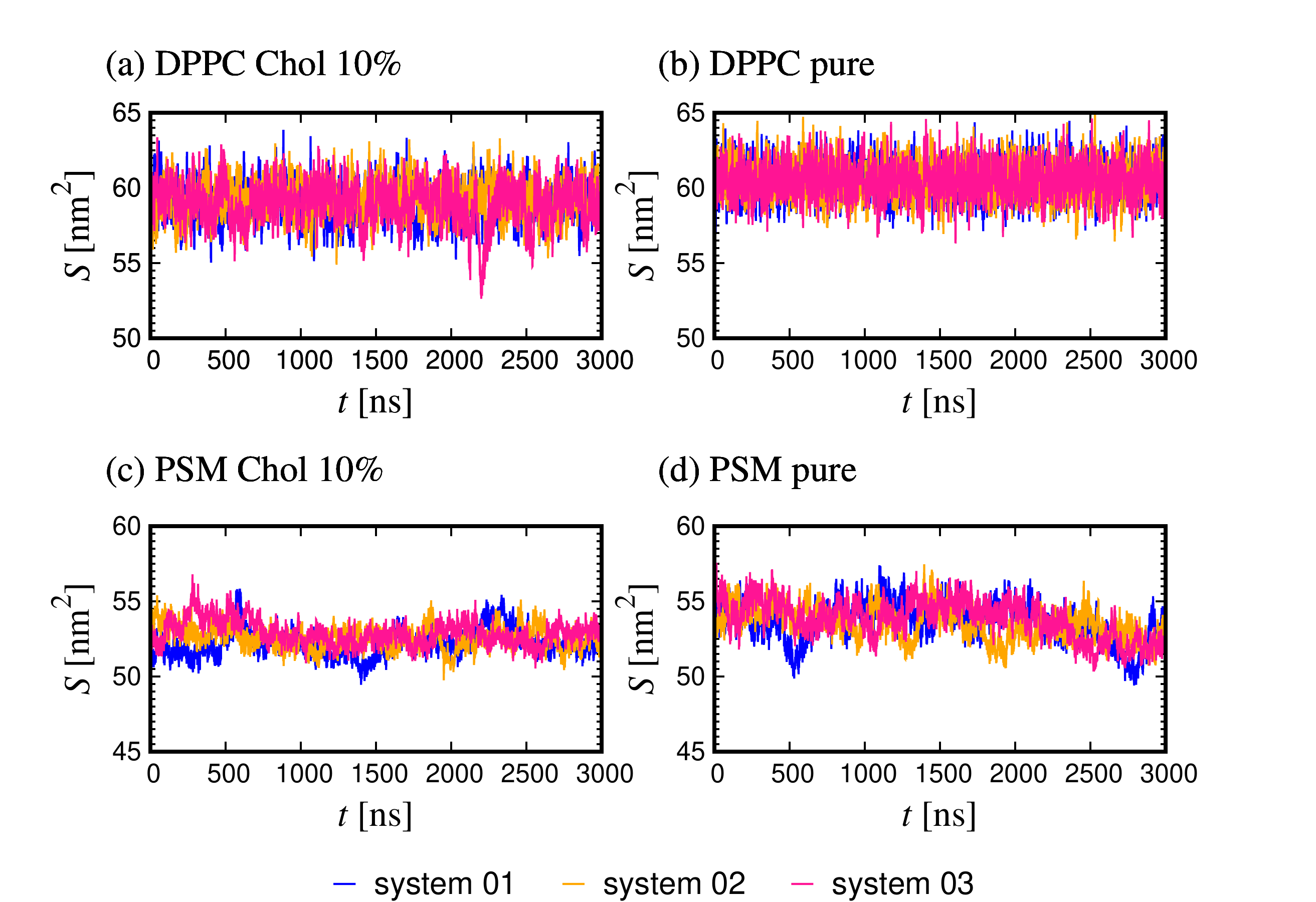}
  \caption{Time evolution of surface area $S$ in the $x$-$y$ plane 
 during the equilibration at 323K. (a) and
 (b): DPPC; (c) and (d): PSM.
(a) and (b) refer to the systems with Chol, while (c) and (d) 
 are for those without Chol.
Systems 01, 02, and 03 correspond to three distinct runs with different initial configurations.
}
  \label{fig:xy323}
\end{figure}

\begin{figure}[H]
  \centering
  \includegraphics[width=0.7\textwidth]{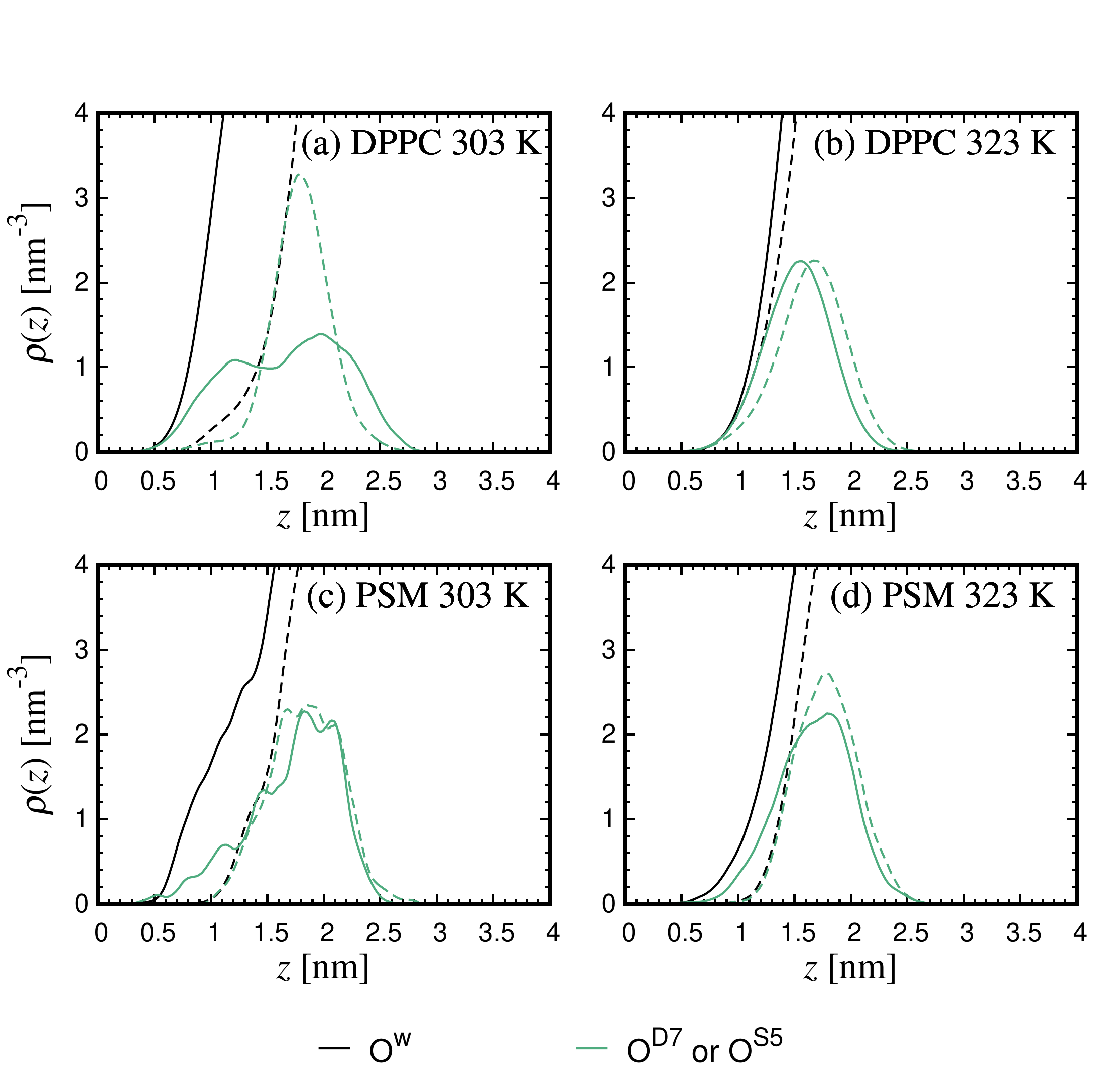}
    \caption{Number density distributions of O$^{\text{D}7}$ in DPPC
 and  O$^{\text{S}5}$ in PSM along the
 $z$-direction. 
For comparison, number density distributions of the water oxygen atoms
 (O$^\text{w}$) are plotted, which are same as those in Fig.~2.
Solid lines represent pure membrane systems, while dashed
 lines represent systems containing Chol. (a) and (b) refer 
 to DPPC, and (c) and (d) refer to PSM.}
  \label{fig:zp_sup2}
\end{figure}

\begin{figure}[H]
  \centering
  \includegraphics[width=0.7\textwidth]{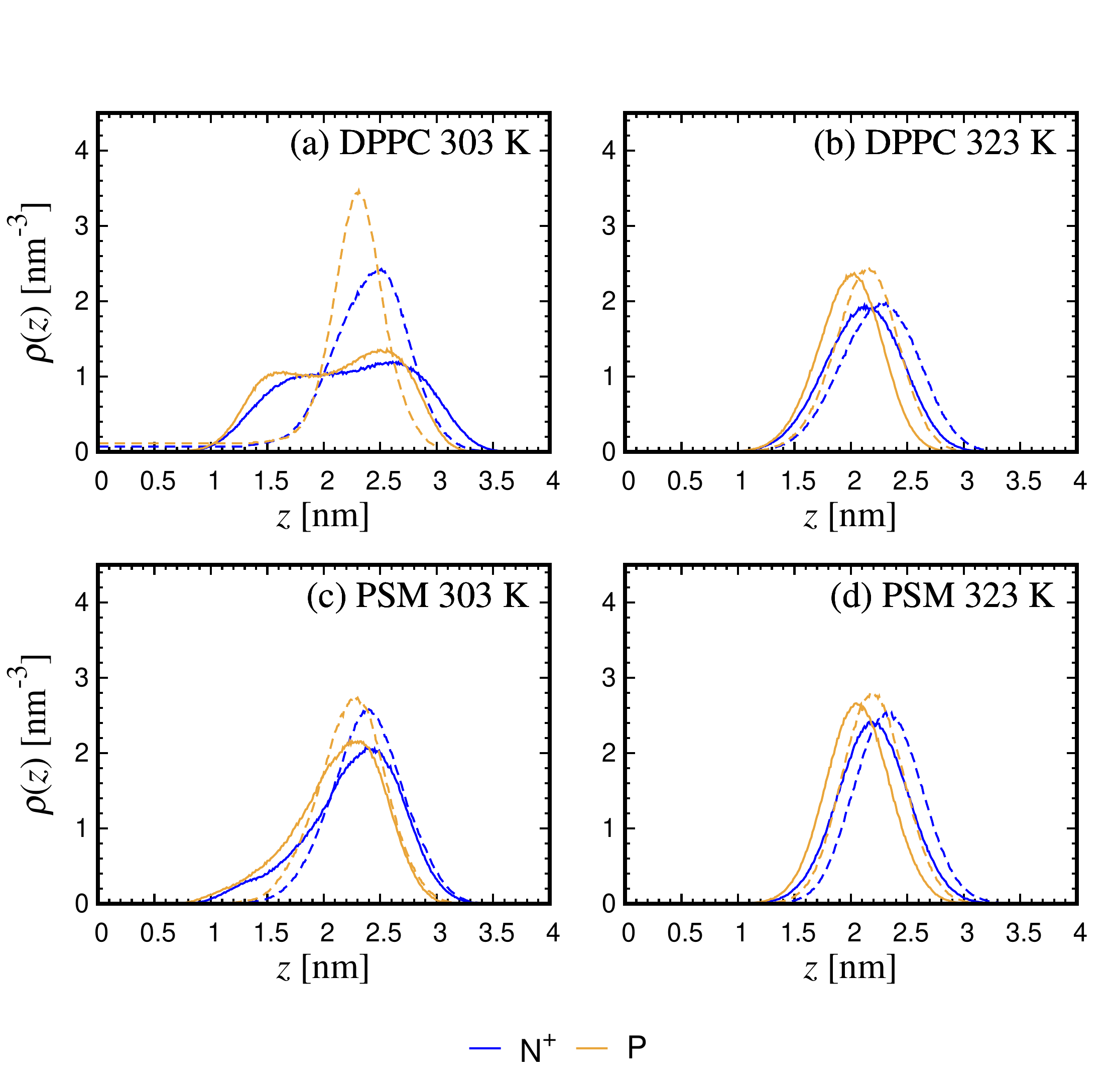}
    \caption{Number density distributions of nitrogen (N$^+$), and phosphorus (P) atoms along the
 $z$-direction. Solid lines represent pure membrane systems, while dashed
 lines represent systems containing Chol. (a) and (b) 
 refer to DPPC, and (c) and (d) refer to PSM.}
  \label{fig:zp_sup}
\end{figure}

\begin{figure}[H]
  \centering
  \includegraphics[width=0.5\textwidth]{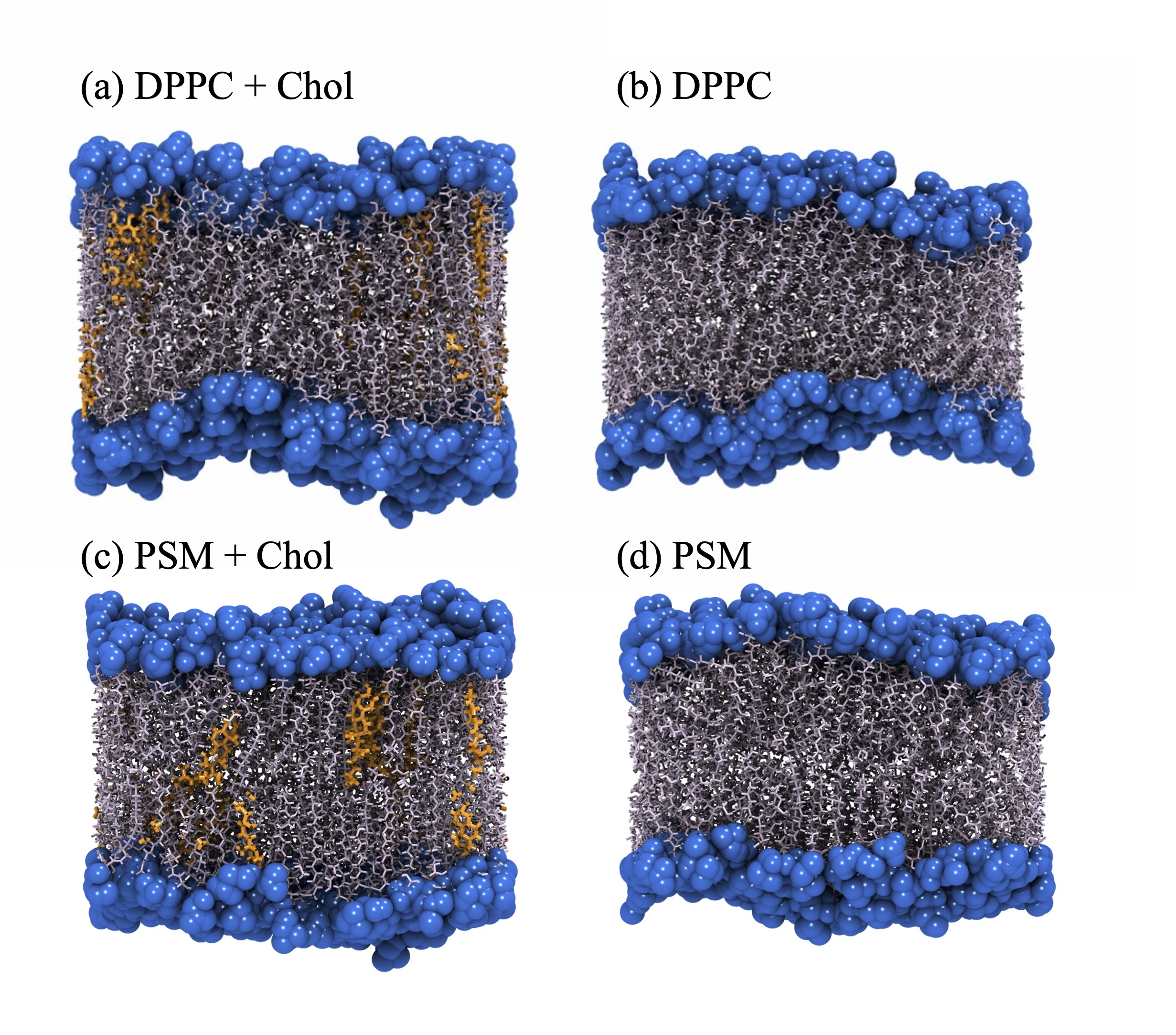}
  \caption{Snapshots at 323 K for (a) DPPC with Chol, (b) DPPC without
 Chol, (c) PSM with Chol, and (d) PSM without Chol.}
  \label{fig:allsnap}
\end{figure}

\begin{figure}[H]
  \centering
  \includegraphics[width=0.75\textwidth]{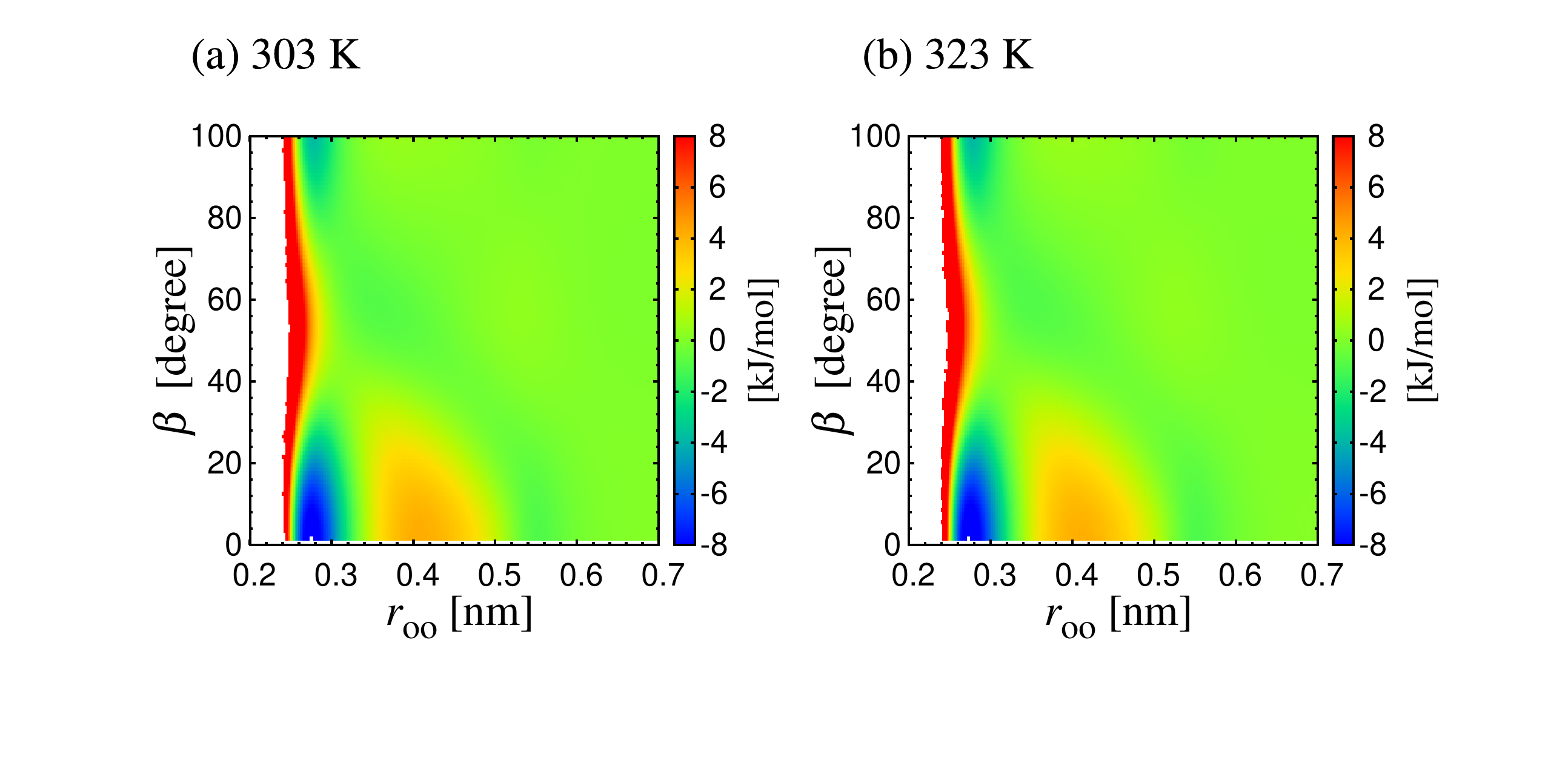}
  \caption{2D PMF, $W(r_\mr{oo},\beta)$, between water molecules
 in bulk water at 1 g/cm$^3$ at (a) 303 K and (b) 323 K.}
  \label{fig:pmfwater}
\end{figure}

\begin{figure}[H]
  \centering
  \includegraphics[width=1.0\textwidth]{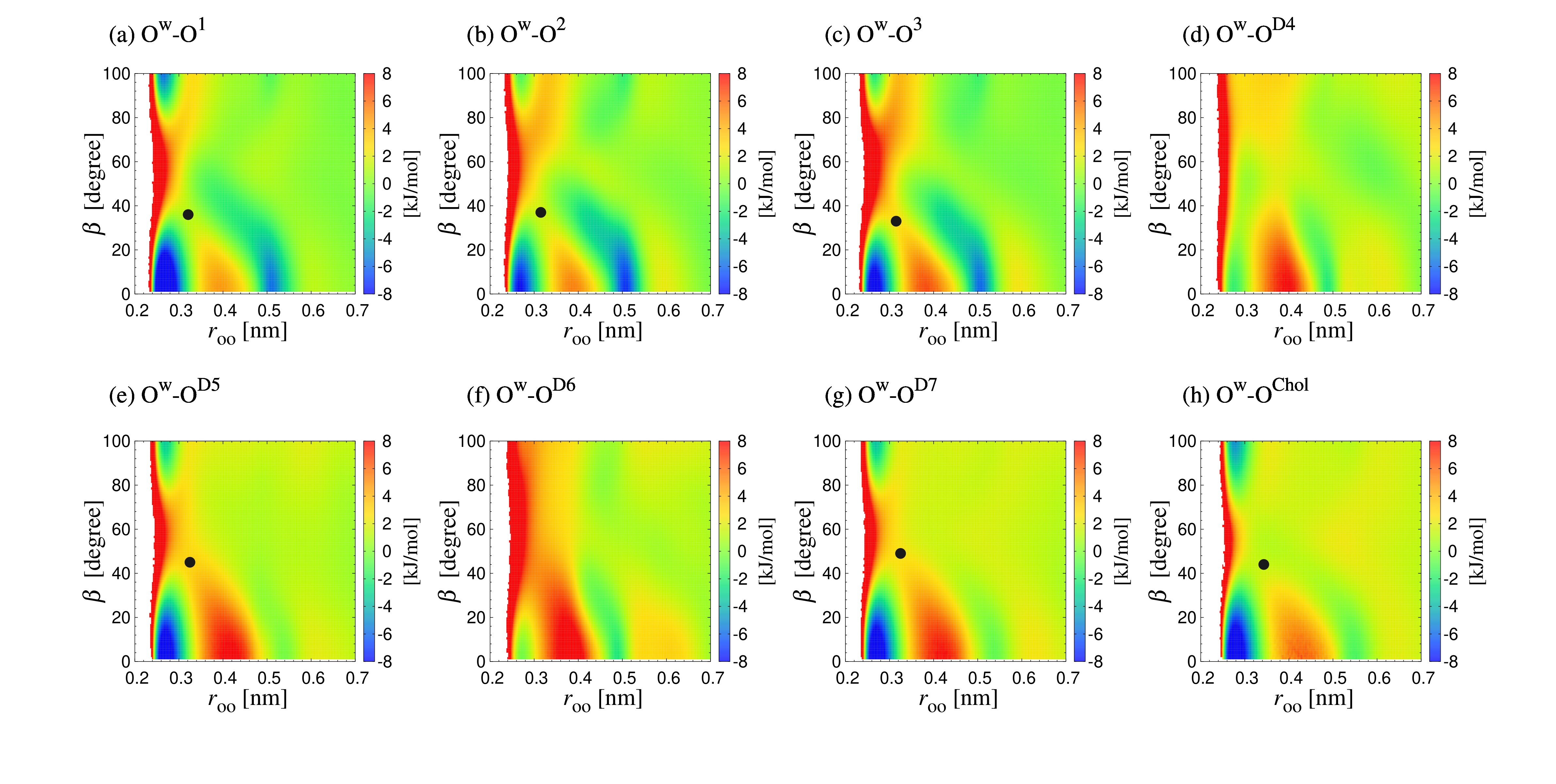}
  \caption{2D PMF, $W(r_\mr{oo},\beta)$,
 between water oxygen (\ow) and each acceptor oxygen in DPPC
 with Chol at 323 K.
Black points represent saddle points.}
  \label{fig:pmfdppc_323chl}
\end{figure}

\begin{figure}[H]
  \centering
  \includegraphics[width=1.0\textwidth]{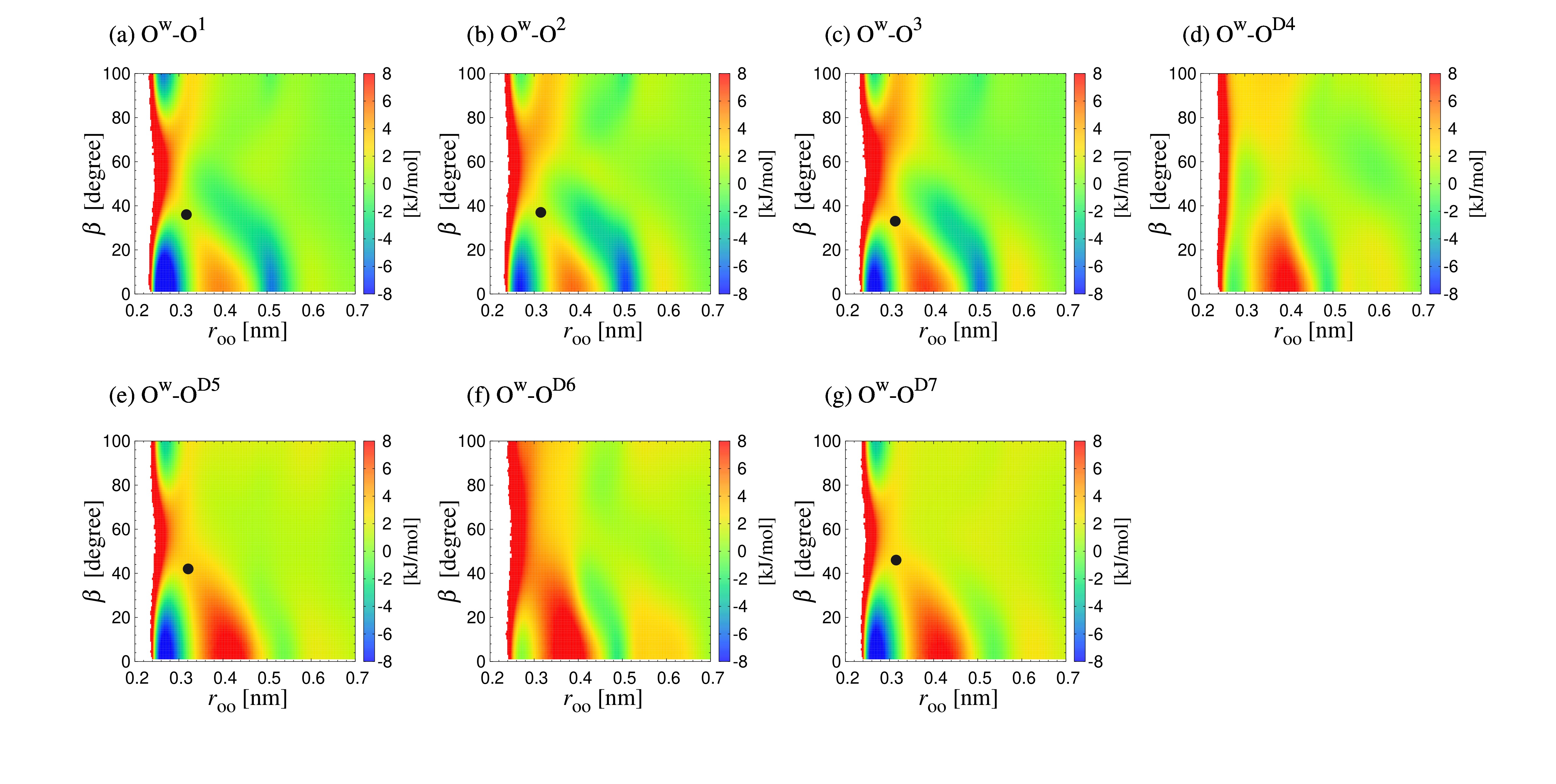}
  \caption{2D PMF, $W(r_\mr{oo},\beta)$,
 between water oxygen (\ow) and each acceptor oxygen in DPPC
 without Chol at 303 K.
Black points represent saddle points.}
  \label{fig:pmfdppc_303pure}
\end{figure}

\begin{figure}[H]
  \centering
  \includegraphics[width=1.0\textwidth]{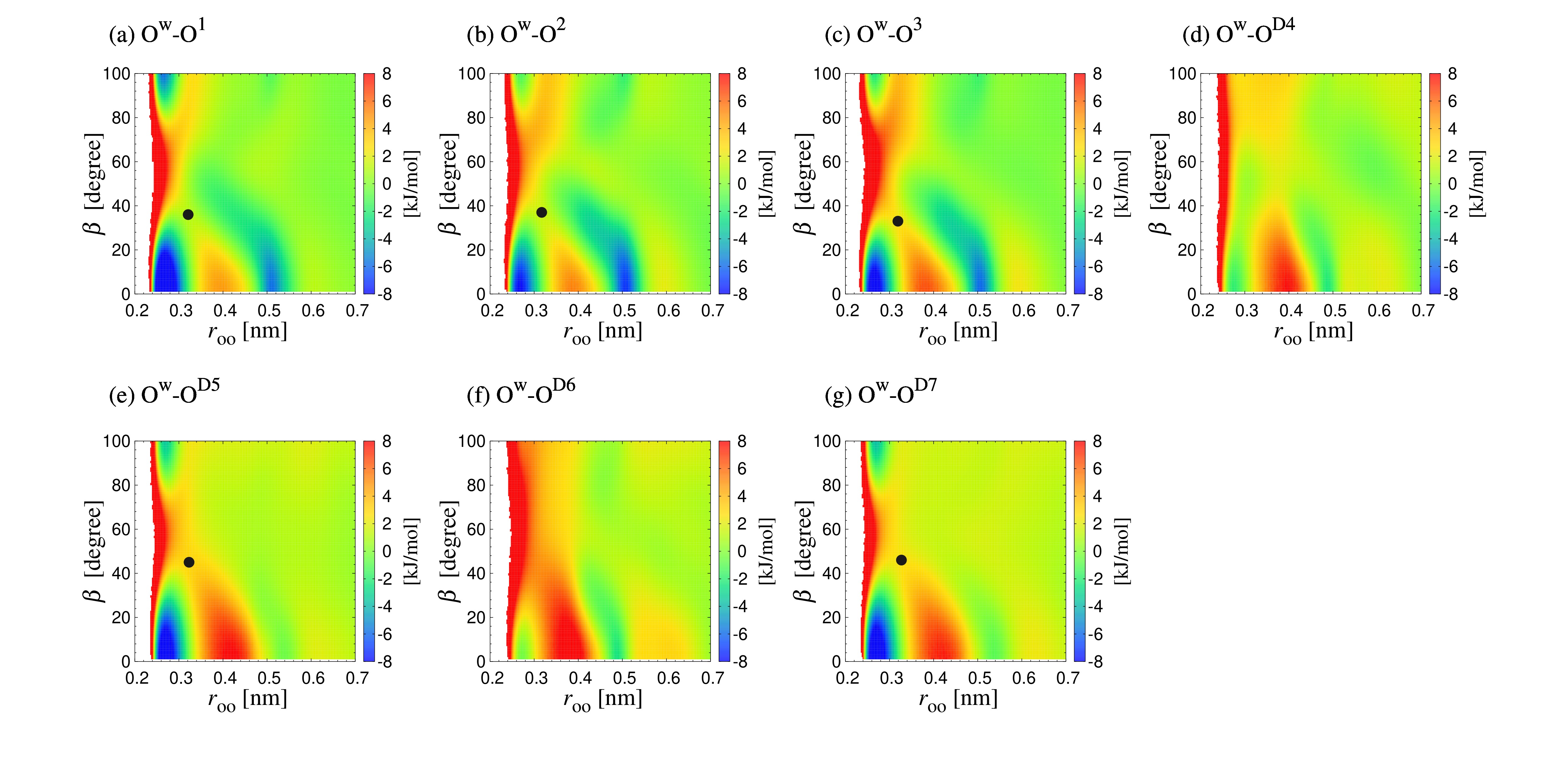}
  \caption{2D PMF, $W(r_\mr{oo},\beta)$,
 between water oxygen (\ow) and each acceptor oxygen in DPPC
 without Chol at 323 K.
Black points represent saddle points.}
  \label{fig:pmfdppc_323pure}
\end{figure}

\begin{figure}[H]
  \centering
  \includegraphics[width=1.0\textwidth]{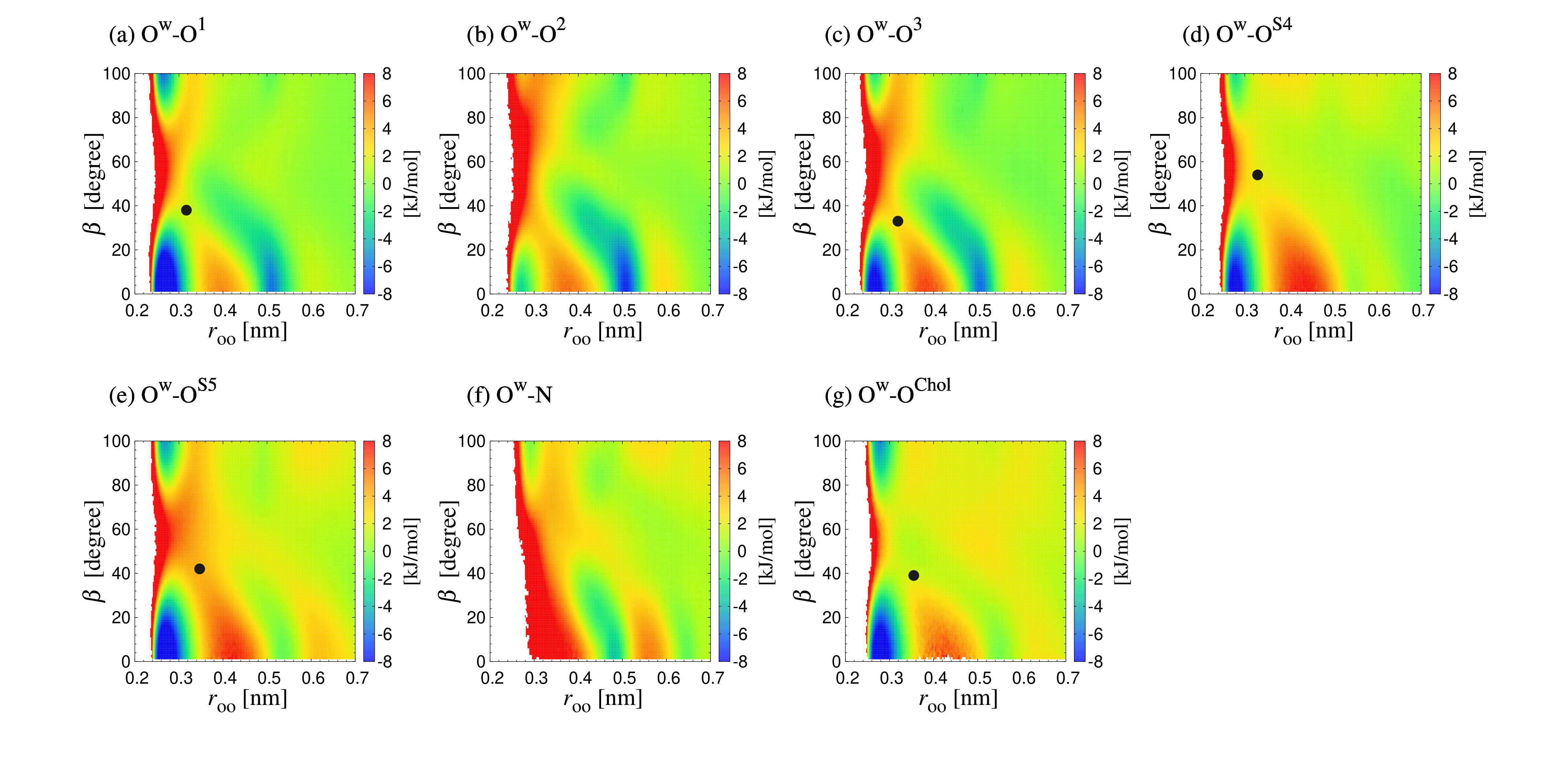}
  \caption{2D PMF, $W(r_\mr{oo},\beta)$,
 between water oxygen (\ow) and each acceptor oxygen in PSM
 with Chol at 323 K.
Black points represent saddle points.}
  \label{fig:pmfpsm_323chl}
\end{figure}

\begin{figure}[H]
  \centering
  \includegraphics[width=1.0\textwidth]{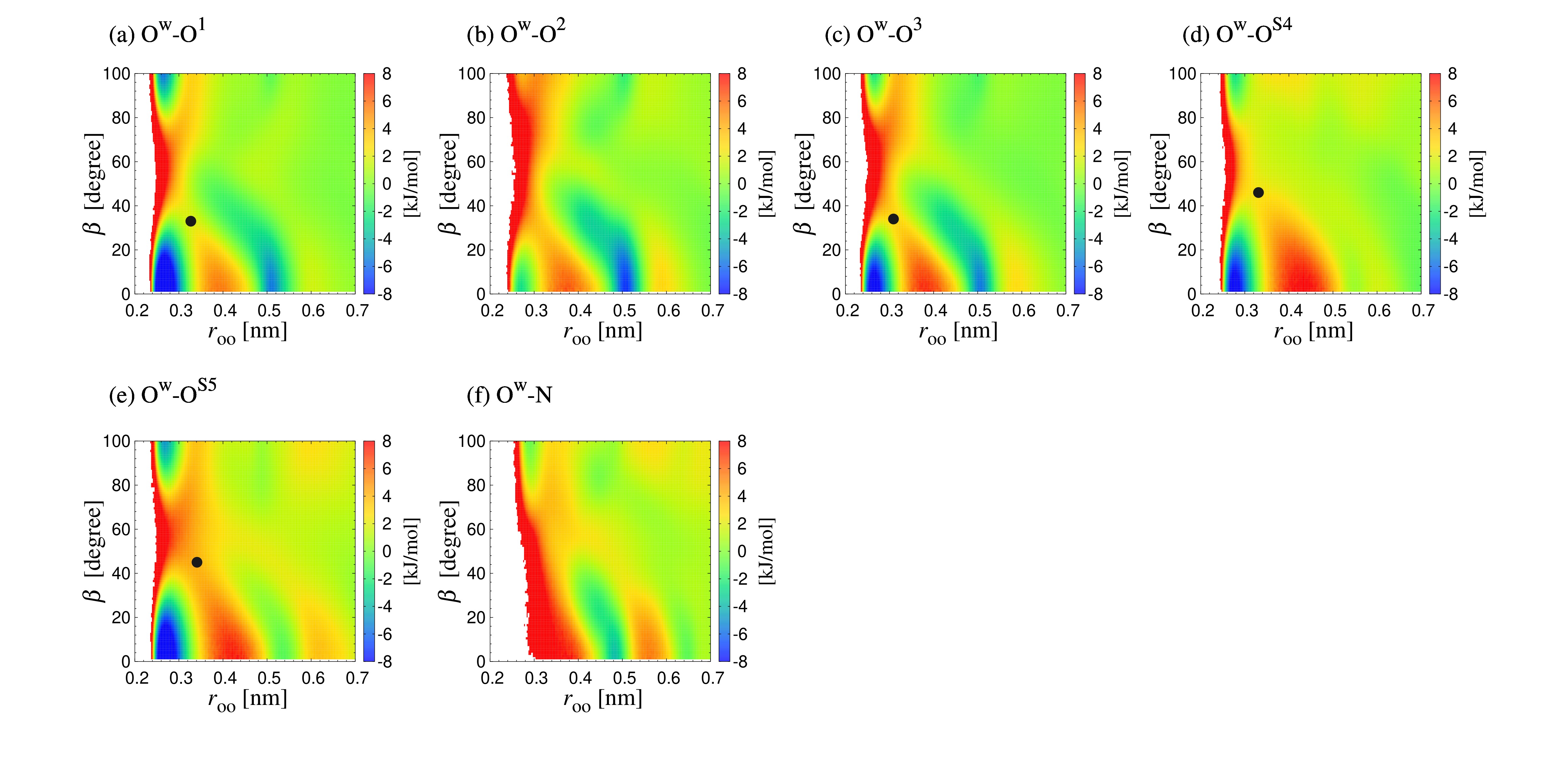}
  \caption{2D PMF, $W(r_\mr{oo},\beta)$,
 between water oxygen (\ow) and each acceptor oxygen in PSM
 without Chol at 303 K.}
  \label{fig:pmfpsm_303pure}
\end{figure}

\begin{figure}[H]
  \centering
  \includegraphics[width=1.0\textwidth]{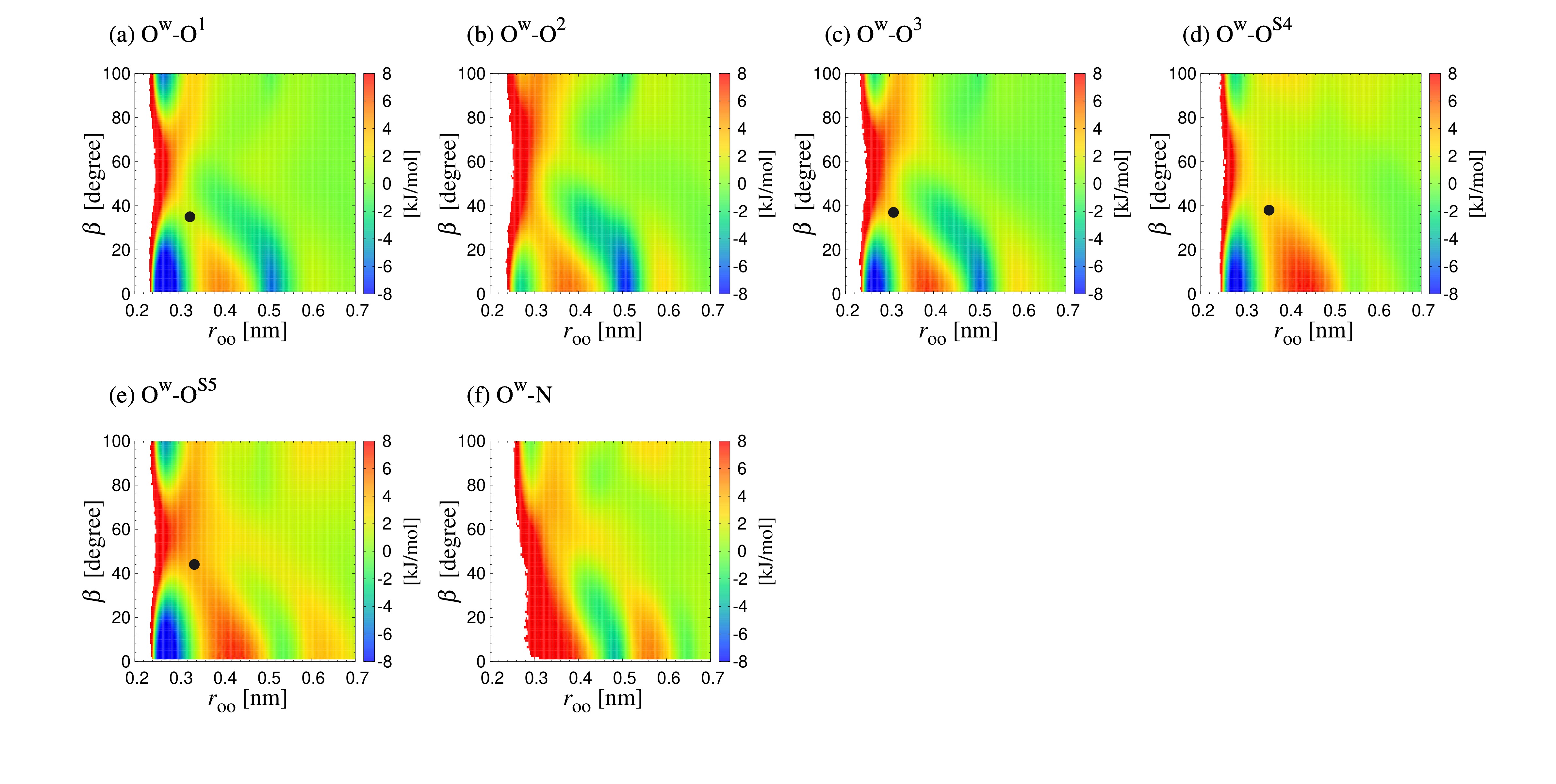}
  \caption{2D PMF, $W(r_\mr{oo},\beta)$,
 between water oxygen (\ow) and each acceptor oxygen in PSM
 without Chol at 323 K.
Black points represent saddle points.}
  \label{fig:pmfpsm_323pure}
\end{figure}

\begin{figure}[H]
  \centering
  \includegraphics[width=0.75\textwidth]{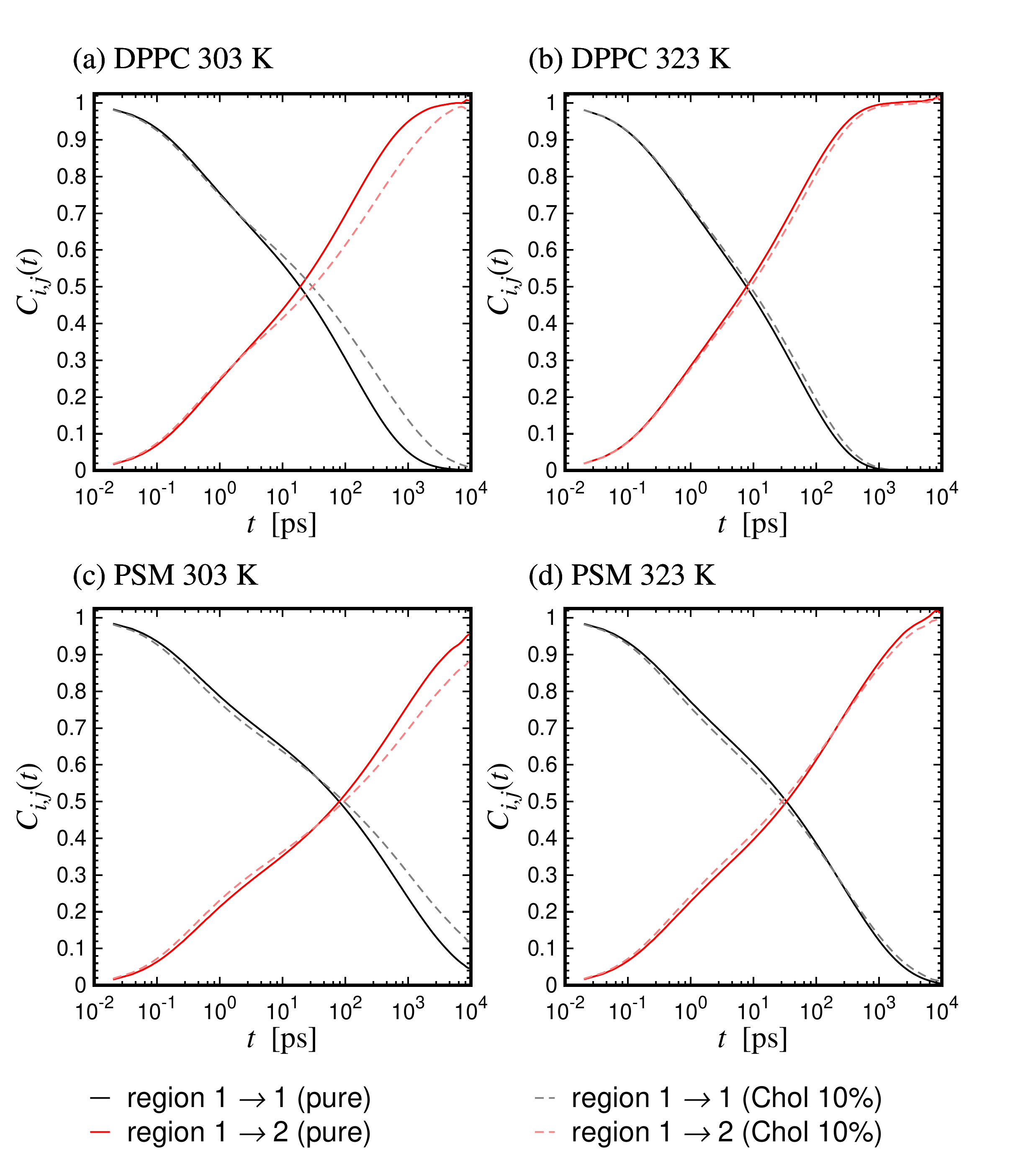}
  \caption{Conditional probability $C_{1,j}(t)$, representing 
transition dynamics from region $1$ at the initial time $t=0$ to 
 region 2 
or remaining within the
same region 1 during the time interval $t$ [(a) DPPC at 303 K, (b) DPPC at 323 K, (c)
 PSM at 303 K, and (d) PSM at 323 K].}
  \label{fig:tcf1}
\end{figure}

\begin{figure}[H]
  \centering
  \includegraphics[width=1.0\textwidth]{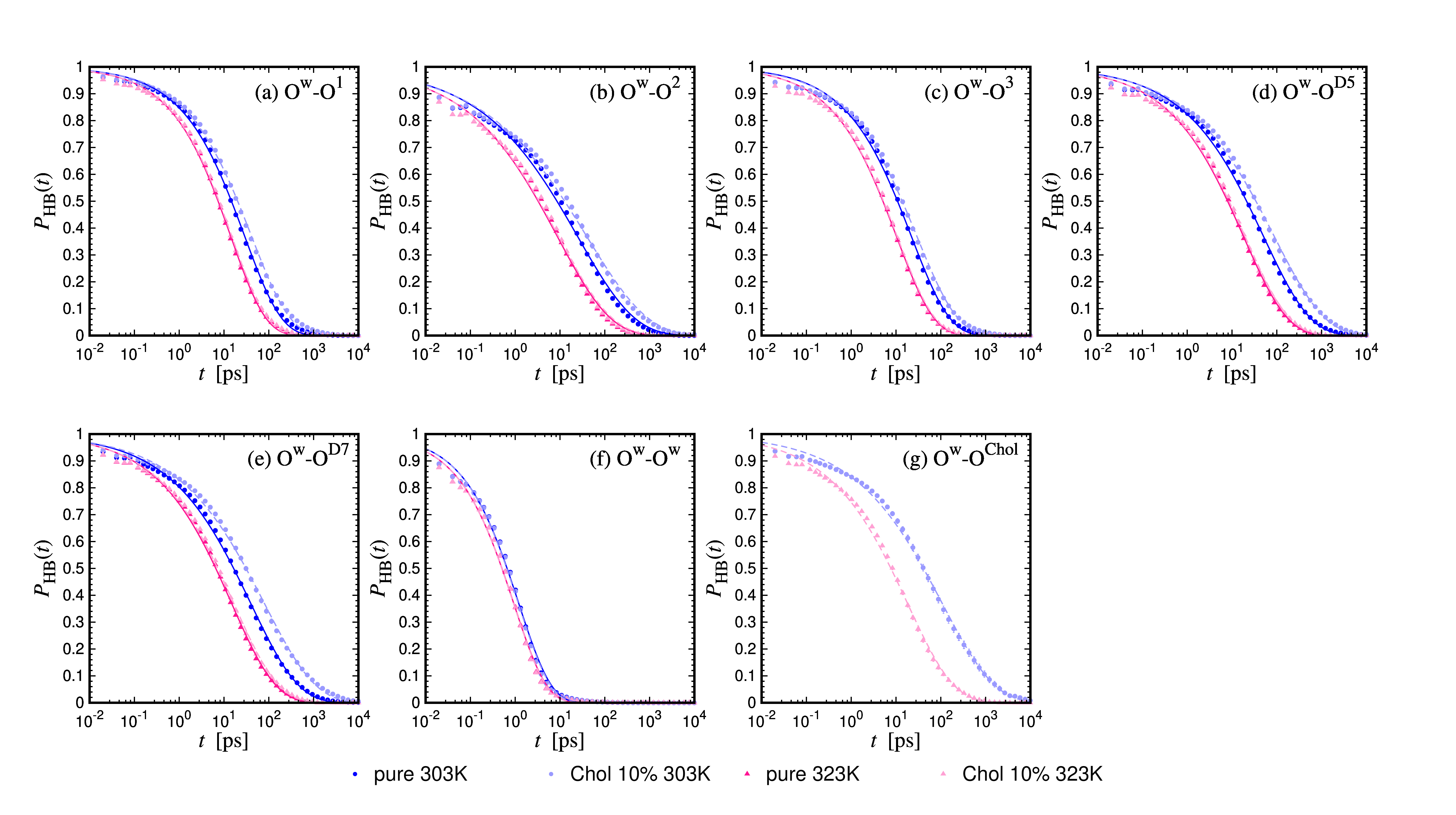}
  \caption{H-bond time correlation function $\phb(t)$ with respect to
 acceptor oxygens in the DPPC systems. 
The solid line represents the result of fitting with the KWW function, $\phb(t)\simeq \exp[-(t/\tau\kww)^{\beta\kww}]$.}
  \label{fig:phbdppc}
\end{figure}

\begin{figure}[H]
  \centering
  \includegraphics[width=0.75\textwidth]{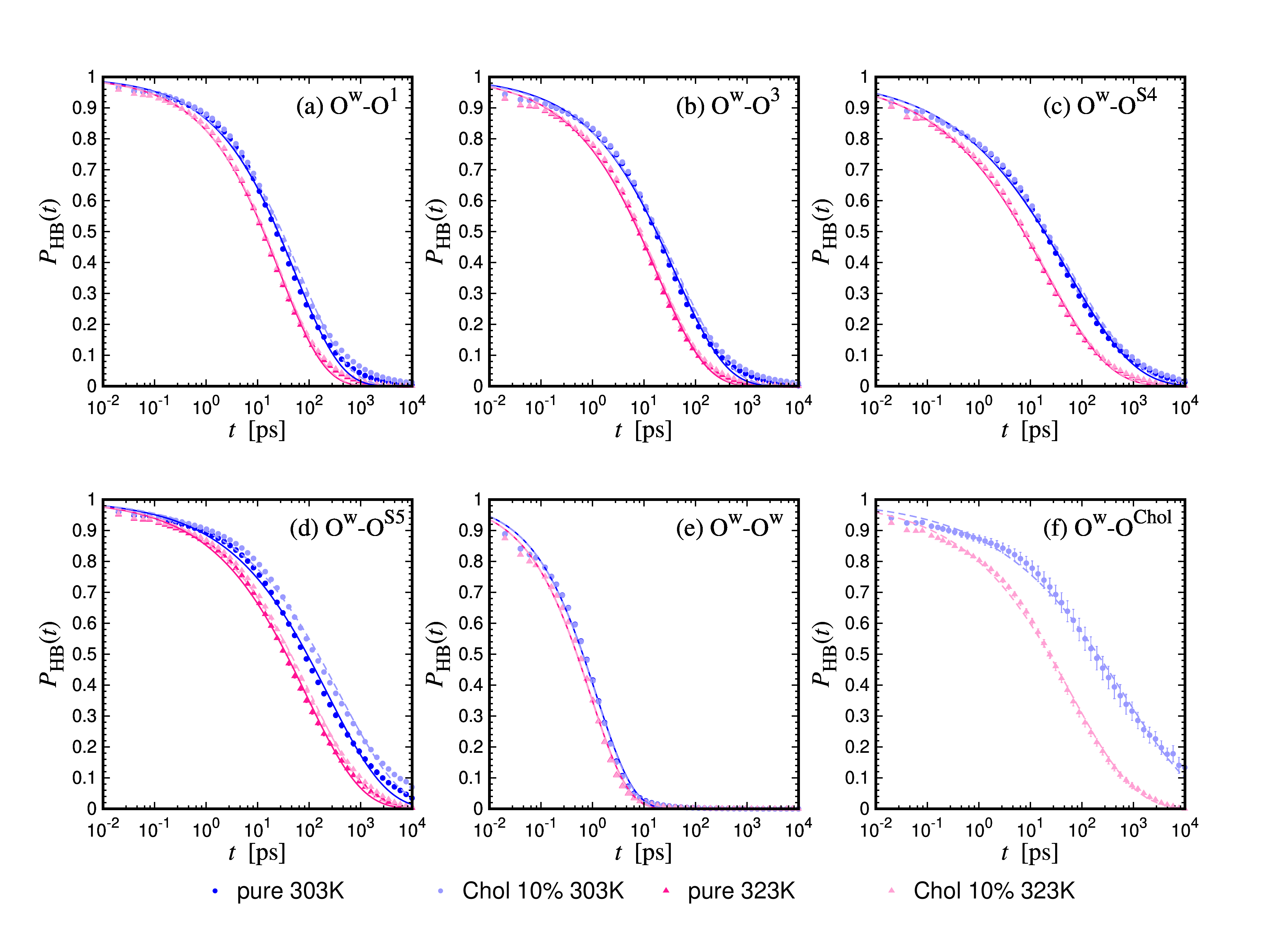}
  \caption{H-bond time correlation function $\phb(t)$ with respect to
 acceptor oxygens in the PSM systems. 
The solid line represents the result of fitting with the KWW function, $\phb(t)\simeq \exp[-(t/\tau\kww)^{\beta\kww}]$.}
  \label{fig:phbpsm}
\end{figure}

\begin{table}[H]
  \caption{$\tau\kww$ of acceptor oxygen atoms in the DPPC systems
 with and without Chol.
 The error is provided at the standard error and is not shown when it is smaller than $0.01$ ps.}
  \label{tab:taukww_dppc}
  \begin{center}
    \begin{tabular}{ccccc} 
      \toprule
      & \multicolumn{2}{c}{303~K~[ps]}& \multicolumn{2}{c}{323~K~[ps]}\\ 
      \cmidrule(lr){2-3}\cmidrule(lr){4-5}
      & \multicolumn{1}{c}{pure} & \multicolumn{1}{c}{Chol 10~\%}& \multicolumn{1}{c}{pure} & \multicolumn{1}{c}{Chol 10~\%}\\
      \midrule
      \oxy{   w} &   $   1.21 $  &$   1.18 $  &$  0.94 $  &$  0.93 $ \\
      \oxy{   1} &   $   29.28 \pm 0.02 $  &$   41.7 \pm      0.2 $ & $    {14.27} $ & $    {15.07\pm 0.01} $ \\
      \oxy{   2} &   $   24.97 \pm 0.04 $  &$   40.0 \pm      0.3 $ & $   8.81 $ & $   9.35 $ \\
      \oxy{   3} &   $   22.26 \pm 0.02 $  &$   29.6 \pm      0.1 $ & $    {10.32} $ & $    {10.70} $ \\
      \oxy{  D5} &   $   51.1 \pm      0.1 $ & $   88.7 \pm      0.5 $ & $    {18.20\pm0.02} $  &$    {19.74 \pm 0.02} $ \\
      \oxy{  D7} &   $   40.2 \pm      0.1 $ & $   81.1 \pm      0.4 $ & $    {15.01\pm0.01} $  &$    {16.74 \pm 0.02} $ \\
      \oxy{Chol} &     \multicolumn{1}{c}{-}  &  $    110 \pm        1 $ &   \multicolumn{1}{c}{-}  &  $   18.3 \pm      0.1 $ \\
             \bottomrule
    \end{tabular}
  \end{center}
\end{table}

\begin{table}[H]
  \caption{$\beta\kww$ of acceptor oxygen atoms in the DPPC systems
 with and without Chol.
 The error is not shown since it is smaller than $0.01$.}
  \label{tab:betakww_dppc}
  \begin{center}
    \begin{tabular}{ccccc} 
      \toprule
      & \multicolumn{2}{c}{303~K}& \multicolumn{2}{c}{323~K}\\ 
      \cmidrule(lr){2-3}\cmidrule(lr){4-5}
      & \multicolumn{1}{c}{pure} & \multicolumn{1}{c}{Chol 10~\%}& \multicolumn{1}{c}{pure} & \multicolumn{1}{c}{Chol 10~\%}\\
      \midrule
      \oxy{   w} &   $  0.60 $  &$  0.61 $ & $  0.60 $  &$  0.60 $ \\
      \oxy{   1} &   $  0.53 $ & $  0.51 $ & $  0.56 $  &$  0.55 $ \\
      \oxy{   2} &   $  0.34 $ & $  0.32 $ & $  0.37 $ & $  0.37 $ \\
      \oxy{   3} &   $  0.51 $ & $  0.48 $ & $  0.52 $ & $  0.52 $ \\
      \oxy{  D5} &   $  0.42 $ & $  0.39 $ & $  0.45 $  &$  0.44 $ \\
      \oxy{  D7} &   $  0.41 $ & $  0.38 $ & $  0.44 $  &$  0.44 $ \\
      \oxy{Chol} &     \multicolumn{1}{c}{-}  &  $  0.37 $ &   \multicolumn{1}{c}{-}  &  $  0.42 $ \\      
       \bottomrule
    \end{tabular}
  \end{center}
\end{table}

\begin{table}[H]
  \caption{$\tau\hb$ of acceptor oxygen atoms in the DPPC systems
 with and without Chol.
 The error is provided at the standard error and is not shown when it is smaller than $0.01$ ps.}
  \label{tab:tauhb_dppc}
  \begin{center}
    \begin{tabular}{ccccc} 
      \toprule
      & \multicolumn{2}{c}{303~K~[ps]}& \multicolumn{2}{c}{323~K~[ps]}\\ 
      \cmidrule(lr){2-3}\cmidrule(lr){4-5}
      & \multicolumn{1}{c}{pure} & \multicolumn{1}{c}{Chol 10~\%}& \multicolumn{1}{c}{pure} & \multicolumn{1}{c}{Chol 10~\%}\\
      \midrule
      \oxy{   w} &   $   1.81 $  &$   1.76 $  &$   1.42 $  &$   1.40 $ \\
      \oxy{   1} &   $   52.5 \pm      0.5 $ & $   81 \pm      4 $ & $   23.8 \pm      0.2 $ & $   25.4 \pm      0.3 $ \\
      \oxy{   2} &   $    135 \pm        3 $ & $    280 \pm       30 $ & $   37.3 \pm      0.5 $ & $   40.9 \pm      0.5 $ \\
      \oxy{   3} &   $   43.0 \pm      0.7 $ & $   64 \pm      3 $ & $   19.2 \pm      0.2 $ & $   20.1 \pm      0.2 $ \\
      \oxy{  D5} &   $    151 \pm        3 $ & $    330 \pm       30 $ & $   46.1 \pm      0.5 $ & $   51.4 \pm      0.7 $ \\
      \oxy{  D7} &   $    125 \pm        3 $ & $    310 \pm       20 $ & $   38.4 \pm      0.4 $ & $   44.4 \pm      0.5 $ \\
      \oxy{Chol} &     \multicolumn{1}{c}{-}  &  $    450 \pm       60 $ &   \multicolumn{1}{c}{-}  &  $   53\pm      3 $ \\
       \bottomrule
    \end{tabular}
  \end{center}
\end{table}

\newpage

\begin{table}[H]
  \caption{$\tau\kww$ of acceptor oxygen atoms in the PSM systems
 with and without Chol.
 The error is provided at the standard error and is not shown when it is smaller than $0.01$ ps.}
  \label{tab:taukww_psm}
  \begin{center}
    \begin{tabular}{ccccc} 
      \toprule
      & \multicolumn{2}{c}{303~K~[ps]}& \multicolumn{2}{c}{323~K~[ps]}\\ 
      \cmidrule(lr){2-3}\cmidrule(lr){4-5}
      & \multicolumn{1}{c}{pure} & \multicolumn{1}{c}{Chol 10~\%}& \multicolumn{1}{c}{pure} & \multicolumn{1}{c}{Chol 10~\%}\\
      \midrule
      \oxy{   w} &   $   1.19 $  &$   1.18 $  &$  0.92 $  &$  0.91 $ \\
      \oxy{   1} &   $   53.2 \pm      0.1 $ & $   65.8 \pm      0.2 $ & $   27.1 \pm      0.1 $ & $    {28.41\pm0.02} $ \\
      \oxy{   3} &   $   39.9 \pm      0.1 $ & $   45.3 \pm      0.1 $ & $    {17.68\pm0.03} $  &$    {19.14\pm0.03} $ \\
      \oxy{  S4} &   $   55.1 \pm      0.3 $ & $   62.9 \pm      0.4 $ & $   20.3 \pm      0.1 $ & $    {21.08\pm0.03} $ \\
      \oxy{  S5} &   $    236 \pm        1 $ & $    400 \pm        2 $ & $   88.5 \pm      0.4 $ & $     {109.4\pm0.4} $ \\
      \oxy{Chol} &     \multicolumn{1}{c}{-}  &  $    670 \pm       20 $ &   \multicolumn{1}{c}{-}  &  $   63.9 \pm      0.7 $ \\
      \bottomrule
    \end{tabular}
  \end{center}
\end{table}

\begin{table}[H]
  \caption{$\beta\kww$ of acceptor oxygen atoms in the PSM systems
 with and without Chol.
 The error is not shown since it is smaller than $0.01$.}
  \label{tab:betakww_psm}
  \begin{center}
    \begin{tabular}{ccccc} 
      \toprule
      & \multicolumn{2}{c}{303~K}& \multicolumn{2}{c}{323~K}\\ 
      \cmidrule(lr){2-3}\cmidrule(lr){4-5}
      & \multicolumn{1}{c}{pure} & \multicolumn{1}{c}{Chol 10~\%}& \multicolumn{1}{c}{pure} & \multicolumn{1}{c}{Chol 10~\%}\\
      \midrule
      \oxy{   w} &   $  0.61 $  &$  0.61 $  &$  0.60 $  &$  0.60 $ \\
      \oxy{   1} &   $  0.48 $ & $  0.46 $ & $  0.50 $  & $ 0.50 $ \\
      \oxy{   3} &   $  0.44 $ & $  0.42 $ & $  0.46 $ & $  0.45 $ \\
      \oxy{  S4} &   $  0.34 $ & $  0.33 $ & $  0.36 $ & $  0.36 $ \\
      \oxy{  S5} &   $  0.39 $ & $  0.37 $ & $  0.41 $ & $  0.41 $ \\
      \oxy{Chol} &     \multicolumn{1}{c}{-}  &  $  0.31 $ &   \multicolumn{1}{c}{-}  &  $  0.36 $ \\
      \bottomrule
    \end{tabular}
  \end{center}
\end{table}

\begin{table}[H]
  \caption{$\tau\hb$ of acceptor oxygen atoms in the PSM systems
 with and without Chol.
 The error is provided at the standard error and is not shown when it is smaller than $0.01$ ps.}
  \label{tab:tauhb_psm}
  \begin{center}
    \begin{tabular}{ccccc} 
      \toprule
      & \multicolumn{2}{c}{303~K~[ps]}& \multicolumn{2}{c}{323~K~[ps]}\\ 
      \cmidrule(lr){2-3}\cmidrule(lr){4-5}
      & \multicolumn{1}{c}{pure} & \multicolumn{1}{c}{Chol 10~\%}& \multicolumn{1}{c}{pure} & \multicolumn{1}{c}{Chol 10~\%}\\
      \midrule
      \oxy{   w} &   $   1.77 $  &$   1.74 $  &$   1.38 $  &$   1.36 $ \\
      \oxy{   1} &   $    113 \pm        4 $ & $    153 \pm        7 $ & $   54\pm      2 $ & $   57.3 \pm      0.7 $ \\
      \oxy{   3} &   $    104 \pm        2 $ & $    129 \pm        6 $ & $   43 \pm      1 $ & $   47 \pm      1 $ \\
      \oxy{  S4} &   $    320 \pm       30 $ & $    390 \pm       30 $ & $   95 \pm      6 $ & $   98 \pm      2 $ \\
      \oxy{  S5} &   $    860 \pm       80 $ & $   1600 \pm      100 $ & $    280 \pm       20 $ & $    350 \pm       20 $ \\
      \oxy{Chol} &     \multicolumn{1}{c}{-}  &  $   6000 \pm     1000 $ &   \multicolumn{1}{c}{-}  &  $    290 \pm       30 $ \\
      \bottomrule
    \end{tabular}
  \end{center}
\end{table}

\begin{figure}[H]
  \centering
  \includegraphics[width=0.75\textwidth]{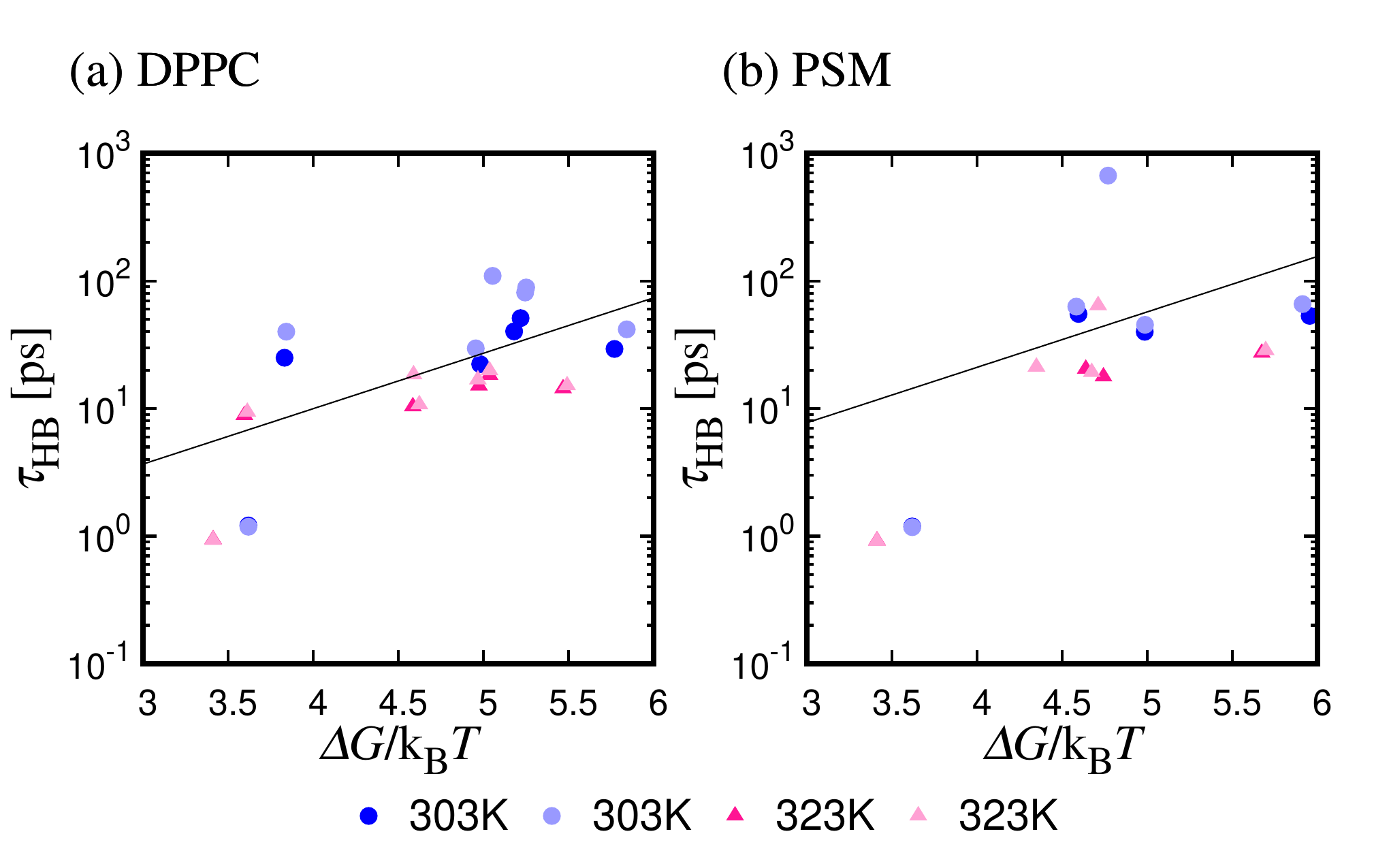}
  \caption{$\tau_\mathrm{HB}$ vs. $\Delta G/k_\mathrm{B}T$ for DPPC (a)
 and PSM (b) systems.
The activation energy $\Delta G$ is determined by the free energy
 difference between the most stable and saddle points on the 2D PMF,
 $W(r_\mathrm{OO}, \beta)$ (see Figs. 5 and 6, and Figs. S7-S12). 
The straight line represents the Arrhenius equation, $\tau_\mathrm{HB} =
 A\exp(\Delta G/k_\mathrm{B}T)$, where $A$ is determined by fitting.
The values of $A$ are 0.183 ps and 0.385 ps for  DPPC and PSM systems, respectively.
  }
  \label{fig:energy}
\end{figure}

\end{document}